\documentclass[utf8]{pepe} 
\usepackage{url,hyperref,lineno,microtype,subcaption,amsmath,bm}
\usepackage[onehalfspacing]{setspace}
\usepackage{color}

\newcommand{\m}{\phantom{-}}
\newcommand{\EDM}{\hat{\boldsymbol{D}}}
\newcommand{\bmr}{\boldsymbol{r}}
\newcommand{\bmx}{\boldsymbol{x}}
\newcommand{\bmK}{\boldsymbol{K}}
\newcommand{\bmP}{\boldsymbol{P}}
\newcommand{\bmJ}{\boldsymbol{J}}

\newcommand{\bmk}{\boldsymbol{k}}
\newcommand{\bmp}{\boldsymbol{p}}
\newcommand{\bmq}{\boldsymbol{q}}

\def\sone{\boldsymbol{\sigma}_1}
\def\stwo{\boldsymbol{\sigma}_2}
\def\stre{\boldsymbol{\sigma}_3}
\def\tone{\vec\tau_1}
\def\ttwo{\vec\tau_2}
\def\ttre{\vec\tau_3}
\def\omk{\omega_k}
\def\Lx{\Lambda_{\chi}}
\def\tri{{{}^3{\rm H}}}
\def\hel{{{}^3{\rm He}}}
\newcommand{\bra}{\langle}
\newcommand{\ket}{\rangle}
\newcommand{\slashed}[1]{#1 \llap{/\kern-0.5pt}}
\newcommand{\bmsi}{\boldsymbol{\sigma}}
\newcommand{\bmna}{\boldsymbol{\nabla}}
\newcommand{\bmnalr}{\overleftrightarrow{\boldsymbol{\nabla}}}
\newcommand{\bmta}{\vec\tau}

\newcommand{\bs}{\boldsymbol}

\definecolor {Blue}                {rgb}{.9,.9,.99}

\definecolor {Cyan}                {rgb}{0.,0.255,0.255}

\def\keyFont{\fontsize{8}{11}\helveticabold }
\def\firstAuthorLast{J.~de~Vries {et~al.}} %use et al only if is more than 1 author
\def\Authors{
  J.~de~Vries\,$^{1,2}$,
  E.~Epelbaum\,$^{3}$,
  L.~Girlanda\,$^{4,5}$,
  A.~Gnech\,$^{6}$,
  E.~Mereghetti\,$^{7}$,
  M.~Viviani\,$^{8,*}$,
}
\def\Address{${}^1$ Amherst Center for Fundamental Interactions, Department of Physics, University of Massachusetts, Amherst, MA, USA\\
${}^2$ RIKEN BNL Research Center, Brookhaven National Laboratory,
  Upton, New York, NY, USA\\
${}^3$ Ruhr-Universit\"at Bochum, Fakult\"at für Physik und Astronomie, Institut f\"ur
  Theoretische Physik II, Bochum, Germany\\
${}^4$  Department of Mathematics and Physics, University of Salento,
  Lecce, Italy \\
${}^5$ INFN-Lecce,  Lecce, Italy \\  
${}^6$ Gran Sasso Science Institute, L'Aquila, Italy\\
$^{7}$ Theoretical Division, Los Alamos National Laboratory, Los
  Alamos, NM, USA\\
  ${}^8$  INFN-Pisa, Pisa, Italy
  }
\def\corrAuthor{M. Viviani\\
INFN-Pisa, Largo B. Pontecorvo 3, I-56127 Pisa, Italy}
\def\corrEmail{michele.viviani@pi.infn.it}

\begin{document}
\onecolumn
\firstpage{1}

%\title[PV and TV Interactions]{Parity-violating nuclear forces
%  in the presence or absence of time-reversal violation} 

\title[PV and TV Interactions]{Parity- and time-reversal-violating nuclear forces} 

\author[\firstAuthorLast ]{\Authors} %This field will be automatically populated
\address{} %This field will be automatically populated
\correspondance{} %This field will be automatically populated

\extraAuth{}% If there are more than 1 corresponding author, comment this line and uncomment the next one.
%\extraAuth{corresponding Author2 \\ Laboratory X2, Institute X2, Department X2, Organization X2, Street X2, City X2 , State XX2 (only USA, Canada and Australia), Zip Code2, X2 Country X2, email2@uni2.edu}

\maketitle

\begin{abstract}
Parity-violating and time-reversal conserving (PVTC) and parity-violating and time-reversal-violating (PVTV)
forces in nuclei form only a tiny component of the total interaction between nucleons.
The study of these tiny forces can nevertheless be of extreme interest because they allow to obtain information
on fundamental symmetries using nuclear systems.
The PVTC interaction derives from the weak interaction between the quarks inside
nucleons and nuclei and the study of PVTC  effects opens a window on the quark-quark
weak interaction. The PVTV interaction is sensitive to more exotic
interactions at the fundamental
level, in particular to strong CP violation in the Standard Model Lagrangian, or even to
exotic phenomena predicted in various beyond-the-Standard-Model scenarios.  The presence of these 
interactions can be revealed either by studying various asymmetries in polarized scattering of nuclear systems, or
by measuring the presence of non-vanishing permanent electric dipole moments of nucleons, nuclei and diamagnetic atoms and molecules.
In this contribution, we review the derivation of the nuclear PVTC and PVTV interactions
within various frameworks. We focus in particular on the application of chiral effective field theory, which allows for a more strict connection
with the fundamental interactions at the quark level. We investigate PVTC and PVTV effects induced by these potential
on several few-nucleon observables, such as the longitudinal asymmetry in  proton-proton scattering and radiative neutron-proton capture, and the electric dipole moments
of the deuteron and the trinucleon system.
\tiny
 \keyFont{ \section{Keywords:}Fundamental symmetries in nuclei, nuclear forces, effective field theory, chiral perturbation theory, few-body systems} %All article types: you may provide up to 8 keywords; at least 5 are mandatory.
\end{abstract}

\section{Introduction}

The interaction between nucleons is at the heart of nuclear
physics and has been a subject of great scientific interest since many
decades. The strong nuclear forces have their origin in the residual
interaction between quarks and gluons inside colorless nucleons and
are described by quantum chromodynamics (QCD). The resulting
parity-conserving, time-reversal-conserving (PCTC) nuclear interactions
are known to exhibit a complicated pattern with a delicate interplay of strongly state-dependent repulsive and attractive pieces. While the nucleon-nucleon
(NN) scattering data below the pion production threshold can nowadays be
accurately described by modern NN potentials, the
(weaker) three-nucleon (3N) forces and the electromagnetic
interactions (EM) between the nucleons, known to play an important role
in the nuclear structure and dynamics, are not so well understood and subject of active research.
The current status of PCTC nuclear forces
is reviewed in other contributions to this topical issue.

In addition to the bulk PCTC interactions mentioned above, nuclear forces
also feature much tinier components, which originate from the weak
forces between quarks and/or physics beyond the standard
model (BSM) and whose strength is smaller than
that of the strong and EM interactions by many orders of magnitude. 
These tiny components are, nevertheless, extremely interesting since
investigation of their effects may shed new light
on fundamental symmetries and BSM physics. 
While effects of such exotic PCTC components are, of course, 
completely overwhelmed by the strong and EM nuclear forces, parity-
(P)  violating and/or time-reversal- (T) violating nuclear
interactions can be determined by measuring specific observables
which would vanish if these symmetries were conserved.  
In this contribution, we review the theory of parity-violating, time-reversal-conserving
(PVTC) and parity-violating, time-reversal-violating (PVTV) 
nuclear forces and discuss selected applications.

Starting from the fifties of the last century, a wide variety of
phenomenological models have been developed to describe nuclear
forces, the most prominent utilizing the one-boson exchange picture,
see Ref.~\cite{Machleidt:2017vls} and references therein. 
More recently, the development of chiral effective field
theory ($\chi$EFT)~\cite{Weinberg:1990rz} has given a new impetus
to the derivation of nuclear interactions~\cite{Ordonez:1995rz,Epelbaum:2008ga,Machleidt:2011zz}.
The $\chi$EFT approach utilizes the spontaneously broken approximate
SU(2)$_L$$\times$SU(2)$_R$
chiral symmetry of QCD\footnote{Here and in what follows, we restrict
  ourselves to the two-flavor case of the light up and down quarks
  unless specified otherwise.}
in order to describe low-energy dynamics of
pions, the (pseudo-) Goldstone bosons of the spontaneously broken axial
generators, in a systematic and model-independent fashion within the
framework of  the effective chiral
Lagrangian~\cite{Weinberg:1966kf,Weinberg:1968de,Weinberg:1978kz,Coleman:1969sm,Callan:1969sn,Gasser:1983yg}, 
see Refs.~\cite{Bernard:1995dp,Bernard:2007zu,Bijnens:2014lea} for review articles. 
Owing to the derivative nature of the Goldstone
boson interactions, the scattering amplitude in the pion- and single-baryon
sectors can be calculated via a perturbative expansion in powers
of $Q/\Lambda_\chi$, where $Q$ refers to momenta of the order of the
pion mass $m_\pi$ and $\Lambda_\chi \sim m_\rho \sim 1$~GeV denotes  the chiral
symmetry breaking scale, with $m_\rho$  the $\rho$-meson mass. The effective Lagrangian involves (an
infinite number of) all
possible hadronic interactions compatible with the
symmetries of QCD, which 
are naturally organized according to the number of derivatives
and/or quark or pion mass insertions.\footnote{In the isospin limit,
  the quark and pion masses are related to each other via $m_\pi^2 = 2 B
  m_q + \mathcal{O} (m_q^2)$, where $B$ is a constant
  proportional to the quark condensate $\langle 0 | \bar u u | 0
  \rangle = \langle 0 | \bar d d | 0
  \rangle$.} Every term in the effective Lagrangian
is multiplied with a coefficient, whose strength is not fixed by the
symmetry. These so-called  low-energy constants (LECs) can be
determined by fits to experimental data and/or obtained from lattice QCD
simulations, see  \cite{Bernard:2007zu,Bijnens:2014lea} and references
therein. At every order in the $Q/\Lambda_\chi$-expansion, only a
finite number of terms from the effective Lagrangian contributes to the
scattering amplitude.  The resulting framework, commonly referred to as
chiral perturbation theory ($\chi$PT), is nowadays widely applied to
analyze low-energy processes in the Goldstone boson and single-nucleon sectors. 
It has also been generalized to study few- and many-nucleon systems, where
certain resummations beyond perturbation theory are 
necessary in order to dynamically generate the
ultrasoft scale associated with nuclear binding. According to 
\cite{Weinberg:1990rz}, the breakdown of the perturbative expansion for the
NN scattering amplitude is traced back to enhanced
contributions of ladder diagrams, i.e.~Feynman diagrams that become
infrared divergent in the static limit of infinitely heavy
nucleons. The simplest and natural way to resum enhanced ladder
diagrams is provided by solving the nuclear Schr\"odinger
equation. The framework therefore essentially reduces to the 
conventional quantum mechanical $A$-body problem. The
corresponding nuclear forces and current operators are defined in
terms of non-iterative parts of the scattering amplitude, which are free from the above
mentioned enhancement. They can be derived from the effective chiral
Lagrangian in a systematically improvable way via a perturbative
expansion in powers of $Q/\Lambda_\chi$~\cite{Epelbaum:2008ga,Machleidt:2011zz}.
Assuming the scaling of few-nucleon contact operators according to
naive dimensional analysis\footnote{Notice that for systems near the
  unitary limit corresponding to the infinitely large 
scattering length (such as e.g.~the NN systems in the S-waves), the
scattering amplitude exhibits a certain amount 
of fine tuning beyond naive dimensional analysis. The expansion of the
scattering amplitude does, therefore, not necessarily coincide with
the expansion of nuclear potentials \cite{Epelbaum:2017byx}.}, the
PCTC interactions are  dominated by the
pairwise NN force, which receives its dominant  contribution
at order $(Q/\Lambda_\chi )^\nu$ with $\nu = 0$, defined to be the
leading order (LO). Parity conservation forbids the
appearance of nuclear forces at order  $\nu = 1$, so that the
next-to-leading order (NLO) contribution
to the PCTC NN potential appears at order $\nu = 2$.
Next-to-next-to-leading order (N$^2$LO) has $\nu=3$ and so on. 
PCTC three- and four-nucleon forces are suppressed and
start contributing at orders $\nu = 3$ (N$^2$LO) and  $\nu = 4$
(N$^3$LO), respectively.   Presently, the chiral
expansion of the PCTC NN force has been pushed to order $\nu = 5$ (N$^4$LO)
\cite{Entem:2014msa,Epelbaum:2014sza,Entem:2017gor,Reinert:2017usi},
while many-nucleon interactions have been worked out 
 up through N$^3$LO, see
\cite{Epelbaum:2008ga,Machleidt:2011zz} and references therein. We
further emphasize that 
a number of alternative formulations of $\chi$EFT for nuclear systems have
been proposed
\cite{Kaplan:1998tg,Nogga:2005hy,Birse:2005um,Valderrama:2009ei,Long:2012ve,Epelbaum:2012ua},
see also
Refs.~\cite{Lepage:1997cs,Epelbaum:2006pt,Epelbaum:2009sd,Valderrama:2016koj,
  Epelbaum:2018zli,Hammer:2019poc} for a related discussion.   

Another framework to analyze nuclear systems at very
low energies is based on the so-called pionless formulation of EFT,
see Refs.~\cite{Bedaque:2002mn,Hammer:2010kp,Hammer:2019poc} for
review articles. It is valid 
at momenta well below the pion mass, at which the pionic
degrees of freedom can be integrated out. In the resulting picture, nucleons
interact with each other solely through short-range contact two- and
many-body forces. This formulation is considerably simpler
than $\chi$EFT both at the conceptual and practical levels, and has
been successfully applied to study e.g.~Efimov physics and universality in
few-body systems near the unitary limit, low-energy properties of
halo-nuclei and reactions of astrophysical relevance, see
Refs.~\cite{Bedaque:2002mn,Hammer:2010kp,Hammer:2019poc} and
references therein.

In this paper we focus on the PVTC and PVTV interactions in the
frameworks of $\chi$EFT and pionless EFT. We also outline
various meson-exchange models frequently adopted to analyze the
results for some PVTC and PVTV observables. In the subsections below,
we briefly discuss the origin of the PVTC and PVTV interactions and
summarize the current experimental and theoretical status of research
along these lines. 

\subsection{The PVTC interaction}
The PVTC component of the nuclear force is governed by 
the weak interaction between the quarks inside
the nucleons (and pions). Studying such effects, therefore,  opens
a window on the so-called ``pure'' hadronic weak interaction
(HWI)~\cite{RamseyMusolf:2006dz,Hertzog:2012zd,Schindler:2013yua,Haxton:2013aca,deVries:2015gea}. This 
part of the weak  interaction is far less known experimentally.

A number of experiments aimed at studying PVTC in low-energy processes
involving few-nucleon systems have been completed/are being  
planned at cold-neutron facilities, such as the Los Alamos
Neutron Science Center (LANSCE), the National Institute of Standards
and Technology (NIST) Center for Neutron Research, the Spallation
Neutron Source (SNS) at Oak Ridge National Laboratory, and the
European Spallation Source (ESS) in Lund. The
primary objective of this experimental program is to determine the
LECs which appear in the PVTC nuclear potentials. 
For a recent review of the current status of experiments along this
line and the impact of anticipated results see Ref.~\cite{Gardner:2017xyl}.

PVTC nuclear forces have already been analyzed in
the framework of $\chi$EFT \cite{Zhu:2004vw,deVries:2013fxa,Viviani:2014zha}. The LO
PVTC NN force is driven by the one-pion-exchange term with
$\nu =-1$, while the NLO terms with $\nu=1$ emerge from
two-pion-exchange diagrams and NN contact interactions\footnote{Notice that PVTC hadronic interactions involve a
  typical suppression factor of $\sim G_F M_\pi^2 \sim 10^{-7}$ as
  compared to PCTC vertices \cite{Adelberger:1985ik}.}.
In Ref.~\cite{Girlanda:2008ts}, it was shown that the PVTC NN
potential involves only five independent
contact operators at this order corresponding to five
S-P transition amplitudes at low energies~\cite{Danilov:1965hc}.
Including the PVTC pion-nucleon coupling
constant $h^1_\pi$, the NN potential at  NLO thus contains six LECs
which need to be determined from experimental data. 
At N$^2$LO one has to take into account five additional LECs, which 
determine the strength of the subleading PVTC pion-nucleon interactions \cite{deVries:2014vqa}.

In pionless EFT, the LO PVTC NN potential is completely
described in terms of the already mentioned five contact
terms~\cite{Schindler:2013yua,Schindler:2015nga}. 
The large-$N_c$ scaling of PVTC NN contact interactions was
analyzed in Refs.~\cite{Phillips:2014kna,
  Schindler:2015nga}.
These studies suggest that three out of five PVTC contact interactions 
are suppressed
by a factor of $(1/N_c)^2$ or by the factor $\sin^2\theta_W
\approx0.23$, see also a related discussion in
~\cite{Vanasse:2018buq}.  If the large-$N_c$ scaling persists to
the physically relevant case of $N_c = 3$, 
the pionless potential at LO should be dominated by only
2 LECs~\cite{Gardner:2017xyl}. Unfortunately, the currently available
experimental data do not allow one to draw definitive conclusions on
whether the suggested large-$N_c$ hierarchy of PVTC contact interactions
is indeed realized in Nature.

Regarding the various meson-exchange models developed to
describe the PVTC interaction, we will mainly discuss
the model proposed by Desplanques, Donoghue, and Holstein (DDH)~\cite{Desplanques:1979hn}
which includes pion and vector-meson exchanges with
seven unknown meson-nucleon PVTC coupling constants.

\subsection{The PVTV interaction}
PVTV nuclear forces originate from more exotic sources at the
fundamental level, which include the so-called $\theta$-term
in the Standard Model (SM) Lagrangian~\cite{tHooft:1976rip}, or even
BSM interactions~\cite{Pospelov:2005pr}. 
Due to the CPT theorem, any PVTV interaction also violates the CP symmetry,
where C refers to  charge conjugation. CP violation is a
key ingredient for the dynamical generation of a matter-antimatter asymmetry
in the Universe ~\cite{Sakharov:1967dj}. The SM with three generations of quarks has a natural source of CP-violation
in the phase of the Cabibbo-Kobayashi-Maskawa (CKM) quark mixing matrix. This
mechanism is however not sufficient for explaining the observed asymmetry~\cite{Cohen:1993nk}.

The phase of the CKM matrix also does not contribute sizably to the nuclear PVTV interaction.
For example, let us consider the electric dipole moment (EDM) of a system of particles.
A nonzero permanent EDM of a particle or a system of
particles necessarily involves the breaking of the parity and
time-reflection symmetries. 
EDMs of the electron, nucleons and nuclei are mostly sensitive to P-
and T-violating flavor-diagonal interactions. To induce a non-zero
EDM, on the other hand, the phase of the CKM requires  contributions
from all three generations of quarks,  including heavy quarks, leading
to a large
suppression~\cite{Pospelov:2005pr,Czarnecki:1997bu,Mannel:2012qk,Mannel:2012hb}. For 
example, the expected size of the nucleon EDM based on the CKM
mechanism in the SM is $|d_N^{\rm CKM}| \sim 10^{-18}\ e$ fm~\cite{Wirzba:2016saz,Seng:2014lea}.
Therefore, any observed permanent EDM of an atomic
or nuclear system larger in magnitude than the expected
size within the SM  would highlight PVTV effects beyond the CKM mixing matrix.
The present experimental upper bounds on the EDMs of neutron and proton
are $|d_n|<3.0\cdot 10^{-13}\ e$ fm~\cite{Baker:2006ts,Afach:2015sja}
and $|d_p|<2.0\cdot 10^{-12}\ e$ fm,
where the proton EDM has been inferred from a measurement of the diamagnetic
${}^{199}{\rm Hg}$ atom~\cite{Graner:2016ses} using a calculation of the nuclear
Schiff moment~\cite{Dmitriev:2003sc}. For the electron, the most
recent upper bound is
$|d_e|<1.1\cdot10^{-16}\ e$ fm~\cite{Andreev:2018ayy},
derived from the EDM of the ThO molecule. In all cases, the current
experimental sensitivities are orders of magnitude away from CKM predictions.

$\chi$EFT allows to derive PVTV nuclear forces in a systematic
and model independent way. To this aim, the PCTC effective chiral
Lagrangian has to be extended to include all possible PVTV terms
classified according to their chiral dimension. Some of these terms
are induced, at the microscopic level, by the SM mechanisms discussed
above. The effective chiral Lagrangian induced by the $\theta$-term is
discussed in Refs.~\cite{Mereghetti:2010tp,Bsaisou:2014zwa}.
BSM theories such as supersymmetry, multi-Higgs
scenarios, left-right symmetric model {\it etc.}
would give rise to additional PVTV sources of dimension six (and higher) in the
quark-gluon Lagrangian~\cite{Grzadkowski:2010es}. 
The $\chi$EFT Lagrangians originating from these sources were derived in
Refs.~\cite{deVries:2012ab,Bsaisou:2014oka}.
Various terms in the resulting effective chiral Lagrangian
possess different scaling with respect to the underlying microscopic
PVTV sources. $\chi$EFT can thus be used to establish relations
between the fundamental PVTV mechanisms and specific terms in the
nuclear potentials and, accordingly, specific pattern in
the corresponding nuclear
observables~\cite{Mereghetti:2010tp,deVries:2012ab,Bsaisou:2014oka}. In
principle, this offers the possibility to identify the fundamental
sources of time-reversal violation and to shed light on some of the
BSM scenarios, provided the corresponding LECs in
the effective Lagrangian can be determined from Lattice QCD calculations or experimental
data~\cite{deVries:2011re,Dekens:2014jka}.   

In the framework of $\chi$EFT,  the PVTV NN potential was
derived up to N$^2$LO including one- and two-pion exchange contributions and
the corresponding contact interactions~\cite{Maekawa:2011vs,Bsaisou:2012rg}. 
Subsequent works showed the presence in the PVTV Lagrangian of a three-pion
term~\cite{deVries:2012ab}, which was for the first time included in the calculations
in Ref.~\cite{Bsaisou:2014zwa}.
This term also generates a PVTV 3N force at NLO, which contributes
to the  ${}^3$H and ${}^3$He EDM. The calculation reported in Ref.~\cite{Bsaisou:2014zwa}
was also the first one carried out using solely the interactions derived in
$\chi$EFT. More precisely, the  PVTV potential at NLO was used in
combination with the  N$^2$LO PCTC potentials from Ref.~\cite{Epelbaum:2004fk}. 
Finally, in Ref.~\cite{Gnech:2019dod}, the EDM of
deuteron and trinucleons was studied using the $\chi$EFT PVTV potential up
to N$^2$LO along with the N$^4$LO PCTC potential of Ref.~\cite{Entem:2017gor}.
In this paper, it was also shown that the N$^2$LO contribution to the
PVTV 3N force generated by the three-pion interaction vanishes. The LO
$\chi$EFT PVTV potential has also been applied in combination with
many-body methods to calculate Schiff moments of heavy nuclei
\cite{Dobaczewski:2018nim}.  

Currently, no direct limits on EDMs of light nuclei have been established.
However, experiments are planned to measure the EDM of 
protons and light nuclei in dedicated storage rings
~\cite{Orlov:2006su,Semertzidis:2011qv,Lehrach:2012eg,Pretz:2013us,Rathmann:2013rqa,Abusaif:2019gry}.
This new approach could reach an accuracy of $\sim 10^{-16}\ e$ fm,
although this has to be established in practice. If successful, these
experiments would lead to a great 
improvement in the hadronic sector of EDM searches.
A measurement 
of a non-vanishing EDM of this magnitude would provide evidence of a
PVTV source beyond the CKM mechanism. However,
a single measurement would be insufficient to identify the source of PVTV.
For this reason, experiments with various light nuclei such as ${}^2$H,
${}^3$H and ${}^3$He are planned. Such measurements would provide a
complementary information needed to impose constraints on 
PVTV sources at the fundamental level.

A brief discussion of the PVTV potentials derived in the framework of the one-meson exchange model
and in the pionless EFT approach will also be reported in this review.

\subsection{Outline of the article}
Our paper is organized as follow. In Section~\ref{sec:micro}, we discuss the
origins of PVTC and PVTV interactions at the fundamental level and
list the relevant terms in the quark-gluon Lagrangian.
In Section~\ref{sec:chieft}, we give the corresponding terms in the effective
chiral Lagrangian and discuss the derivation of the  PVTC and PVTV
potentials in $\chi$EFT. In Section~\ref{sec:pionless}, we specifically focus on
the contact few-nucleon interactions which enter the potentials in both
chiral and pionless EFT formulations. We also discuss the expected hierarchy of
the corresponding LECs as suggested by the large-$N_c$ analysis. Next,
in Section~\ref{sec:obep}, the various meson-exchange models developed to
describe the PVTC and PVTV interactions will be summarized. 
Then, in Section~\ref{sec:res}, we report on a selected set of results for PVTC and PVTV
observables in light  nuclei up to $A=3$. Finally, the main
conclusions of this paper and future perspectives are summarized in section~\ref{sec:conc}.

\section{Parity violation and time-reversal violation at the microscopic level}
\label{sec:micro}
Parity is violated in the SM of particle physics
because of the different gauge interactions of left- and
right-handed fermion fields. Only left-handed particles interact via
$SU(2)_L$ gauge interactions such that this part of the SM violates
parity maximally. The remaining color and electromagnetic interactions conserve parity
modulo the QCD vacuum angle which is discussed below. 
Parity violation was first observed in semileptonic charged
current interactions in 1957 \cite{Wu:1957my}. 
Twenty years later, in the late `70s, PVTC was observed in neutral current
electron-nucleus scattering \cite{Prescott:1978tm}, providing a strong
confirmation of the SM. 
Subsequent PVTC electron scattering experiments have quantitatively
confirmed the SM picture \cite{Androic:2018kni}.  
In addition to PVTC in $\beta$ decays and semileptonic neutral current
processes, the SM predicts PVTC in weak interactions between
quarks. At energies smaller than the masses of the $W$  
and $Z$ bosons, these interactions can be represented by four-fermion operators.
Right below the electroweak (EW) scale, and limiting ourselves to the lightest $u$
and $d$ quarks, the four-fermion Lagrangian is 
\begin{eqnarray}
\mathcal L_W &=& -\frac{G_F}{\sqrt{2}}\left\{ \left(1 - \frac{2}{3}
s_w^2 \right) \bar q_L \gamma^\mu \tau_a q_L \, \bar q_L \gamma_\mu
\tau_a q_L  
-\frac{2 s_w^2}{3} \bar q_L \gamma^\mu \tau_3  q_L \, (\bar q_L
\gamma_\mu q_L + \bar q_R \gamma_\mu q_R ) 
\right.
\nonumber \\
& & \left.
- 2 s^2_w \left( \bar q_L \gamma^\mu \tau_3 q_L \, \bar q_L \gamma_\mu
\tau_3 q_L - \frac{1}{3}\bar q_L \gamma^\mu \tau_a q_L \, \bar q_L
\gamma_\mu \tau_a q_L 
\right) + \dots \right\}\,, \label{eq:SM1}
\end{eqnarray}
where $G_F$ is the Fermi coupling constant and
$s^2_w \equiv \sin^2 \theta_W=0.231$, with $\theta_W$ the Weinberg 
mixing angle. $q_L$ and $q_R$ denote the left-handed and right-handed
doublets  $q_{L}^T = (u_L, d_L)$ and $q_R^T = (u_R, d_R)$, and the dots denote
terms that conserve parity\footnote{Here $u$ and $d$ denote the $u$-
  and $d$-quark Dirac fields, respectively. Moreover $u_{R,L}=
  {1\pm\gamma^5\over2}u$, etc.}. Eq. \eqref{eq:SM1} was obtained assuming the CKM matrix  to be the
identity, that is $V_{ud}=1$. The three operators in Eq. \eqref{eq:SM1} all break
parity, but have different  transformation properties under
chiral symmetry and  isospin.
We note that the isovector and isotensor terms (the second and
third operators) given in Eq.~(\ref{eq:SM1}) are suppressed by a
factor $s^2_w$ with respect to the isoscalar one.

The operators in Eq. \eqref{eq:SM1} need to be evolved
using the renormalization group equations (RGE)
from the EW scale down to the QCD scale, and in this process they mix with
additional PVTC operators \cite{Tiburzi:2012hx}. 
After the RGE evolution, the PVTC Lagrangian assumes the form
\begin{eqnarray}
\mathcal L^{\rm SM}_{\rm PVTC} &=& -\frac{G_F}{\sqrt{2}}\Bigg\{ C^{\rm SM}_1\,  \bar q_L
\gamma^\mu \tau_a q_L \, \bar q_L \gamma_\mu \tau_a q_L 
+ C^{\rm SM}_2\, \bar q_L \gamma^\mu  q_L \, \bar q_L \gamma_\mu  q_L + C^{\rm SM}_3\,
\bar q_L \gamma^\mu \tau_3  q_L \, \bar q_L \gamma_\mu q_L  
\nonumber \\
& & 
+ C^{\rm SM}_4 \left( \bar q_L \gamma^\mu \tau_3  q_L \, \bar q_R \gamma_\mu
q_R - \bar q_L \gamma^\mu   q_L \, \bar q_R \gamma_\mu \tau_3 q_R
\right) 
\nonumber \\
& & 
+ C^{\rm SM}_5 \left( \bar q^{\,\alpha}_L \gamma^\mu \tau_3  q^{\,\beta}_L \,
\bar q^{\,\beta}_R \gamma_\mu q^{\,\alpha}_R - \bar q^{\,\alpha}_L
\gamma^\mu   q^{\,\beta}_L \,  
\bar q^{\,\beta}_R \gamma_\mu \tau_3 q^{\,\alpha}_R   \right)
\nonumber \\ 
& & + C^{\rm SM}_6 \left( \bar q_L \gamma^\mu \tau_3 q_L \, \bar q_L \gamma_\mu
\tau_3 q_L - \frac{1}{3}\bar q_L \gamma^\mu \tau_a q_L \, \bar q_L
\gamma_\mu \tau_a q_L 
\right)  - (L \leftrightarrow R)\Bigg\}\,,
\label{eq:SM2}
\end{eqnarray}
where in the SM, the coefficients $C^{\rm SM}_i$ are known functions of SM
parameters as $s_w$, the strong coupling constant $g_s$, etc. Greek indices $\alpha$ and
$\beta$ appearing as superscript in some of the quark fields in Eq. \eqref{eq:SM2}
    specify color indices. They are only shown for cases where the color contractions are not obvious.  
Notice that the QCD evolution does not remedy the $s_w^2$ suppression of the isospin-one and -two operators \cite{Tiburzi:2012hx}.
BSM physics that arises at scales well above the EW can be represented  
at the EW scale via gauge-invariant higher-dimensional
operators \cite{Buchmuller:1985jz,Grzadkowski:2010es}. This framework is usually called the SM Effective Field Theory (SM-EFT). 
SM-EFT operators can induce new PVTC couplings of the  $W$ and $Z$ bosons to left-
and right-handed quarks, and new PVTC four-fermion operators. After evolving the
effective operators from the EW to the QCD scale,  
the net effect of BSM PVTC SM-EFT operators is to modify the
coefficients $C^{\rm SM}_{i}$ in Eq. \eqref{eq:SM2} with respect to their SM
values, namely in Eq.~(\ref{eq:SM2}) one substitutes $C^{\rm SM}_{i}\rightarrow C^{\rm SM+BSM}_{i}$. 
We have focused so far on operators involving only the $u$ and $d$ quarks.  
Flavor-conserving ($\Delta F=0$) operators involving the $s$ quark can
also generate interesting contributions to hadronic P violation \cite{Kaplan:1992vj,Tiburzi:2012hx},
such as contributions to isospin-one operators that are not suppressed by $s^2_w$.

While P and C are maximally broken by the $V-A$ structure of the SM, 
the breaking of CP is much more delicate. In the SM with three generations of quarks, CP
is broken by the phase of the CKM matrix, which explains all the
observed CP violation in the kaon
\cite{Christenson:1964fg,Abouzaid:2010ny,Batley:2002gn},  and $B$ meson systems
\cite{Aubert:2001nu,Abe:2001xe}. Theoretical uncertainties are at the moment too large to definitively conclude whether 
the recently discovered CP violation in $D$ decays \cite{Aaij:2019kcg} is compatible with the SM.
The phase of CKM gives, on the other hand, unobservable
contributions to flavor-diagonal CP violation, 
in particular to the neutron
\cite{Khriplovich:1981ca,Czarnecki:1997bu,Seng:2014lea} and electron
EDMs
\cite{Pospelov:1991zt,Booth:1993af,Pospelov:2013sca}.

The second source of CP violation in the SM is the QCD $\theta$ term
\cite{Callan:1976je,tHooft:1976rip,tHooft:1976snw} 
\begin{eqnarray}
  \mathcal L_{PVTV}^\theta = -\theta\frac{g_s^2}{64 \pi^2}
  \varepsilon^{\mu\nu\alpha\beta} \, G^{a}_{\mu\nu} G^{a}_{\alpha
  \beta}\,, 
\end{eqnarray}
where $g_s$ is the strong coupling constants and $G^{a}_{\mu\nu}$ the
gluon field tensors ($a$ is a color index).
The $\theta$ term is a total derivative, but it contributes to
physical processes through extended, spacetime-dependent field
configurations known as instantons. 
CP violation from the QCD $\theta$ term is intimately related to the quark
masses.  All phases of the  quark  mass  matrix  can  be  eliminated
through  non-anomalous $SU(2)$  vector  and  axial rotations, except
for a common phase $\rho$. The mass plus QCD $\theta$ terms which
are left are 
\begin{equation}
\mathcal L_{PVTV}^{{\rm mass+}\theta} = -\left( e^{i\rho} \bar q_L \mathcal M q_R +
e^{-i\rho}\, \bar q_R \mathcal M q_L \right) -\theta\frac{g_s^2}{64
  \pi^2} \varepsilon^{\mu\nu\alpha\beta} \, G^{a}_{\mu\nu}
G^{a}_{\alpha \beta}\,, 
\end{equation}  
where $\mathcal M = \textrm{diag}(m_u,m_d)$. The parameters
$\rho$ and $\theta$ are not independent. 
In $\chi$EFT, it is convenient to rotate $\mathcal L_{PVTV}^{{\rm
    mass+}\theta}$ into a complex mass term
with an anomalous $U(1)_A$ rotation,
obtaining, after vacuum alignment \cite{Baluni:1978rf},
\begin{eqnarray}\label{thetarotated}
\mathcal L_{PVTV}^{{\rm mass+}\theta} = m_* \bar\theta\, \bar q i \gamma_5 q\,,
\end{eqnarray}
where 
\begin{eqnarray}\label{mstar}
\bar\theta =  \theta + n_f \rho\,,  \qquad m_* = \frac{m_u m_d }{ m_u +
  m_d } = \frac{\bar m (1 - \epsilon^2)}{2}\,. 
\end{eqnarray}
$n_f=2$ is the number of light flavors, and the combinations of light
quarks masses $\bar m$ and $\epsilon$ are $2 \bar m = m_u + m_d$,
$\epsilon = (m_d - m_u)/(m_d + m_u)$.  
Eqs. \eqref{thetarotated} and \eqref{mstar} can be easily generalized
to include strangeness. $\bar\theta$ is a free parameter in the QCD
Lagrangian, and one would expect $\bar\theta = \mathcal O(1)$. This
would however lead to a large neutron EDM $|d_n| \sim 10^{-3}
\bar\theta$ $e$ fm \cite{Crewther:1979pi,Dragos:2019oxn}, ten orders
of magnitude larger than the current 
limits, $d_n < 3.0 \cdot 10^{-13}$ $e$ fm \cite{Afach:2015sja}.  
Therefore $\bar\theta \lesssim 10^{-10}$, the so-called strong CP problem.   

The phase of the CKM and the QCD $\bar\theta$ term are the only CP-violating
parameters in the SM Lagrangian. They are however not sufficient to
explain the observed matter-antimatter asymmetry of the Universe
\cite{Gavela:1993ts,Gavela:1994ds,Gavela:1994dt,Huet:1994jb}, 
and it is therefore natural to think about CP-violating sources induced by BSM physics.
The low-energy CP-violating operators relevant for EDMs have been
cataloged in several works,
e.g. Refs. \cite{Pospelov:2005pr,Khriplovich:1997ga, Dekens:2013zca,Engel:2013lsa}. 
Ref. \cite{deVries:2012ab} considered all the low-energy operators
that are induced by SM-EFT operators at tree level, retaining the two
lightest quarks. Generalization to three flavors 
are given, for example, in Refs. \cite{Jenkins:2017jig,Mereghetti:2018oxv}.
The  most relevant $SU(3)_c \times U(1)_{\rm em}$-invariant purely
hadronic operators induced by dimension-six SM-EFT operators are  
\begin{eqnarray}\label{quark}
\mathcal L_{PVTV}^{6, \rm hadr} &=&  \frac{g_s \tilde{C}_{G}}{6 v^2} f^{a b
  c} \epsilon^{\mu \nu \alpha \beta}  G^a_{\alpha \beta} G_{\mu
  \rho}^{b} G^{c\, \rho}_{\nu}    -\frac{ 1}{2 v^2} \left(\bar{q}
\left[d_{E}\right] i \sigma^{\mu\nu} \gamma_5 q \; e
F_{\mu\nu}   
+ \bar{q} [d_{CE}] i \sigma^{\mu\nu} \; g_s G_{\mu\nu}
\gamma_5  q \right) \nonumber\\ 
& & 
- \frac{4 G_F}{\sqrt{2}} \Bigg\{ \Sigma^{(ud)}_1 (\bar d_L u_R \bar
u_L d_R - \bar u_L u_R \bar d_L d_R )  + \Sigma^{(ud)}_2 (\bar
d^\alpha_L u^\beta_R\, \bar u^\beta_L d^\alpha_R - \bar u^\alpha_L
u^\beta_R\, \bar d^\beta_L d^\alpha_R )  \Bigg\} \nonumber \\ 
& &- \frac{4 G_F}{\sqrt{2}} \Bigg\{  \Xi^{(ud)}_1 \, \bar d_L
\gamma^\mu  u_L\, \bar u_R \gamma_\mu  d_R   + 
\Xi^{(ud)}_2 \, \bar d^\alpha_L \gamma^\mu  u^\beta_L\, \bar u^\beta_R
\gamma_\mu  d^\alpha_R  
\Bigg\}\,,
\end{eqnarray}
where $f^{abc}$ are the structure constants of the Lie algebra of the
color $SU(3)$ group, $[d_{E}]$
and $[d_{CE}]$ are matrices in flavor space, $[d_E] = \textrm{diag}(m_u \tilde c_\gamma^{(u)}, m_d \tilde c_\gamma^{(d)})$
and $[d_{CE}] = \textrm{diag}(m_u \tilde c_g^{(u)}, m_d \tilde c_g^{(d)})$.
The coefficients $\tilde C_G$, $\tilde c^{(q)}_{\gamma, g}$,
$\Sigma^{(ud)}_{1,2}$ and $\Xi^{(ud)}_{1,2}$ are dimensionless and
scale as $(v/\Lambda_X)^2$, where $v=246$ GeV is the Higgs vacuum
expectation value, and $\Lambda_X$ is the scale of new physics. 
The Weinberg three-gluon, the quark EDM (qEDM), and
the chromo-EDM (qCEDM) operators (the first, second, and third term 
given in Eq.~(\ref{quark}), respectively)
have received the most attention in the literature
\cite{Pospelov:2005pr,Weinberg:1989dx}. They can be written directly in
terms of $SU(3)_c \times SU(2)_L \times U(1)_Y$-invariant operators at
the EW scale, and receive corrections by a variety of CP-violating operators in
the SM-EFT, involving heavy SM fields.  The four-quark operators,
given in the second line of Eq.~(\ref{quark}), can also be
expressed in terms of gauge-invariant operators at the EW scale, and
they arise, for example, in leptoquark models,
see~\cite{Fuyuto:2018scm, Dekens:2018bci}.  The four-quark operators,
given in the third line of Eq.~(\ref{quark}), 
are on the other hand induced by right-handed couplings of
quarks to the $W$ boson \cite{Ng:2011ui,deVries:2012ab}, and are
generated, for example, in left-right symmetric models.  

While all operators in Eqs.~\eqref{thetarotated} and \eqref{quark}
violate P and CP symmetry, they  transform differently under
isospin and chiral rotations. As such, the operators induce different
$\chi$EFT Lagrangians at lower energies, and different hierarchies of
CP-violating hadronic and nuclear observables such as EDMs or
scattering observables.

\section{PVTC and PVTV chiral potentials}
\label{sec:chieft}
In this section, we discuss the derivation of the PVTC and PVTV NN and 3N
potentials within the framework of $\chi$EFT. In the first and second
subsections we briefly review the properties of the PVTC and PVTV
chiral Lagrangians. 
In subsection~\ref{sec:lag2pot}, we present briefly two methods used to derive 
the potentials starting from a Lagrangian. Finally, in the last
two subsections, we present the PVTC and PVTV chiral potentials, respectively.

In order to discuss hadronic observables such as nuclear EDMs or PVTC asymmetries
in $pp$ scattering, the quark-level PVTC and PVTV Lagrangians of
Eqs. \eqref{eq:SM2} and \eqref{quark} need to be matched onto 
nuclear EFTs, such as chiral EFT and pionless EFT. 
Due to the nonperturbative nature of QCD at low energy,
this matching cannot be done in perturbation theory. Nevertheless, the approximate
chiral and isospin symmetries of the QCD Lagrangian  
provide an organizing principle for low-energy
interactions, see~\cite{Bernard:1995dp,Bernard:2007zu,Bijnens:2014lea} for review
articles. 

Let us first introduce the nucleon and pion fields. The (relativistic) nucleon field 
$N(x)$ is considered to be an isospin doublet 
\begin{equation}
N(x)=\left(\begin{array}{c} 
         p(x)\\
         n(x)
         \end{array}\right)\,, 
\end{equation}
where $p(x)$ ($n(x)$) is the proton (neutron) field. The pion fields
are given in ``Cartesian'' coordinates $\pi_a$, $a=1,2,3$, where
\begin{equation}
  \pi_1(x)= \frac{\pi^{(+)}(x)+\pi^{(-)}(x)}{\sqrt{2}}\,, \quad
  \pi_2(x)=\frac{i \left(\pi^{(+)}(x)-\pi^{(-)}(x)\right)}{\sqrt{2}}\,, \quad
  \pi_3(x)=\pi^{(0)}(x)\,,\label{eq:pifield}
\end{equation}
$\pi^{(+)}(x)$, $\pi^{(-)}(x)$, and $\pi^{(0)}(x)$ being the fields
associated to the three charge states of the pion. The
pion fields in Cartesian coordinates are collectively denoted by
$\vec\pi(x)$. We use the $2\times 2$ matrices $\tau_a$,
$a=0,\ldots,3$, where $\tau_0$ is the identity matrix,
while $\tau_a$, $a=1,\ldots,3$ are the Pauli matrices acting on
the isospin degrees of freedom (often indicated cumulatively as
$\vec\tau$). For example, $\vec\tau\cdot\vec\pi(x)=\sum_{a=1}^3 \tau_a
\pi_a(x)$. Sometimes the $a=3$ component will be denoted as the
``$z$'' component, i.e. $\pi_3\equiv \pi_z$, etc., in our notation.
Finally, we denote the nucleon (pion) mass by $M$ ($m_\pi$).

In some cases, we will
%use
%the non-relativistic reduction of the nucleonic field, denoted as $N_r(x)$,
perform a non-relativistic reduction of the nucleon field $N(x)$ and
use $N_r(x)$ 
\begin{equation}
   N_r(x)=\left(\begin{array}{c} 
          p_r(x)\\
          n_r(x)
          \end{array}\right)\,, 
\end{equation}
where $p_r(x)$ ($n_r(x)$) is the two component Pauli spinor
representing the static proton (neutron) field. Effects of the anti-nucleon
degrees of freedom are taken into account in the form of $1/M$
relativistic corrections to the vertices. The coefficient  
of the annihilation operator reduces to $\chi_s \exp(i\bmp\cdot\bmx)$,
where $\chi_s$ is a spinor describing a spin state with $z$-projection
$s=\pm{1\over2}$. 
    
The main ``building block'' to construct the chiral Lagrangian is the
SU(2) pionic matrix field $U(x)$, often written as (but its
definition is not unique)~\cite{Bernard:1995dp}
\begin{equation}
  U(x)=e^{\frac{i}{f_\pi}\vec\pi(x)\cdot \vec\tau} \, , 
\end{equation}
where $f_\pi\approx 92.4$ MeV is the pion decay constant. Another low
energy constant frequently entering the chiral Lagrangian is
the axial coupling constant $g_A\approx 1.29$. 
Following the standard convention, we give here the effective value
that takes into account the Goldberger-Treiman discrepancy and is
extracted from the empirical value of the pion-nucleon coupling
constant. The effective chiral Lagrangian
is constructed in terms of $N(x)$ and $U(x)$ and therefore
contains vertices with arbitrary number of pion fields. In the
following, we will retain explicitly only relevant terms with the minimum number of pion
fields, obtained by expanding $U(x)$ in powers of the pion field. Additional terms with a
larger number of pion fields will only contribute to the PVTC and PVTV potential at higher orders in the chiral
expansion. For an introduction to
the chiral Lagrangians and their building blocks, the reader is
referred to Refs.~\cite{Bernard:1995dp,Bernard:2007zu,Bijnens:2014lea} and
references therein.

Each term of the chiral Lagrangian will be classified by the so-called
``chiral order''. Each four-gradient of the pion matrix field or a
multiplication by a pion mass increases the order of the term by one.
Four-gradients acting on nucleon fields are more difficult to classify, since
the time derivative brings down a factor proportional to the nucleon
mass. An easier counting is obtained using the non-relativistic heavy
baryon perturbation theory~\cite{Jenkins:1990jv,Bernard:1995dp}, which was used in the derivation of the 
PVTC potential in \cite{deVries:2014vqa} and of the PVTV potential in \cite{Maekawa:2011vs}. 
In the following, we will use both the relativistic and nonrelativistic nucleon fields.

For the sake of completeness, we report first of all the terms of the
PCTC Lagrangian that contribute to the PVTC and PVTV potentials up to
the order we are interested in. In $SU(2)$ $\chi$PT, the PCTC
Lagrangian can be conveniently organized in sectors with different
numbers of pions and nucleons (below we give the explict expression
for the relevant terms in the $\pi N$ Lagrangian only). 
\begin{eqnarray}
  \mathcal L_{PCTC} &=& \mathcal L_{PCTC,\pi N} + \mathcal L_{PCTC,NN}+
  \mathcal L_{PCTC,\pi\pi}+\cdots\,, \\
  \mathcal L_{PCTC,\pi N} &=& \overline{N } \bigg [ -\frac{1}{4\,
      f_\pi^2} ({\vec \tau}\times{\vec \pi}) 
   \cdot \partial_\mu {\vec \pi}\, \gamma^\mu -\frac{g_A}{2\,
     f_\pi}({\vec \tau}\cdot  \partial_\mu {\vec \pi})\, \gamma^\mu \gamma^5\nonumber \\
   && + 4\, c_1\, m_\pi^2\left(1-\frac{{\vec \pi}^2}{2\, f_\pi^2}\right) 
      +\frac{c_2}{f_\pi^2} \left( \partial_0 {\vec \pi}  \cdot  \partial_0{\vec \pi}
         +\frac{1}{M} \partial_0{\vec \pi} \cdot  \partial_i{\vec \pi}
            \gamma^0 \, i\overleftrightarrow{\partial}^i \right)
            \nonumber\\
   && + \frac{c_3}{f_\pi^2} \, \partial_\mu{\vec \pi} \cdot
            \partial^\mu{\vec \pi} 
      - \frac{c_4}{2 f^2_\pi}\, ({\vec
        \tau}\cdot\partial_\mu{\vec\pi}\times\partial_\nu{\vec \pi})\, \sigma^{\mu\nu}
      +\cdots\bigg] N \label{eq:lagpctc}      
\end{eqnarray}
where ``$\cdots$'' in the previous expression denotes terms of higher order and/or more pions
fields of no interest here. Above $\overleftrightarrow{\partial}^\mu
\equiv \overrightarrow{\partial}^\mu- \overleftarrow{\partial}^\mu$
and $\sigma^{\mu\nu}= {i\over 2}[\gamma^\mu,\gamma^\nu]$. 
The parameters $c_{i=1-4}$ are LECs
appearing in the Lagrangian of order $Q^2$. They have dimension of the
inverse of an energy. For a complete discussion of the 
terms appearing in the Lagrangians $\mathcal L_{PCTC,\pi N}$,
$\mathcal L_{PCTC,NN}$, and $\mathcal L_{PCTC,\pi\pi}$, etc.,    
see Refs.~\cite{Bernard:1995dp,Baroni:2015uza}. 

\subsection{The PVTC chiral Lagrangian}
\label{sec:chipv}

The effective chiral Lagrangian that involves contributions from the
weak sector of the SM was first
discussed in the seminal paper by Kaplan and Savage~\cite{Kaplan:1992vj} and subsequently
revisited in Refs.~\cite{Kaplan:1998xi,Zhu:2004vw,deVries:2014vqa,Viviani:2014zha}. 
Also the PVTC Lagrangian can be conveniently organized in sectors
with different  numbers of pions and nucleons, explictly
\begin{eqnarray}
  \mathcal L_{PVTC} &=& \mathcal L_{PVTC,\pi N} + \mathcal L_{PVTC,NN}+
  \mathcal L_{PVTC,\pi\pi\pi}+\cdots\,, \\
  \mathcal L_{PVTC,\pi N} &=& \mathcal L_{PVTC,\pi N}^{(0)}+ \mathcal L_{PVTC,\pi N}^{(1)}+\cdots  \,,\\
  \mathcal L_{PVTC,NN} &=& \mathcal L_{PVTC,N N}^{(1)}+ \mathcal L_{PVTC,NN}^{(3)}+\cdots  \,,\\
  \mathcal L_{PVTC,\pi\pi\pi} &=& \mathcal L_{PVTC,\pi\pi\pi}^{(2)}+ \cdots  \,,
\end{eqnarray}
where the superscript $(n)$ denotes the chiral order of each piece.
The pion-nucleon interaction terms are collected in $\mathcal L_{PVTC,\pi N}$
and those entering the PVTC potential up to
$Q^1$ are the following~\cite{Kaplan:1992vj,deVries:2014vqa} 
\begin{eqnarray}
  \mathcal{L}_{PVTC,\pi N}^{(0)}&=& \frac{h^1_{\pi}}{\sqrt{2}}\overline{N}
     (\vec \pi\times\vec \tau)_z N\,, \label{eq:LagrangianaPV0}\\
  \mathcal{L}_{PVTC,\pi N}^{(1)}&=&-{h^0_V\over 2f_\pi}
  \overline{N}\gamma^\mu \partial_\mu (\vec \tau\cdot\vec \pi) N
  -  \frac{h_V^1}{f_\pi}\overline{N}\gamma^{\mu}N \partial_\mu \pi_z
  \nonumber\\ 
    &&-\frac{2h_V^2}{f_\pi} \sum_{a,b} \mathcal{I}_{ab} \partial_\mu
  \pi_a \, \overline{N} \gamma^\mu \tau_b N- \frac{h_A^1}{f_\pi^2}\overline{N}\gamma^{\mu}\gamma^5 N
   (\vec\pi\times\partial_\mu\vec\pi)_z \nonumber \\
  &&  +{h^2_A\over f_\pi^2}  \sum_{a,b=1}^3 \mathcal{I}_{ab} \overline{N}
  \Bigl( (\vec\pi\times\partial_\mu\vec\pi)_a \tau_b +
         \partial_\mu \pi_a (\vec\pi\times\vec\tau)_b\Bigr)
         \gamma^{\mu}\gamma^5 N\,,  \label{eq:LagrangianaPV1}
\end{eqnarray}
where
\begin{equation}
  \mathcal I_{ab}=
\left(
\begin{array}{ccc}
-1 & 0 & 0\\
0 & -1& 0 \\
0 & 0  &  +2 \\
\end{array}
\right)
\,.\label{eq:Iab}
\end{equation}
The parameters $h^1_{\pi}$ and $h^{\Delta I}_{V,A}$ are unknown LECs.
The superscript $\Delta I$
% is a label of the isorank of the operator.
labels the rank of the corresponding isospin tensor.
The LECs can be estimated by naive dimensional analysis
(NDA)~\cite{Kaplan:1992vj,Zhu:2004vw,deVries:2014vqa,Viviani:2014zha}   
\begin{equation}
  h^1_{\pi}  \sim G_F f_\pi \Lambda_\chi \sim10^{-6}\,,\qquad
  h^{\Delta I}_{V,A} \sim \frac{f_\pi}{\Lambda_\chi} h^1_{\pi} \sim 10^{-7}\,,\label{eq:pvestim}
\end{equation}
where $\Lambda_\chi=4\pi f_\pi\sim 1.2$ GeV is the typical scale of the strong interaction. 
Eq.~(\ref{eq:pvestim}) shows the order-of-magnitude estimates of the PVTC interactions.
These estimates do not take into account factors of $s_w^2$ and $N_c$ that could modify the
expected scaling of the LECs.

The contact terms entering the Lagrangian $\mathcal L_{PVTC,NN}$ are
products of a pair of bilinears of nucleon fields that are odd under P and
even under CP. The most general bilinear product reads 
\begin{equation}\label{LagPV:NN}
 \widetilde{O}_{AB}=\sum_{a,b=0}^3 F_{ab}
(\overline{N}\,\tau_a
 \Gamma_A \, N )\, 
(\overline{N}\, \tau_b \Gamma_B\,   N )\,, 
\end{equation}
where $\Gamma_A$ and $\Gamma_B$ are elements of the Clifford algebra
with the possible addition of 4-gradients and $ F_{ab}$ are unknown
parameters.
To violate P but conserve CP, at least one 4-gradient is required.
We must build isoscalar, isovector and isotensor terms as
discussed in Section~\ref{sec:micro}. The operators moreover have to conserve the
electric charge and thus commute with the third component
of the isospin operator. The terms with only one gradient operator are
collected in $\mathcal L^{(1)}_{PVTC,NN}$ (i.e. of chiral order 1). Only five
independent terms can be written~\cite{Girlanda:2008ts}, corresponding
to the five possible $S\leftrightarrow P$ transitions in NN
scattering~\cite{Danilov:1965hc}. It is more convenient to give the
Lagrangian using the non-relativistic reduction of the nucleon fields $N_r$:
\begin{eqnarray}
  \mathcal L_{PVTC, NN}^{(1)} &=& \frac{1}{\Lambda_\chi^2 f_\pi}\biggl[
    \frac{C_1}{2}\bmna \times (N_r^\dagger\bmsi N_r) \cdot  N_r^\dagger \bmsi N_r +
    \frac{C_2}{2}\bmna \times (N_r^\dagger \bmsi \tau_a N_r) \cdot  N_r^\dagger \bmsi\tau_aN_r \nonumber \\
    &+& C_3 \epsilon_{ab3} \bmna \cdot (N_r^\dagger \bmsi \tau_a N_r)   N_r^\dagger \tau_b N_r        +
    C_4 \bmna \times (N_r^\dagger \bmsi \tau_3 N_r) \cdot  N_r^\dagger \bmsi N_r \nonumber\\
    &+ & \frac{C_5}{2} {\mathcal I}_{ab} \bmna \times  (N_r^\dagger \bmsi \tau_a N_r) \cdot  N_r^\dagger \bmsi
       \tau_b N_r\biggr]\,. \label{eq:LPVNN1}
\end{eqnarray}
The factor $\frac{1}{\Lambda_\chi^2 f_\pi}$ has been chosen to ensure
$C_i$ are dimensionless and for convenience in the power counting. 
The construction of Eq. \eqref{eq:LPVNN1} and the elimination of redundancies 
will be discussed in more details in Section \ref{sec:pionless}.
The operators multiplying the LECs $C_{1,2}$ are isoscalar, those
multiplying $C_{3,4}$ change isospin by one unit, while that
multiplying $C_5$ is an isotensor. The scaling of the LECs from naive
dimensional analysis \cite{Manohar:1983md} is given by 
\begin{eqnarray}
 C_i \sim G_F \Lambda_\chi f_\pi\,,
\end{eqnarray}
which once again does not take into account the suppression by $s_w^2$ affecting,
for example, the isovector operators.
The operators in Eq. \eqref{eq:LPVNN1} contribute to the PVTC potential at NLO (suppressed by $(Q/\Lambda_\chi)^2$ with respect to LO),
and we will give the potential derived from them in Eq. \eqref{eq:potctpv}. The terms appearing in
$\mathcal L_{PVTC,NN}^{(3)}$ contain two additional gradients and contribute to the PVTC
potential at higher order. They have not been considered so far. 

Finally, there are some terms with $3\pi$ vertices appearing in $\mathcal
L^{(2)}_{PVTC,\pi\pi\pi}$ as discussed in Ref.~\cite{Viviani:2014zha}.
These terms would contribute to the $Q^2$ PVTC potential, but their
contributions at the end vanishes as discussed in Subsect.~\ref{sec:pvtcpot}. 

\subsubsection{Connection to the underlying PVTC sources}
\label{sec:chipvsources}

Attempts to estimate the values of the coupling constants were performed
mainly in the framework of the meson exchange models (which will
be discussed in Section~\ref{sec:obep}). However, since in both
$\chi$EFT and meson exchange frameworks the lowest order pion-nucleon
Lagrangian term is the same as given in Eq.~(\ref{eq:LagrangianaPV0}),
we can report here the values for $h^1_\pi$ estimated from the
underlying fundamental theory also before the advent of $\chi$EFT~\cite{Michel:1964zz,Donoghue:1992dd,
McKellar:1967mxj,Fischbach:1968zz,Tadic:1969xx,Kummer:1968ra,Mckellar:1973rr}. One of the most
comprehensive calculation including all previous results was performed in 1980 by
Desplanques, Donoghue, and Holstein (DDH)~\cite{Desplanques:1979hn} using the valence
quark model. Additional calculations have been performed later~\cite{Dubovik:1986pj,Feldman:1991tj,Meissner:1998pu},
using similar or other methods and finding qualitatively similar results.
These estimates, however, are based on a series of
rather uncertain assumptions (see, for example, Ref.~\cite{Haeberli:1995uz}). 
For example, DDH presented not a single value for $h^1_\pi$ but rather a
{\it range} inside of which it was extremely likely that this parameter
would be found~\cite{Desplanques:1979hn}.
In addition they presented also a single number called the ``best value"
but this is described simply as an educated guess in view of all
the uncertainties.  The values of $h^1_\pi$ were~\cite{Desplanques:1979hn}
\begin{equation}
  \textrm{DDH:}\quad h^1_\pi= 4.56\times 10^{-7}\quad \textrm{(``best\ value'')}\,,\qquad
  h^1_\pi= 0-11.4\times 10^{-7}\quad \textrm{(``reasonable\ range'')}\,.\label{eq:ddhbest}
\end{equation}
Some years ago, a lattice QCD calculation of $h^1_\pi$ has also been made~\cite{Wasem:2011zz}, resulting in the following
estimate
\begin{equation}
  \textrm{Lattice:}\quad h^1_\pi= (1.1\pm0.5)\times 10^{-7}\,,\label{eq:lattice}
\end{equation}
where the theoretical uncertainty is related to the statistical Monte
Carlo error. While the systematic errors are expected to be within
the quoted statistical error~\cite{Wasem:2011zz}, 
we stress that the calculation was performed at a heavy pion mass and not extrapolated to the physical point,
disconnected diagrams were not included, and operator renormalization was neglected.

Regarding the other LECs entering the contact Lagrangian given in Eq.~(\ref{eq:LPVNN1}),
no direct estimates were reported in literature. These LECs were estimated by
comparing the expression of contact potential with the potential
developed using the exchanges of heavy mesons, as for example, in the
DDH potential~\cite{deVries:2014vqa,Viviani:2014zha}
(this issue will be considered in more detail in
Sect.~\ref{sec:obep}). However, since also the DDH estimates are
rather uncertain, here we will not discuss this issue.

\subsection{The PVTV Lagrangian}
\label{sec:chitv}
The PVTV chiral Lagrangian taking into account the QCD $\bar\theta$
term was first considered in the seminal paper by Crewther, di
Vecchia, Veneziano and Witten \cite{Crewther:1979pi}, and consequently
revisited in 
Refs. \cite{Cheng:1990pi,Pich:1991fq,Cho:1992rv,Borasoy:2000pq,Ottnad:2009jw}. Subleading 
terms in the chiral expansion were systematically constructed in
Refs. \cite{Mereghetti:2010tp,Bsaisou:2014oka}. The chiral Lagrangian
induced by the dimension-six operators in Eq. \eqref{quark} were
derived in Refs. \cite{deVries:2012ab,Bsaisou:2014oka}. 

As before, in $SU(2)$ $\chi$PT, the PVTV Lagrangian can be organized in sectors
with different  numbers of pions and nucleons
\begin{eqnarray}
  \mathcal L_{PVTV} &=&  \mathcal L_{PVTV,\pi N} + \mathcal
  L_{PVTV,NN}+\mathcal L_{PVTV,\pi\pi\pi} +\cdots \,,\\
  \mathcal L_{PVTV,\pi N} &=& \mathcal L_{PVTV,\pi N}^{(0)}+ \mathcal L_{PVTV,\pi N}^{(1)}+\cdots  \,,\\
  \mathcal L_{PVTV,NN} &=& \mathcal L_{PVTV,N N}^{(1)}+ \mathcal L_{PVTV,NN}^{(3)}+\cdots  \,,\\
  \mathcal L_{PVTV,\pi\pi\pi} &=& \mathcal L_{PVTV,\pi\pi\pi}^{(0)}+ \cdots  \,.  
\end{eqnarray}
As in the previous subsection, we report here only the most important
interactions for each sector, focusing on the terms with the minimum
number of pion fields entering in the final expression of the
potential. Terms with additional pions are not universal for the
different PVTV sources at the quark level, but instead
depend on their chiral-symmetry breaking pattern. These differences only
enter at higher order in the potentials than we consider here.

In the PVTV case, the simultaneous violation of P, T, and
isospin symmetry allows for a pion tadpole  linear in the pion field $\sim \pi_3$ with a
corresponding LEC proportional to the symmetry-violating source terms
at the quark level. Such tadpoles can always be removed by
appropriate field redefinitions of the pion and nucleon fields
\cite{Mereghetti:2010tp,deVries:2012ab,Bsaisou:2014oka}. At LO in the
chiral expansion, the tadpole removal is the same as the vacuum
alignment procedure at the quark level \cite{Baluni:1978rf}. While tadpoles
can be removed, the corresponding field redefinitions affect other
couplings in the chiral Lagrangian. In particular, for
chiral-symmetry-breaking CP sources that do not transform as a quark mass term, a PVTV three-pion
vertex of chiral order $Q^0$ is left behind \cite{deVries:2012ab,Bsaisou:2014oka}. 
\begin{equation}\label{LagTV:pi}
\mathcal L^{(0)}_{PVTV,\pi\pi\pi} = M \bar\Delta {\pi_3 \vec \pi^2} \,,
\end{equation}
where $\bar\Delta$ is a LEC. 
Other three-pion vertices will appear at N${}^2$LO, but they
will contribute to high orders of the PVTV potential.

Arguably the most important interactions appear in the pion-nucleon
sector. Simultaneous violation of P, T, and chiral symmetry allows
for non-derivative single-pion-nucleon interactions, something which
is not possible in the PCTC Lagrangian. In principle, three different
interactions can be written  
\begin{equation}\label{LagTV:piN0}
  \mathcal L^{(0)}_{PVTV,\pi N} = {\bar g_0} \overline{N} \vec \pi
  \cdot \vec \tau N  +{\bar g_1} \overline{N}  \pi_3 N
  +{\bar g_2} \overline{N} \pi_3 \tau_3 N\,, 
\end{equation}
corresponding, respectively, to an isospin singlet, vector, and tensor
interaction. As discussed below, the relative size of the LECs
$\bar g_{0,1,2}$ strongly depends on the quark-level PVTV source under
consideration. In the case of CP-violation from chiral invariant operators, such as the gCEDM,
$\bar g_i$ are suppressed by powers of the pion masses, and the pion-nucleon Lagrangian contains 
chiral-invariant, derivative couplings as important as those in Eq. \eqref{LagTV:piN0} \cite{deVries:2012ab}.
These can however always be absorbed into a shift of $\bar g_{0}$ and of the $\Delta I=0$
NN operators discussed below.

The NLO Lagrangian contains several two-pion two-nucleon PVTV
interactions \cite{Mereghetti:2010tp,deVries:2012ab,Bsaisou:2014oka,Gnech:2019dod},
but, for all CP-violating sources, they contribute to the two- and three-body PVTV potentials 
at N$^3$LO and N$^2$LO, respectively. We therefore ignore these couplings.
Isospin-breaking sources also generate a single-pion-nucleon NLO coupling. 
The coupling involves a time derivative of the pion field,
thus inducing a relativistic 
correction in the $\mathcal O(Q)$ PVTV potential.
At N$^2$LO the number of interactions proliferates significantly and
there are also new pure pionic interactions. These contributions can either be absorbed into LO LECs or appear 
at high orders in the PVTV potential considered here.

Apart from pionic and pion-nucleon interactions, there appear PVTV NN
contact interactions. As in the PVTC case, at least one gradient is
required such that these operators start at order $Q$.  Terms with
three or more gradients have not been considered so far. At order $Q$,
only five independent interactions of this kind can be written,
corresponding to the five possible $S\leftrightarrow P$ 
transitions (see Section 4 for a general discussion of
this kind of interaction terms). Neglecting terms with multiple pions,
the Lagrangian reads (again, it is convenient to write it in terms of
the non-relativistic nucleon field $N_r$)
\begin{eqnarray}
  \mathcal L_{PVTV, NN}^{(1)} &=& \frac{1}{\Lambda^2_\chi f_\pi}\biggl[
    \bar C_1\bmna \cdot (N_r^\dagger\bmsi N_r) \,  N_r^\dagger  N_r +
    \bar C_2\bmna \cdot (N_r^\dagger \bmsi \tau_a N_r) \,  N_r^\dagger \tau_a N_r \nonumber \\
 &+& \bar C_3 \bmna \cdot (N_r^\dagger \bmsi \tau_3 N_r)   N_r^\dagger  N_r
    + \bar C_4\bmna \cdot (N_r^\dagger \bmsi  N_r)   N_r^\dagger \tau_3  N_r \nonumber\\
    &+&\bar C_5 {\mathcal I}_{ab} \bmna \cdot  (N_r^\dagger \bmsi \tau_a N_r) \,  N_r^\dagger \tau_b N_r
    \biggr]\,. \label{eq:LTVNN1}
\end{eqnarray}
As suggested by the factor of $\Lambda_\chi^2$ which we pulled out of the definition of the LECs,
in $\chi$EFT these operators contribute in general at N$^2$LO
and are suppressed with respect to the PVTV one-pion exchange (OPE) potential. The only exception,
as discussed in Sec. \ref{sec:chitvsources}, are quark-level operators that do not
break chiral symmetry, for which $\bar C_{1,2}$ are as important as the contributions
from $\bar g_{0,1}$.

Finally, the calculation of EDMs or other PVTV electromagnetic moments
requires the inclusion of electromagnetic currents. Nucleon EDMs are
induced by pion loops involving the interactions in $\mathcal
L^{(0,1)}_{TV,\pi N}$. The renormalization of these loops requires the
inclusion of short-distance counter terms contributing to the nucleon
EDMs. Such counter terms indeed appear in the chiral Lagrangian 
\begin{eqnarray}\label{LagTV:edm}
 \mathcal L_{PVTV,N \gamma} &=&{1\over 4} \overline{N} \left( \bar{d}_{0 }+ \bar{d}_{1 }
 \tau_3 \right) \epsilon^{\mu\nu\alpha\beta} \sigma_{\mu\nu} N\, F_{\alpha\beta}\,,
\end{eqnarray}
where $F_{\alpha\beta}$ is the electromagnetic field strength and
$\bar d_0$ and  $\bar d_1$ are LECs related to the proton and neutron
EDMs, respectively. The above interactions are sufficient for calculations of hadronic and
nuclear PVTV scattering observables and EDMs up to
NLO in the chiral expansion. Calculations of higher PVTV moments, such as
magnetic quadrupole moments, can depend on additional LECs \cite{Liu:2012tra}. 

\subsubsection{Connection to the underlying PVTV sources}
\label{sec:chitvsources}
\begin{table}
\begin{tabular}{c|cccccccc}
	   &  $(4\pi \epsilon_{m_\pi}) \, \bar\theta$ &  $ (4\pi
  \epsilon_{m_\pi}) \epsilon_v \tilde c^{(u,d)}_{g}$  & $(4\pi
  \epsilon_{m_\pi}) \, \epsilon_v \tilde c^{(u,d)}_{\gamma}$  &
  $4\pi\epsilon_v \tilde C_G$ & $\epsilon_v \Xi_{1,2}^{(ud)}/(4\pi)$ &
  $ \epsilon_v \Sigma_{1,2}^{(ud)}/(4\pi)$\\ 
	   \hline
$\bar\Delta $ 		&  $\epsilon_{m_\pi}  $ & $\epsilon_{m_\pi}$& --& $\varepsilon \epsilon_{m_\pi}^2$ & $1$ &
           $\epsilon_{m_\pi} $  \\	    
$\bar g_0$ 		&  $1$ & $1$ & --  & $
           \epsilon_{m_\pi}$ & $\varepsilon \epsilon_{m_\pi}$ &
           $\epsilon_{m_\pi}$  \\ 
$\bar g_1$ 		&  $\varepsilon\epsilon_{m_\pi}$  &
           $1$ & --  & $\varepsilon \epsilon_{m_\pi}$ & $1$ &
           $\varepsilon\epsilon_{m_\pi}$ \\ 
$\bar g_2$ 		&  $\varepsilon^2 \epsilon^2_{m_\pi}$
           & $\varepsilon \epsilon_{m_\pi}$ & --  & $\varepsilon^2
           \epsilon^2_{m_\pi}$ & $\varepsilon \epsilon_{m_\pi}$ &
           $\varepsilon^2 \epsilon^2_{m_\pi}$ \\ 
$\bar d_{0,1} f_\pi $ 		&  $e\, \epsilon_\chi$ & $e
           \,\epsilon_\chi$ & $e \, \epsilon_\chi$  & $e \,
           \epsilon_\chi$ & $e\, \epsilon_\chi$ & $e\,
           \epsilon_\chi$\\ 
$\bar C_{1,2} $ 		& $1$  & $1$ &
           -- & $1$ & $\varepsilon 
           \epsilon_{m_\pi}$ & $1$\\ 
$\bar C_{3,4} $ 		& $\varepsilon 
           \epsilon_{m_\pi}$ & $1$ & -- & $\varepsilon
            \epsilon_{m_\pi}$ & $1$ &
           $\varepsilon  \epsilon_{m_\pi}$ \\ 
$\bar C_{5} $ 		& $\varepsilon^2 
           \epsilon^2_{m_\pi}$ & $\varepsilon 
           \epsilon_{m_\pi}$ & -- & $\varepsilon^2 
           \epsilon^2_{m_\pi}$ & $\varepsilon 
           \epsilon_{m_\pi}$ & $\varepsilon^2 
           \epsilon^2_{m_\pi}$ \\
\hline
\end{tabular}
\caption{Scaling of the LECs in the chiral Lagrangian in dependence of
  the microscopic CP violation sources. We introduced the counting parameters 
$\epsilon_v \equiv {\Lambda^2_\chi}/{v^2}$, $\epsilon_{m_\pi} \equiv
  m^2_\pi/\Lambda_\chi^2$, 
$\epsilon_{\chi} \equiv f^2_\pi/\Lambda_\chi^2$. With
  $\epsilon_{m_\pi} \sim \epsilon_{\chi}$, we introduced two different
  parameters to explicitly track insertions of the light quark masses 
from the QCD Lagrangian. $\varepsilon$ is the isospin breaking
parameter  $\varepsilon = (m_d-m_u)/(m_d + m_u) \simeq 1/3$.  
The scaling of the LECs induced by dimension-six sources assume a
Peccei-Quinn mechanism. A ``$-$'' implies the interaction is only induced
at higher order than considered here. The parameters $\bar C_{1,2}$,
$\bar C_{3,4}$, and $\bar C_{5}$ are the LECs entering the
contact PVTV potential, respectively of isoscalar, isovector, and
isotensor type.}\label{Tab:LECs}
\end{table}

In the previous section we listed the PVTV hadronic interactions
relevant for observables of experimental interest. However, for a
given PVTV source at the quark-gluon level, a specific hierarchy among
the various interactions appear.  The relative importance of the LECs
in Eqs. \eqref{LagTV:pi}, \eqref{LagTV:piN0}, \eqref{eq:LTVNN1} and
\eqref{LagTV:edm} for the different microscopic sources of CP violation  is
summarized in Table \ref{Tab:LECs}. 
These estimates are based on NDA~\cite{Manohar:1983md}.
NDA is valid in the regime in which the strong coupling $g_s$ is non-perturbative, 
and, as done for NDA estimates of the chiral-invariant PCTC
interactions,  we will take $g_s \simeq 4\pi$.  In addition, for
dimension-six sources, we assumed that a Peccei-Quinn 
mechanism \cite{Peccei:1977hh} relaxes $\bar\theta$ to an induced
$\bar\theta_{\rm ind}$, which depends on the coefficients and vacuum
matrix elements of the operators in Eq. \eqref{quark}
\cite{Pospelov:2005pr,Mereghetti:2015rra,Cirigliano:2016yhc}. The
scaling of the couplings without this assumption can be found in
Ref. \cite{deVries:2012ab}. To make the power counting explicit, we
introduced three ratios of scales  
\begin{equation}
  \epsilon_v \equiv \frac{\Lambda^2_\chi}{v^2}, \qquad \epsilon_{m_\pi}
  \equiv \frac{m_\pi^2}{\Lambda_\chi^2}, \qquad \epsilon_\chi \equiv
  \frac{f_\pi^2}{\Lambda_\chi^2} = \frac{1}{(4\pi)^2}\,.  
\end{equation}
Numerically, $\epsilon_\chi \sim \epsilon_{m_\pi}$, but we define two
different parameters to track the dependence of the LECs on the quark
masses. 
To assess the size of the contribution of different CP violating sources to the nucleon
and nuclear EDMs, the scaling of the LECs in Table \ref{Tab:LECs} can be combined with a
naive estimate of these observables.
As we will discuss in detail in Sections \ref{sec:chiedm} and \ref{sec:resedm}, the nucleon EDM receives tree level contributions 
from $\bar d_{0,1}$ and loop contributions by $\bar g_0$ and $\bar g_1$, leading to
\begin{eqnarray}\label{dnNDA}
d_{n,p} &\sim&  \frac{\bar d_0 \mp \bar d_1}{2} + \frac{e}{f_\pi}
\epsilon_{\chi} \left( \alpha_0 \bar g_0  + \alpha_1 \bar g_1 \epsilon_{m_\pi}^{1/2} +
\ldots \right)\,, 
\end{eqnarray}
where $e$ is the electric charge and the coefficients of the
  loops $\alpha_{0,1}$ will be given explicitly in 
Section \ref{sec:chiedm}. The additional suppression of $\bar g_1$ is
due to the fact that this coupling  only involves neutral pions, which
do not interact with a single photon at LO. Nuclear EDMs, on the other
hand, receive tree level contributions from the single nucleon EDM,
and from pion-nucleon and nucleon-nucleon couplings, 
\begin{eqnarray}\label{dANDA}
d_A &=&  a_n d_n + a_p d_p + e \left( a_\Delta \bar
\Delta + \sum^{2}_{i=0} a_i \bar g_i  + \epsilon_\chi  \sum_{i=1}^5 A_i  \bar
C_i\right) \,. 
\end{eqnarray}
The coefficients $a_{n,p}$, $a_{\Delta,0,1,2}$ and $A_{1,\ldots,5}$
depend on the nucleus under consideration, and in
Section \ref{sec:resedm} we will present results for their  calculation
in chiral EFT for the deuteron, $^3$H and $^3$He. By power counting, 
they are expected to be $\mathcal O(1)$ (measured in units of fm in the case of the dimensionful $a_{\Delta,0,1,2}$ and $A_{1,\ldots,5}$),
barring isospin selection rules, which for example suppress 
the contributions of the isoscalar operators $\bar g_0$ and $\bar C_{1,2}$
in nuclei with $N=Z$, such as the deuteron \cite{Liu:2004tq,deVries:2011an} \footnote{$\bar g_0$
and $\bar C_{1,2}$ contribute to the deuteron EDM in conjunction with isospin breaking in the strong interaction, 
or via the spin-orbit coupling of the photon to the nucleons \cite{deVries:2011an}. Both contributions are beyond the accuracy we work at in this paper.}.

The reader should be aware that the dimensionless Wilson coefficients
of the dimension-six operators, $\tilde c^{(u,d)}_{g}$, $\tilde
c^{(u,d)}_{\gamma}$, $ \tilde C_G$,  $\Xi_{1,2}^{(ud)}$ and
$\Sigma_{1,2}^{(ud)}$ also come with intrinsic suppression
factors. These arise from the typical loop and chiral factors that
appear in BSM models.  
For example,  quark and gluon dipole operators are typically induced
at the one-loop level, and the quark EDM and chromo-EDM coefficients come
with explicit factors of the quark mass (already included in
Eq.~\eqref{quark}). This implies that one can expect $\{ \tilde
c^{(u,d)}_{g} , \tilde c^{(u,d)}_{\gamma} , \tilde C_G \} = \mathcal
O( \epsilon_\Lambda/(4\pi)^2 )  $, where $\epsilon_\Lambda =
v^2/\Lambda_X^2$. Of course this is just an estimate and certainly
models exist where these operators appear only at the two- or
higher-loop level. On the other hand, the four-quark operators  $\Xi$
and $\Sigma$ can be induced at tree level, so that $\{ \Xi, \Sigma \}
= \mathcal O( \epsilon_\Lambda)$.  
Once the matching coefficients are calculated in a given model, Table
\ref{Tab:LECs} and Eqs. \eqref{dnNDA}-\eqref{dANDA}  allow to identify the dominant low-energy operator and
to get a rough idea of the EDM constraints. 

Table \ref{Tab:LECs} highlights that the chiral and isospin properties
of the quark-level CP-violating sources induce very specific hierarchies
between different low-energy couplings. 
These hierarchies in turn imply different relations between the EDMs
of the nucleon, deuteron, and three-nucleon systems, which, if
observed, would allow to disentangle the various CP-violating sources.  
From Table \ref{Tab:LECs} we see that chiral-symmetry-breaking
sources, such as $\bar\theta$, $\tilde c_g^{(u,d)}$ and
$\Xi_{1,2}^{(u,d)}$, induce relatively large PVTV pion-nucleon couplings.
These couplings appear in the table with entry $1$, indicating no
further suppression. In particular, the isoscalar $\bar\theta$ term
and isovector $\Xi^{(u,d)}$ predominantly induce, respectively, $\bar
g_0$ and $\bar g_1$, while a qCEDM would yield both couplings with
similar strengths. The consequence is that
for these sources light nuclear EDMs are enhanced with respect to the
nucleon EDM. For these chiral-symmetry-breaking sources, the  contact nucleon
interactions proportional to $\bar C_{i}$ are suppressed in the
chiral expansion because these operators involve an explicit
derivative. The suppression can be explicitly seen combining the scaling in Table \ref{Tab:LECs}
with the explicit factor of $\epsilon_\chi$ in Eqs. \eqref{eq:LTVNN1} and \eqref{dANDA}.

Chiral invariant sources such as the Weinberg operator $\tilde C_{G}$
and the four-quark operators $\Sigma_{1,2}^{(u,d)}$, on the other
hand, require additional chiral-symmetry breaking to generate $\bar g_{0,1}$,
as indicated by extra powers of $\epsilon_{m_\pi}$.  
In this case, EDMs of light-nuclei are expected to be of similar size
as the nucleon EDM.  Furthermore, the contact nucleon operators
proportional to $\bar C_{1,2}$ now contribute to the PVTV potential at
the same order as $\bar g_{0,1}$. Finally, the qEDM mostly induces $\bar d_{0,1}$, all other
couplings being suppressed by $\mathcal O(\alpha_{\rm em})$,
where $\alpha_{\rm em}$ is the fine structure constant $\sim 1/137$. In this
case one expects nuclear EDMs to be dominated by the constituent
nucleon EDMs. 

While most statements are source-dependent, there is an important
general message hidden in Table \ref{Tab:LECs}. There is no PVTV
source for which the couplings $\bar g_2$ and $\bar C_{3,4,5}$ appear at
LO. For all sources they appear with a relative suppression of
$\varepsilon \epsilon_{m_\pi}$ or $\epsilon_\chi$ compared to other
PVTV interactions. For most calculations one can simply neglect the
associated interactions, reducing the number of LECs entering the
expression of hadronic and nuclear observables. The suppression of the
LECs $\bar g_2$ and $\bar C_{3,4,5}$ ultimately is a consequence of
imposing gauge invariance on the dimension-six PVTV sources. 

Table \ref{Tab:LECs} relies on NDA estimates for
hadronic matrix elements \cite{Manohar:1983md}.  
A more quantitative assessment of the discriminating power of EDM
experiments necessitates to replace the NDA estimates in Table
\ref{Tab:LECs} with solid nonperturbative 
calculations of the LECs.   
At the moment, there exist controlled estimates only of a few LECs. 
The pion-nucleon couplings $\bar g_0$ induced by the QCD $\bar\theta$
term is related by  chiral symmetry to modifications in the baryon
spectrum \cite{Crewther:1979pi}. In particular, in $SU(2)$ $\chi$PT  
$\bar g_0$ is related to the quark mass contribution to the nucleon
mass splitting \cite{Mereghetti:2010tp,deVries:2015una}, up to N${}^2$LO
corrections. Using Lattice QCD evaluations of the nucleon mass
splitting \cite{Borsanyi:2014jba,Brantley:2016our}, one finds 
\begin{equation}\label{g0theta}
  \bar g_0(\bar\theta) = \left( 15.5 \pm 2.6 \right)
  \cdot 10^{-3} \, \bar\theta\,, 
\end{equation}
where the $15\%$ error includes both the Lattice QCD error on $m_n - m_p$,
and an estimate of the error from N$^2$LO chiral corrections. 
Unfortunately, chiral-symmetry-based relations do not allow to extract  $\bar
g_1$ and $\bar d_{0,1}$. $\bar g_1$ has been estimated with resonance
saturation leading to $\bar g_1( \bar \theta)/\bar g_0(\bar
\theta)\simeq -0.2$,
somewhat larger than expected from NDA \cite{Bsaisou:2012rg}. The
LECs $\bar d_{0,1}$ 
are usually estimated by naturalness arguments and considered to be of
similar size to non-analytic contributions to the isoscalar and
isovector nucleon EDM, see the Subsect.~\ref{sec:chiedm}.

The relation between PVTV pion-nucleon couplings and corrections to the
nucleon and pion masses is not specific to the QCD $\bar\theta$ term,
but can be generalized to all chiral-symmetry-breaking sources, 
such as for example the qCEDM \cite{deVries:2012ab,deVries:2016jox}
and $\Xi^{(ud)}_{1,2}$ \cite{Seng:2016pfd,Cirigliano:2016yhc}. Since
corrections to spectroscopic quantities should be easier to compute  
on the lattice, these chiral relations allow a calculation of $\bar
g_{0,1}$ in Lattice QCD. While promising, this strategy has yet to lead to
controlled results. The best estimate of 
$\bar g_{0,1}$ induced by the qCEDM comes from QCD sum rules
\cite{Pospelov:2000bw,Pospelov:2005pr} 
\begin{equation}
  {\bar g_0} = \left( 0.1 \pm 0.2 \right)  \left( 0.7 \tilde
  c_g^{(u)} - 1.5 \tilde c_g^{(d)} \right) \cdot 10^{-6}, \qquad   
  {\bar g_1} = \left( 0.4^{+0.8}_{-0.2} \right)  \left( 0.7
  \tilde c_g^{(u)} - 1.5 \tilde c_g^{(d)} \right) \cdot 10^{-6}\,. 
\end{equation}
These estimates agree with NDA, especially for $\bar g_1$. However, $\bar g_0$
seems to be slightly suppressed, in agreement with large-$N_c$
expectations \cite{Samart:2016ufg}.

Only for the four quark operators proportional to $\Xi_{1,2}^{(ud)}$
of Eq.~(\ref{quark}) does the three-pion vertex with LEC $\bar\Delta$
appear at LO in the chiral Lagrangian. For this case,  
the LEC $\bar\Delta$ is related by $SU(3)$ symmetry to $K \rightarrow \pi
\pi$ matrix elements and $K - \bar K$ matrix elements that have been
calculated on the lattice. We obtain 
\begin{equation}
\bar \Delta =  \frac{f_\pi}{M v^2}  \left( \mathcal A_{1\, LR} \,
\textrm{Im}\,  \Xi^{(ud)}_{1}  
+ \mathcal A_{2\, LR} \, \textrm{Im}\,  \Xi^{(ud)}_{2}
\right )\,,
\end{equation}
with 
\begin{equation}\label{Xi}
\mathcal A_{1\, LR}(\mu = 3\, {\rm GeV}) = \left( 2.2 \pm 0.13 \right)
\, \rm{GeV}^2, \quad \mathcal A_{2\, LR}(\mu = 3\, {\rm GeV}) = \left(
10.1 \pm 0.6 \right) \rm{GeV}^2\,. 
\end{equation}
The matrix elements in Eq. \eqref{Xi} are in good agreement with
NDA. The value of $\bar\Delta$ also determines the tadpole component
of $\bar g_1$,  which again is in line with NDA.

Most of the remaining LECs are undetermined at present. The focus
of the Lattice QCD community has been on the matrix elements connecting the
nucleon EDMs to the $\bar\theta$ term \cite{Guo:2015tla, Abramczyk:2017oxr, Dragos:2019oxn},
the qEDMs \cite{Gupta:2018qil,Aoki:2019cca}, the qCEDMs \cite{Abramczyk:2017oxr,Kim:2018rce}, and the Weinberg
operator \cite{Rizik:2018lrz}. Some results are given in next subsection.

\subsubsection{The nucleon EDM in chiral perturbation theory}
\label{sec:chiedm}

The PVTV LECs defined in the previous section can be used to calculate
the nucleon PVTV electric dipole form factor (EDFF). At zero momentum
transfer, the EDFFs are identified with the nucleon EDMs. In dimensional
regularization with modified minimal subtraction  up to NLO in the
chiral expansion, the EDMs are given by \cite{Ottnad:2009jw,Mereghetti:2010kp}
\begin{eqnarray}\label{chirallog}
d_n &=& \bar d_0  (\mu)-\bar d_1 (\mu) + \frac{e g_A \bar g_0}{(4\pi)^2
  f_\pi} \left( \log \frac{m^2_\pi}{\mu^2} - \frac{\pi m_\pi}{2
  M} \right)  \,, 
\label{eq:dn}
\\
d_p &=& \bar d_0  (\mu) + \bar d_1 (\mu)  
- \frac{e g_A \bar g_0}{(4\pi)^2 f_\pi} \left[ \left( \log
  \frac{m^2_\pi}{\mu^2} - \frac{2 \pi m_\pi}{M} \right)   
- \frac{\bar g_1}{\bar g_0} \frac{ \pi m_\pi}{2 M}   \right]\,, 
\label{eq:dp}
\end{eqnarray}
where $\mu$ is the dimensional regularization scale.
The leading loops proportional to $\bar g_0$ are divergent and
 renormalized by the $\mu$-dependent LECs $ 
\bar d_{0,1}$. The NLO corrections proportional to $m_\pi/M$ are
finite.
The LEC $\bar \Delta$ does not contribute at this order for
any of the PVTV sources. As standard in $\chi$PT, the loops are associated
to inverse powers of $(4\pi f_\pi)^2 = \Lambda_\chi^2$. Combined with
the scaling of the LECs in Table~\ref{Tab:LECs}, we conclude that for
the $\bar\theta$ term and the qCEDMs the leading loop proportional to $\bar
g_0$ and the counter terms $\bar d_{0,1}$ appear at the same order.
For all other PVTV sources, the short-range counter terms $\bar
d_{0,1}$ are expected to dominate the nucleon EDMs. In no scenario can
the EDMs be calculated solely from the pion-nucleon LECs $\bar
g_{0,1}$ as is often assumed in the literature. Estimates for the nucleon EDMs are often obtained
by setting $\mu = M$ and $\bar d_{0,1}(\mu = M) =0$ such that EDMs depend on the value of $\bar g_{0,1}$, which for some PVTV sources is better known. 

The separation between the short-range and loop contributions is
scheme dependent and therefore not physical. Lattice QCD calculations
can therefore only calculate the total nucleon EDMs $d_n$ and
$d_p$. In recent years, significant efforts have been made towards
calculating the nucleon EDMs in terms of the underlying PVTV
sources. Most efforts have focused on the QCD $\bar\theta$ term and the
qEDM. The most recent results for the $\bar\theta$ term \cite{Dragos:2019oxn} give 
\begin{eqnarray}
d_n =  -(1.5 \pm 0.7)\cdot 10^{-3}\,\bar
\theta\,e\,\mathrm{fm}\,,\qquad d_p =  (1.1 \pm 1.0)\cdot
10^{-3}\,\bar \theta\,e\,\mathrm{fm}\,,
\end{eqnarray}
in good agreement, but with sizeable uncertainties, with expectations
from the chiral logarithm in Eq.~\eqref{chirallog} using Eq.~\eqref{g0theta}.
In the case of the qEDM, the nucleon EDM is related to the tensor
charges, which have been computed with good accuracy
\cite{Gupta:2018qil,Aoki:2019cca}. 
Using the FLAG average \cite{Aoki:2019cca}, we get
\begin{eqnarray}\label{qEDM}
d_n &=&   g_T^{d} \frac{Q_u m_u}{v^2} \tilde c_\gamma^{(u)} +  g_T^{u}
\frac{Q_d m_d}{v^2} \tilde c_\gamma^{(d)}  =  \left(  -(0.96 \pm 0.22
)  \tilde c_\gamma^{(u)} - (4.0 \pm 0.4 ) \tilde c_\gamma^{(d)}
\right) \cdot 10^{-9}  e \, \textrm{fm}\,,  \nonumber \\ 
d_p &=&  g_T^{u} \frac{Q_u m_u}{v^2} \tilde c_\gamma^{(u)} +  g_T^{d}
\frac{Q_d m_d}{v^2} \tilde c_\gamma^{(d)}=  \left(  ( 3.7 \pm 0.8 )
\tilde c_\gamma^{(u)} + (1.0 \pm 0.1 ) \tilde c_\gamma^{(d)}   \right)
\cdot 10^{-9}  e \, \textrm{fm}\,, 
\end{eqnarray}
where $Q_{u,d}$ are the $u$ and $d$-quark charges in units of the
electric charge, and $g^{u,d}_T$ the $u$ and $d$-quark tensor charges of the
proton, and the error on the r.h.s. of Eq. \eqref{qEDM} is dominated
by the uncertainty on the light quark masses.  

On a longer time-scale, calculations of the qCEDMs and the Weinberg
operator are also targeted. For now, the best results come from
calculations using QCD sum rules \cite{Pospelov:2005pr, Haisch:2019bml}.

\subsection{From the Lagrangian to the potential}
\label{sec:lag2pot}
In this subsection, we briefly present two methods that have been used to derive
nucleon-nucleon potentials starting from a Lagrangian. We first introduce
the notation used here and in the next subsections.

The process under consideration is the scattering of two
nucleons from an initial state $|\bmp_1\bmp_2\ket$ to the final state
$ | \bmp'_1 \bmp'_2\ket$ (hereafter the dependence on the
spin-isospin quantum numbers is understood). It is convenient to define the momenta 
\begin{equation}
  \bmK_j={\bmp_j'+\bmp_j\over 2}\,, \quad
  \bmk_j=\bmp_j'-\bmp_j\,,\label{eq:Kk}
 \end{equation}
 where $\bmp_j$ and $\bmp_j'$ are the initial and the final momenta of the nucleon $j$.
 Furthermore it is useful to define
 \begin{equation}
  \bmsi_j\equiv(\bmsi)_{s_j',s_j}\equiv \bra \frac{1}{2}s'_j|\bmsi |\frac{1}{2} s_j \ket \,,\quad
  \bmta_j\equiv(\bmta)_{t_j',t_j}\equiv \bra \frac{1}{2}t'_j|\bmta |\frac{1}{2} t_j \ket \,, 
\end{equation}
which are the spin (isospin) matrix element between the final state
$s_j'$ ($t_j'$) and the initial state $s_j$ ($t_j$) of the nucleon
$j$. 

Because $\bmk_1=-\bmk_2\equiv\bmk$ from the overall momentum
conservation $\bmp_1+\bmp_2=\bmp_1'+\bmp_2'$, the momentum-space
potential $V$ is a function of the momentum variables $\bmk$, $\bmK_1$
and $\bmK_2$, namely 
\begin{equation}
  \bra \bmp_1' \bmp_2' | V | \bmp_1 \bmp_2 \ket =V(\bmk,\bmK_1
  , \bmK_2) (2 \pi)^3 \delta( \bmp_1+\bmp_2 -  \bmp_1'-\bmp_2') \,.
  %\delta_{\bmp_1+\bmp_2,\bmp_1'+\bmp_2'} \,.
\end{equation}
Moreover, we can write in general 
\begin{equation}
  V(\bmk,\bmK_1,\bmK_2)=V^{(\text{CM})}(\bmk,\bmK)+V^{(\bmP)}(\bmk,\bmK) \,,
\end{equation} 
where $\bmK=(\bmK_1-\bmK_2)/2$, $\bmP=\bmp_1+\bmp_2=\bmK_1+\bmK_2$,
and the term $V^{(\bmP)}(\bmk,\bmK)$ represents a boost
correction to $V^{(\text{CM})}(\bmk,\bmK)$, the potential in the
center-of-mass frame (CM). Below we will ignore the boost correction
and provide expressions for $V^{(\text{CM})}(\bmk,\bmK)$ only.
Note that in the CM we define also $\bmp_1=-\bmp_2\equiv\bmp$
and $\bmp_1'=-\bmp_2'\equiv\bmp'$. So we have $\bmk=\bmp'-\bmp$ and
$\bmK=(\bmp'+\bmp)/2$, so in the following we also write
$V^{(\text{CM})}$ as $V^{(\text{CM})}(\bmp,\bmp')$.
From now on, we will suppress the superscript ``(CM)'' for
simplicity. 

In order to derive the potential, two methods have been frequently
used, the method of unitarity transformation (UT), and the method of the time-ordered
perturbation theory (TOPT). They are briefly introduced below.

{\it The time-ordered perturbation theory method.\ }
Let us consider the matrix element of the $T$-matrix,
$T_{fi}=\langle \bmp_1'\bmp_2'|T|\bmp_1\bmp_2\rangle$,
the ``amplitude'' of a process of scattering of two nucleons.
Its square modulus $|T_{fi}|^2$ is directly related to the cross section
of the process. The conventional perturbative expansion for this
matrix element is given as
\begin{equation}
 T_{fi}= 
 \langle \bmp_1'\bmp_2' \!\mid H_I \sum_{n=1}^\infty \left( 
 \frac{1}{E_i -H_0 +i\, \epsilon } H_I \right)^{n-1} \mid\! \bmp_1\bmp_2 \rangle \,,
\label{eq:pt}
\end{equation}
where $E_i$ is the energy of the initial state, $H_0$ is the Hamiltonian
describing free pions and nucleons, and $H_I$ is the Hamiltonian
describing interactions among these particles. These operators are
defined to be in the Schr{\"o}dinger picture and they can be derived from the
Lagrangian constructed in terms of pions and nucleons as described,
for example, in Refs.~\cite{Epelbaum:2002gb,Baroni:2015uza}. The evaluation
of $T_{fi}$ is carried out in practice by inserting complete sets of $H_0$ eigenstates
between successive $H_I$ factors. Power counting is then used to
organize the expansion in powers of $Q/\Lambda_\chi \ll 1$,
where  $Q$ stands for either an external momenta or the pion mass.
We will use the ``naive'' Weinberg counting rules~\cite{Weinberg:1990rz}, 
namely, we will count simply the powers of both the external momenta
and pion mass insertions (we will consider low energy
processes only). Each term will be of some order $(Q/\Lambda_\chi)^\nu$.  The
terms with the lowest power of $\nu$ will be the LO,
and so on.

In the perturbative series given in Eq.~(\ref{eq:pt}), a generic
contribution will be characterized by a certain number
of vertices coming from the interaction Hamiltonian $H_I$ and
energy denominators, and it can be visualized also as a
diagram (hereafter referred to as a TOPT diagram). Each vertex will
give a ``vertex function'' and a $\delta$ conservation of the
momenta of the particles involved in the vertex.
The vertex functions are the results of the
matrix elements of terms appearing in $H_I$  and are given as
products of Dirac four-spinors, momenta, etc. A sum over
the momenta of the particles entering the intermediate states is also
present. When a diagram includes one or more loops, the $\delta$'s
are not sufficient to eliminate all the sums over the momenta of the
intermediate states.  The energy denominators
come from the factors $1/(E_i-E_\alpha+i\epsilon)$, where $E_\alpha$ is
the (kinetic) energy of a specific intermediate state entering the calculation. The chiral
order of each diagram can be calculated as follows.  One needs to consider:
\begin{enumerate}
\item The chiral order of the vertex functions, which can be
  calculated from the non-relativistic (NR) expansion  of the nucleon Dirac four-spinors
  ($1/M$ expansion), and from various other factors. 
  Typically, the powers of $p/M$ coming from the NR expansion of the
  nucleon Dirac four-spinors are counted as $\sim Q^2$ \cite{Weinberg:1990rz,Epelbaum:2008ga,Krebs:2016rqz}. In other
  approaches however they are considered to be of order $Q$ \cite{Kaplan:1998tg,Pastore:2008ui,Pastore:2011ip}. In
  this paper, we will follow the first prescription.
  
\item The energy denominators. We note that typical momenta $\bmp$ of the nucleons
  are much smaller than the mass of the nucleons, so we can treat them non relativistically. Namely
   $\sqrt{p^2+M^2}\simeq M+\frac{p^2}{2M}  \sim O(Q^0)+O(Q^2)$.
  Regarding the pion energies, $\omega_{\mathbf{k}}=\sqrt{m_\pi^2 +k^2}\sim O(Q)$.
  Usually in the energy denominator all the nucleon masses $M$ cancel out
  and therefore we have two cases:
 \begin{itemize}
 \item if there are no pions in the intermediate state,
   the energy denominator has only  nucleon energy terms so it results of order $1/Q^2$.
 \item if there are pions in the intermediate states, the energy denominator reads
  \begin{equation}
  {1\over \Delta E-\omega_k}\sim -{1\over \omega_k}\left(1+{\Delta
  E\over \omega_k} + \cdots\right)\,,\label{eq:NRexpansionpropagator}
 \end{equation}
  where the term $\Delta E = E_1+E_2+\cdots-E_i$ where $E_1,\dots$ are the energies of the nucleons
  in the intermediate state and $E_i$ is the initial scattering
  energy. In the Taylor expansion the first term is of order 
  $Q^{-1}$, while the other terms  are usually called ``recoil corrections''. For the sake
      of consistency with the choice discussed above regarding the NR
      expansion of the Dirac 4-spinors, here we will count the
      $p/M$ terms coming from recoil corrections as $Q^2$ as well.
 \end{itemize}
\item The number of loops, or better the number of the sums over the intermediate state momenta that
  remain after using the conservation $\delta$'s. Each loop at the end will give a
  contribution of order $Q^3$.
\item The number of disconnected parts of the diagram. For each of these parts, a $\delta$
  factor expressing the momentum conservation of each part is present.
  Then, if there are $N_D$ disconnected parts, one of the $\delta$
  simply gives the total momentum conservation, a factor common to all diagrams and therefore not
  relevant. Each of the remaining $N_D-1$ $\delta$'s at the end will
  ``block'' a sum over an external three-momentum, each  one
  therefore reducing the chiral order by $3$ units.
\end{enumerate}

Once the $T$-matrix has been calculated, one would obtain in general
\begin{equation}
 T_{fi}=\sum_{n=n_{\text{min}}} T_{fi}^{(n)}\,,\label{eq:t}
\end{equation}
where $T_{fi} ^{(n)}\sim Q^n$. In all cases the sum starts from a minimum value
$n_{\text{min}}$, $n_{\text{min}}=0$ for the PCTC and $n_{\text{min}}=-1$
for the PVTC and PVTV amplitudes. The idea now is to ``define'' the potential
acting between the two nucleons so that it can reproduce the same amplitude
$T_{fi}$, namely, so that (for more details, see Ref.~\cite{Baroni:2015uza})
\begin{eqnarray}
 T_V=V+V\frac{1}{E_i-H^{(NN)}_0+ i\epsilon}T_V \equiv T_{fi}\,,\label{eq:v}
\end{eqnarray}
where $H^{(NN)}_0$ is the non-interacting Hamiltonian of two nucleons. Clearly,
this procedure is not unique, since usually one imposes 
the relation $T_V=T_{fi}$  to hold ``on shell'', namely by requiring
the conservation of the energy between initial and final states.
This induces an ambiguity, as discussed  for example in
Ref.~\cite{Pastore:2011ip}. However, the obtained potentials
are expected to be equivalent by means of a unitary or at least
a similarity transformation \cite{Krebs:2020rms}.

Finally, to invert Eq.~(\ref{eq:v}), one assumes that $V$ has the
same $Q$ expansion as the $T$ matrix,  
\begin{equation}
 V=\sum_{n=n_{\text{min}}} V^{(n)}\,,\qquad V^{(n)}\sim Q^n\,,\label{eq:vn}
\end{equation}
and Eq.~(\ref{eq:v}) can be solved for $V^{(n)}$ order-by-order
(see, for example, Ref.~\cite{Baroni:2015uza} for more details).
This procedure can be generalized to the $A=3$ case to define a three-nucleon
potential and so on.

{\it The method of unitarity transformation.\ }
The method of unitary transformation (MUT) has been pioneered in the
fifties of the last century to derive nuclear potentials in the
framework of pion field theory \cite{FST,Okubo:1954zz}. In the context of chiral EFT,
this approach was formulated in
Refs.~\cite{Epelbaum:1998ka,Epelbaum:2007us}.
Similarly to TOPT, the
MUT is applied to the pion-nucleon Hamiltonian which can be obtained from the
effective Lagrangian in a straightforward way using the standard
canonical formalism.
Let $\eta$ and $\lambda$ denote the projection operators on the
purely nucleonic subspace and the rest of the Fock space involving
pion states with the usual properties $\eta^2 = \eta$, $\lambda^2 =
\lambda$, $\eta \lambda = \lambda \eta = 0$ and $\eta + \lambda = 1$.    
To derive
nuclear forces and/or current operators, the Hamiltonian needs to be brought
into a block-diagonal form with no coupling between the $\eta$- and
$\lambda$-subspaces, which can be achieved via a suitably
chosen unitary transformation $U$. Following Okubo, a unitary operator can
be conveniently parametrized in terms of the operator $A= \lambda  A
\eta$ that mixes the two subspaces via  
\begin{equation}
\label{U}
U = \left( \begin{array}{cc} \eta (1 +  A^\dagger  A )^{- 1/2} & - 
 A^\dagger ( 1 +  A A^\dagger )^{- 1/2} \\
 A ( 1 +  A^\dagger  A )^{- 1/2} & 
\lambda (1 +  A  A^\dagger )^{- 1/2} \end{array} \right)\,.
\end{equation}
One then obtains the nonlinear decoupling equation for the operator
$A$:
\begin{equation}
\label{decoupling}
\tilde H \equiv U^\dagger H U \stackrel{!}{=} \left( \begin{array}{cc} \eta \tilde H \eta  & 0 \\ 0 & 
\lambda \tilde H \lambda \end{array} \right)\quad \Longrightarrow
\quad
\lambda \left( H - \left[ A, \; H \right] - A H A \right) \eta = 0\,.
\end{equation}
The solution of the decoupling equation and the calculation of the unitary
operator $U$ and the nuclear potential $\eta \tilde H \eta$ is
carried out in perturbation theory by employing the standard chiral
expansion. The resulting expressions for the operators
$A$, $U$ and $\eta \tilde H \eta$ have a form of a sequence of
vertices from the pion-nucleon Hamiltonian $H$ and energy denominators 
involving the kinetic energies of particles in the intermediate states with one or more
virtual pions. They are thus similar to the expressions emerging in
the context of TOPT, see e.g.~the operator in Eq.~(\ref{eq:pt}), and the
corresponding matrix elements can also be interpreted in terms of TOPT-like
diagrams. Notice that contrary to Eq.~(\ref{eq:pt}), the expressions
in the MUT do, per construction, not involve energy denominators that
vanish in the static limit of infinitely heavy nucleons and correspond
to iterative contributions to the scattering amplitude.    
As explained in \cite{Epelbaum:2007us}, in order to
implement the chiral power counting in the algebraic approach
outlined above it is convenient to rewrite it in terms of different
variables. Using the rules given in the description of the TOPT
approach and counting
the powers of the soft scale $Q$ for a given irreducible (i.e.~of
non-iterative type) connected $N$-nucleon TOPT-like diagram without
external sources, one obtains for the chiral order $n$ \cite{Weinberg:1990rz,Epelbaum:2007us}
\begin{equation}
\label{pow_mod}
n = -4 + 2 N + 2 L + \sum_i V_i \Delta_i \,,
\end{equation}
where $L$ is the number of loops, $V_i$ is the number of vertices of
type $i$. Further, the vertex dimension $\Delta_i$ is given by
$\Delta_i = d_i + 1/2 n_i - 2$
with $d_i$ and $n_i$ being the number of derivatives and/or  $m_\pi$-insertions and the number of
nucleon fields, respectively. The above expression is convenient to
use for 
estimating the chiral dimension of TOPT-like diagrams. For the
MUT, it is, however,  advantageous to rewrite it in the equivalent form
\begin{equation}
\label{pow_fin}
n = -2 + \sum_i V_i \kappa_i \,, \quad \quad \kappa_i = d_i + \frac{3}{2} n_i + p_i - 4\,,
\end{equation}
where $p_i$ is the number of pionic fields.
The parameter $\kappa_i$ obviously corresponds to the inverse overall mass
dimension of the coupling constant(s) accompanying a vertex of type $i$. 
In this form, chiral expansion becomes formally equivalent to the
expansion in powers of the coupling constants, and it is
straightforward to employ perturbation theory for solving the decoupling
equation (\ref{decoupling}) and deriving the nuclear potentials  $\eta \tilde H \eta$. 

One nontrivial issue that emerges when applying chiral EFT to nuclear
potentials concerns their renormalization. While on-shell
scattering amplitudes, calculated in chiral EFT, can always be made
finite by including the counterterms from the effective Lagrangian
(provided one uses a chiral-symmetry preserving regularization scheme
such as dimensional regularization), nuclear potentials represent
scheme-dependent quantities, which correspond to non-iterative parts
of the scattering amplitude. There is no a priori reason to expect all
ultraviolet divergences emerging from TOPT-like diagrams, which give
rise to nuclear forces, to be absorbable into a redefinition of the
LECs. Indeed, it was found that the static PCTC three-nucleon force at
order $Q^4$ of the two-pion-one-pion exchange type cannot be
renormalized if one uses the unitary transformation given in
Eq.~(\ref{U}) \cite{Epelbaum:2006eu}. On the other hand, the employed
parametrization of the operator $U$ is clearly not the most general
one and represents just one possible choice. The freedom to change the
off-shell behavior of the nuclear potentials, already mentioned in the
context of TOPT, has been exploited in a systematic way in the PCTC
sector in order to enforce renormalizability of nuclear forces (using
dimensional regularization)
\cite{Epelbaum:2007us,Bernard:2007sp,Bernard:2011zr,Krebs:2012yv,Krebs:2013kha}. The MUT has also been successfully
applied to the effective Lagrangian in the presence of external
classical sources in order to derive the corresponding nuclear current
operators, see \cite{Krebs:2016rqz} and references therein.

\subsection{The PVTC potential up to order $Q^2$}
\label{sec:pvtcpot}

In this subsection we will discuss in detail the derivation of the
PVTC potential up to N$^2$LO using the TOPT approach.
We consider diagrams contributing to the $T$-matrix with one
vertex coming from the PVTC Lagrangian, while all other vertices coming
from the PCTC interaction. Diagrams with two or more PVTC
vertices can be safely neglected. 

\begin{figure}
  \centering
  \includegraphics[scale=0.7]{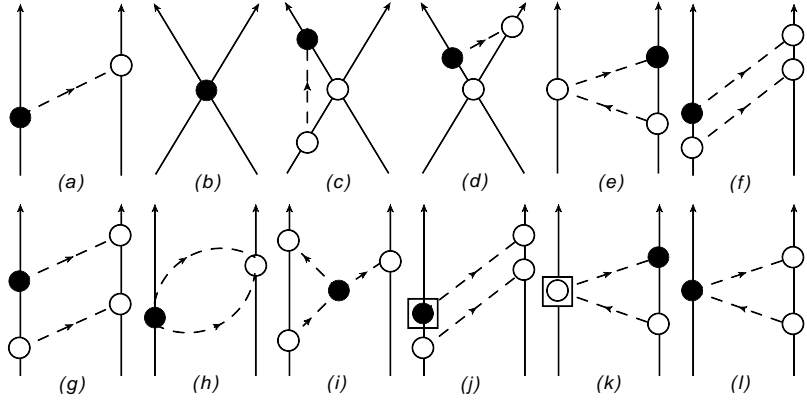}
\caption{ TOPT diagrams contributing up to N$^2$LO to the PVTC
  amplitude. Nucleons and pions are denoted by solid and dashed lines,
  respectively. The open (solid) circles represent LO PCTC (PVTC)
  vertices.
  The vertex depicted by a square surrounding an solid circle 
  denotes the contribution of the subleading PVTC $\pi NN$ terms 
  coming from the Lagrangian given in Eq.~(\ref{eq:LagrangianaPV1}).
  The vertex depicted by a square surrounding an open circle 
  denotes the contribution of the subleading PCTC $\pi\pi NN$ (PVTC $\pi NN$) terms 
  coming from the Lagrangian given in Eq.~(\ref{eq:lagpctc}). 
  }
\label{fig:diagN2LO}
\end{figure}

The TOPT diagrams contributing to the PVTC T-matrix
up to N$^2$LO are shown in Fig.~\ref{fig:diagN2LO} in
panels (a)-(l).  
The one pion exchange diagram (a) gives a contribution to the
$T$-matrix of order $Q^{-1}$ (that will be our LO).  The diagram (b)
represents a PVTC contact interaction of order $Q$; also the
diagrams (c) and (d) with the PCTC contact vertex and one pion exchange are
of order $Q$. The triangle diagram (e) with a PCTC
$\pi \pi NN$ vertex is of order $Q$, while if we consider the PVTC
$\pi \pi NN$ vertex as in panel (l) the diagram is of order $Q^2$.
The box diagrams (f) and (g) includes contribution of order
$Q^0$ and $Q$; the contribution of order $Q^0$ is exactly canceled
when inverting Eq.~(\ref{eq:v}). Finally, the ``bubble'' diagram (h),
the three-pion vertex diagram (i), the box diagram (j) with
the $\pi NN $ vertex coming from the subleading PVTC Lagrangian
terms proportionals to the LECs $h_V^i$, and also the diagram (k) with
the $\pi\pi NN $ vertex coming from the subleading PCTC Lagrangian
terms proportionals to the LECs $c_i$, are of order $Q^2$.
These latter diagrams were considered for the first time
in~\cite{deVries:2015pza} using the MUT, and using TOPT
in~\cite{Gnech:Thesis:2016}.

Contributions
proportional to $1/M$ coming from the NR expansion of the vertex
functions or from recoil corrections in this work are considered to be
at least of order N$^3$LO. 

\begin{figure}
\centering
\includegraphics[scale=0.5]{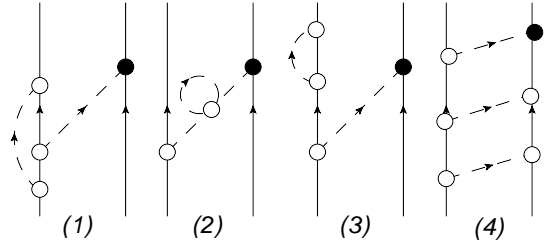}
\caption{\label{fig:otherdiag} Other diagrams that would contribute
  at NLO. These diagrams contribute to the renormalization of the
  LECs (panels $(1)$, $(2)$ and $(3)$) or give a vanishing
  contribution to the potential (panel $(4)$) due to the inversion
  of Eq.~(\ref{eq:v}). Notation as in
  Fig.~\ref{fig:diagN2LO}.} 
\end{figure}

Other types of diagrams like those shown in panels $(1)$, $(2)$, $(3)$ of
Fig.~\ref{fig:otherdiag} simply contribute to a renormalization of the
coupling constants and masses, see Ref.~\cite{Viviani:2014zha} for
more details. In the following, we will disregard these diagrams, but
it should be taken into account  that the formulas below are given in
terms of the renormalized (physical) LECs and masses. The contribution
of diagram $(4)$ is cancelled when inverting Eq.~(\ref{eq:v}). 

Let us now consider each kind of diagram separately:
\begin{itemize}
  \item {\it One pion exchange (OPE) diagram.} 
Diagram~(a) of Fig.~\ref{fig:diagN2LO} gives the LO contribution
($Q^{-1}$) to the potential
\begin{equation}
V^{(-1)}_{PVTC}(a)=  \frac{g_A h^1_\pi}{2\sqrt{2}f_{\pi}} \left(\bmta_1\times\bmta_2\right)_z
        \frac{i\bs{k}\cdot(\bs{\sigma}_1+\bs{\sigma}_2)}
         {\omega_k^2}
        \,, \label{eq:pota_pv}
\end{equation}
where $\omega_k=\sqrt{k^2+m_\pi^2}$, and it 
comes directly from the LO expansion of the vertices and energy denominators. 
Derived from the same diagram, there are terms coming from the NR
expansion of the vertices, the first correction being of order
$(p/M)^2$. However, as discussed previously, they are counted to be of
order $Q^4$, and thus the corresponding terms are considered to be
suppressed by four orders with respect to $V^{(-1)}_{PVTC}$.

\item {\it Contact terms (CT) diagrams.}
The diagrams as that one depicted in panel (b) of Fig.~\ref{fig:diagN2LO} derive
from the interaction terms appearing in $\mathcal L^{(1)}_{PVTC,NN}$.
They give a contribution to the potential of order $Q^1$. As
discussed in Chapter~4, this contribution can be written in various
equivalent forms due to the Fierz identities~\cite{Girlanda:2008ts}.
We have chosen to write this part as follows~\cite{Viviani:2014zha}
\begin{eqnarray}
   V_{PVTC}^{(1)}(b)&=& \frac{1}{\Lambda_{\chi}^2 f_\pi} [
   C_1  i (\bmsi_1\times\bmsi_2)\cdot\bmk   
   +C_2 (\bmta_1\cdot\bmta_2)i(\bmsi_1\times\bmsi_2)\cdot\bmk       \nonumber\\
 &&+C_3 (\bmta_1\times\bmta_2)_z i(\bmsi_1+\bmsi_2)\cdot\bmk   
   +C_4 (\tau_{1z}+\tau_{2z}) i(\bmsi_1 \times\bmsi_2)\cdot\bmk
    \nonumber\\
 &&+C_5 {\mathcal I}_{ab}\tau_{1a} \tau_{2b} i (\bmsi_1\times\bmsi_2)\cdot\bmk]
    \,. \label{eq:potctpv}
\end{eqnarray}
where $\Lambda_\chi=4\pi f_\pi\approx 1.2$ GeV. The parameters $C_i$,
$i=1,\ldots,5$ are LECs. Different (but equivalent) forms of this part were used in
Refs.~\cite{deVries:2013fxa,deVries:2015pza}.

\item {\it Contact plus OPE diagrams.} The diagrams (c) and (d) in
  Fig.~\ref{fig:diagN2LO} are representative of diagrams with a
  contact term and an OPE. However all these diagrams vanish after the
  integration over the loop variable. 

\item {\it NLO two pions exchange: triangle diagrams.}
There are 6 different time-orderings of diagrams given in panel (e) in
Fig.~\ref{fig:diagN2LO}.  After summing them, the total contribution
from the triangle diagrams results to be~\cite{Zhu:2004vw,Kaiser:2007zzb} 
\begin{equation}
  V^{(1)}_{PVTC}(e)=\frac{g_Ah_{\pi}^1}{8\sqrt{2}f_{\pi}^3}
  (\bmta_1 \times\bmta_2)_z i \bs{k} \cdot(\bmsi_1+\bmsi_2)
  \int \frac{d^3q}{(2\pi)^3} \frac{1}{\omega_{+}\omega_{-}(\omega_{+}+\omega_{-})}
  \,, \label{eq:pottriN2LO}
\end{equation}
where $\omega_\pm = \sqrt{(\bmq\pm\bmk)^2+4m_\pi^2}$. 
The integral is singular and has to be regularized using some method.
We will discuss this issue later. 

\item {\it NLO two pions exchange: box diagrams.}
There are 48 diagrams represented by the diagrams of type (f) and  (g)
of Fig.~\ref{fig:diagN2LO} when we consider all possible time
orderings. The final contribution is~\cite{Zhu:2004vw,Kaiser:2007zzb} 
\begin{eqnarray}
  V_{PVTC}^{(1)}(f,g) & = & \frac{h^1_\pi g_A^3}{8\sqrt{2}f_{\pi}^3}\int 
        \frac{d^3q}{\left(2\pi\right)^3}
           \frac{\omega_+^2+\omega_+\omega_-+\omega_-^2}
           {\omega_+^3\omega_-^3\left(\omega_++\omega_-\right)}\nonumber\\
           & & \{-2i\left(\tau_{1z}+\tau_{2z}\right)[ \bs{q}\cdot
             \sone(\bs{q}\times\bs{k})\cdot\stwo-\bs{q}\cdot\stwo
          (\bs{q}\times\bs{k})\cdot\sone]\nonumber\\ 
          && -2i\left(\tau_{1z}-\tau_{2z}\right)[\bs{q}\cdot\stwo(\bs{q}\times\bs{k})\cdot\sone
          + \bs{q}\cdot\sone(\bs{q}\times\bs{k})\cdot\stwo ]  \nonumber\\
        && +i\left(\tone\times\ttwo\right)_z\left(k^2-q^2\right)\bs{k}\cdot
          \left(\sone+\stwo\right)  \}  \,, \nonumber \label{eq:cdpv}\nonumber\\
\end{eqnarray}
and it is of order $Q^1$. Again the integral is singular. 
In this case, in the amplitude $T_{fi}$ there appears a term of order
$Q^0$ coming from diagram (g), but it cancels out when inverting Eq.~(\ref{eq:v}).

\item {\it Bubble diagrams.}
We now turn to the diagrams contributing at order
$Q^2$, that is at N$^2$LO. The sum of ``bubble'' diagrams depicted in panel  (h) of  Fig.~\ref{fig:diagN2LO}
mutually cancel and these diagrams do not give any contribution to the PVTC potential.

\item {\it Diagrams with three pion vertices.}
The expansion of the PVTC Lagrangian in terms of pions gives rise to
two terms proportional to $(\vec\pi)^3$ which would contribute to $T_{fi}$
via the diagram depicted in panel (i) of  Fig.~\ref{fig:diagN2LO}.
However, after summing over all possible time orderings, the
corresponding final contribution vanishes.

\item {\it N$^2$LO two pion exchanges: box diagrams.}
  The box diagrams contributes also at N$^2$LO via diagrams of type (j)
  where the PVTC vertex comes from the subleading Lagrangian terms
  proportional to the LECs $h_0^V$, $h_1^V$, and $h_2^V$
  in Eq.~(\ref{eq:LagrangianaPV1}). We have~\cite{deVries:2015pza,Gnech:2019dod}
\begin{eqnarray}
   V^{(2)}_{PVTC}(j)&=&\frac{g_A^3}{32f_{\pi}^4}\Big[\Big(h^0_V(3+2 \ttwo
  \cdot
  \tone)-\frac{4}{3}h^2_V{\mathcal I}_{ab}\tau_{1b}\tau_{2b}\Big)i\int
  \frac{d^3q}{(2\pi)^3}\frac{1}{\omega_+^2\omega_-^2 }
       [(\bs{q}\cdot\sone(\bs{q}\times\bs{k})\cdot\stwo)\nonumber \\ 
   &&-(\bs{q}\cdot\stwo
          (\bs{q}\times\bs{k})\cdot\sone)]-2i h^1_V\int
       \frac{d^3q}{(2\pi)^3} \frac{1}{\omega_+^2\omega_-^2 }
            [(\bs{q}\cdot\sone(\bs{q}\times\bs{k})\cdot\stwo)\tau_{1z}\nonumber
              \\ 
   &&-(\bs{q}\cdot\stwo
          (\bs{q}\times\bs{k})\cdot\sone)\tau_{2z}]+i
            h^1_V(\tone\times\ttwo)_z\bs{k}\cdot(\sone+\stwo)\int
            \frac{d^3q}{(2\pi)^3} \frac{q^2-k^2}{\omega_+^2\omega_-^2
            }\Big] \,.\label{eq:potPVn} 
 \end{eqnarray}

\item {\it N$^2$LO two pion exchanges: triangle diagrams.}
  The diagram depicted in panels (k) derives from a
  subleading $\pi\pi NN$ vertices in the PCTC
  Lagrangians~\cite{deVries:2015pza,Gnech:2019dod}, see
  Eq.~(\ref{eq:lagpctc}),  
\begin{eqnarray}
  V^{(2)}_{PVTC}(k) &=& -
  i\frac{\mathit{c}_4h_{\pi}^1g_A}{2\sqrt{2}f_{\pi}^3} \int
  \frac{d^3q}{(2\pi)^3}\frac{1}{\omega_+^2\omega_-^2 } \times  
  \nonumber \\ 
  &&[(\bs{q}\cdot\sone(\bs{q}\times\bs{k})\cdot\stwo)\tau_{2z}-(\bs{q}\cdot\stwo
          (\bs{q}\times\bs{k})\cdot\sone)\tau_{1z}] \,. \label{eq:potPVk}
\end{eqnarray}
Note in Eq.~(\ref{eq:potPVk}) the presence of the
LEC $\mathit{c}_4$, which belong to the PCTC sector~\cite{Bernard:1995dp}.

The expression for the diagrams $(l)$ comes from the LO PCTC
and PVTC vertex functions. The final result is~\cite{deVries:2015pza,Gnech:2019dod}
\begin{eqnarray}
  V^{(2)}_{PVTC}(l)&=&-\frac{g_A^2}{8 f_{\pi}^4}\int
  \frac{d^3q}{(2\pi)^3}\frac{1}{\omega_+^2\omega_-^2 } \times
  \nonumber \\ 
  &&\{2h^1_A[(\bs{q}\cdot\sone(\bs{q}\times\bs{k})\cdot\stwo)\tau_{2z}-(\bs{q}\cdot\stwo 
          (\bs{q}\times\bs{k})\cdot\sone)\tau_{1z}] + \nonumber \\
  &&        h^2_A
  {\mathcal I}_{ab}\tau_{1a}\tau_{2b}[(\bs{q}\cdot\sone(\bs{q}\times\bs{k})\cdot\stwo)-(\bs{q}\cdot\stwo 
          (\bs{q}\times\bs{k})\cdot\sone)] \} \,,\label{eq:potPVl}
\end{eqnarray}
where $h_A^1$ and $h_A^2$ are two of the LECs that appear
in the Lagrangian terms given in Eq.~(\ref{eq:LagrangianaPV1}).
\end{itemize}

\begin{figure}
 \begin{center}
 \includegraphics[scale=0.7]{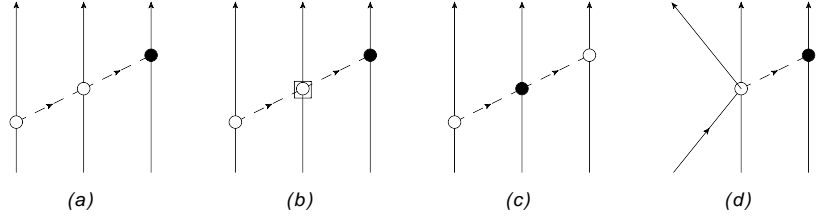}
 \end{center}
 \caption{TOPT diagrams that would contribute to the PVTC 3N force
   For the notation see Fig.~\ref{fig:diagN2LO}.}
   \label{fig:3nPV}
 \end{figure}

   Finally, we conclude this section by mentioning that at N$^2$LO,
    one should also include PVTC 3N forces. Examples of diagrams
    contributing to this 3N force are reported in Fig.~\ref{fig:3nPV}.
    The chiral order of diagrams with more than two nucleons
    is discussed in detail in Ref.~\cite{Epelbaum:2007us}.
    The diagram depicted in panel (a) with a LO PCTC $\pi\pi NN$
    vertex would contribute at NLO but summing over all the time-orderings it
    vanishes. The other three diagrams (the one in panel (b) has a
    subleading PCTC $\pi\pi NN$ vertex proportional to $c_i$,
    $i=1,\ldots,4$ \cite{Bernard:1995dp}) are N$^2$LO  and therefore they have to
    be considered in order to perform fully consistent
    calculations in $A\ge 3$ systems. These kind of diagrams have not
    yet been considered in literature. Note that diagrams with a 3N PVTC contact
    vertex are highly suppressed, so no new LEC has to be introduced
    in this case.

\subsubsection{Regularization of the PVTC potential}\label{sec:regular}
In this section we deal with the divergences in the loop diagrams. We will
briefly present three methods
frequently used in literature, namely the
dimensional regularization (DR)  method used e.g.~in~\cite{Pastore:2008ui}, the spectral function regularization
(SFR) \cite{Epelbaum:2003gr}, and the novel (semi-)local momentum-space
regularization approach of Ref.~\cite{Reinert:2017usi}.
\begin{itemize}
\item {\it Dimensional regularization method.}\ 
This technique is well known for dealing with divergences of loop integrals
present in Feynman diagrams, where the integration is performed
over four-momenta. In case of time-ordering diagrams,
the loops involve integrations over 3-dimensional momenta. 
To deal with the singularities, the integrals are re-defined in
$d$ dimensions and successively one takes the limit $d\rightarrow 3$.
The singular part is singled out by terms  $\sim 1/(3-d)$, which
then can be reabsorbed in some of the LECs.
As usual, we define  $\epsilon=3-d$, and we assume that $\epsilon \rightarrow 0$.
When we use the DR, it is better to ``rescale" all the dimensional quantities
with an energy  scale $\mu$. Therefore we define $q=\tilde{q}\mu$,
$m=\tilde{m}\mu$, etc., where the ``tilde" quantities are adimensional.
We can now go to $d$ dimension and manipulate the integrals as discussed in detail
in Ref.~\cite{Pastore:2008ui}, see also Ref.~\cite{Friar:1996tj}. Here we limit ourselves to list the
results needed to regularize the loop integrals we have encountered. 
Regarding the loop integrals appearing at NLO in Eqs.~(\ref{eq:pottriN2LO})
and~(\ref{eq:cdpv}), we have
\begin{eqnarray}
 && \int {d^3q\over{(2\pi)^3}} \,
     \,\frac{1}{\omega_+\,\omega_-\, (\omega_+ + \omega_-)} =
     -\,\frac{1}{4\,\pi^2}\Bigl(L(k) -d_\epsilon+2\Bigl)\,,\\
 && \int {d^3q\over{(2\pi)^3}} \, \frac{\omega_+^2+\omega_+\,\omega_-+\omega_-^2}
  {\omega_+^3\,\omega_-^3(\omega_++\omega_-)} =
\frac{1}{16\,\pi^2}\, {H(k)\over m_\pi^2} \,,
    \label{eq:j2ij}     
\end{eqnarray}
where
\begin{equation}
  L(k)={1\over 2}\frac{s}{k}\ln\frac{s+k}{s-k}\,,\qquad
  H(k)= {4m_\pi^2\over s^2}L(k)\,,\qquad
  s=\sqrt{4\, m_\pi^2+k^2}\,, \label{eq:funk}
\end{equation}
and
\begin{equation}
  d_\epsilon= \frac{2}{\epsilon}-\gamma+\ln{\pi}-\ln
  \frac{m_\pi^2}{\mu^2}\,,
\end{equation}
which contains the divergent part (above $\gamma$ is the Euler–Mascheroni constant).

The loop integrals appearing in the N$^2$LO diagrams
as in Eqs.~(\ref{eq:potPVk}) and~(\ref{eq:potPVl}) are of the form
\begin{eqnarray}
  &&\int \frac{d^3q}{(2\pi)^3}\frac{1}{\omega_+^2\omega_-^2 } \,,\label{eq:ww1}\\
  &&\int \frac{d^3q}{(2\pi)^3}\frac{1}{\omega_+^2\omega_-^2 } q_i q_j\,,\label{eq:ww2}
\end{eqnarray}
The first integral is finite, but 
the second integrand diverges linearly for $q\rightarrow\infty$.
The finite contribution to potential can be obtained using
    the DR method. Alternatively, one can  impose an ultraviolet
    cut-off $\Lambda_C$ on the integrals. The integrals then yield 
    diverging parts as $\Lambda_C\rightarrow\infty$, 
which can be again reabsorbed by some LECs, finite parts independent on
$\Lambda_C$ that are exactly the same as obtained using the DR method,
and a number of other terms which can be expressed in power series
of $Q/\Lambda_C$, where $Q$ is either $k$ or $m_\pi$.  Taking the
limit $\Lambda_C$ to infinity
these latter parts would disappear. Since, in general we must fix $\Lambda_C$ at
a value greater than the typical energies of the $\chi$EFT, then  
these additional terms carry  at least an additional power of $Q$
which means they give contributions at N$^3$LO (or beyond) to the potential.
Therefore, for the integral in Eq.~(\ref{eq:ww2}),
we have followed the prescription to absorb the divergent parts in
some LEC's, disregard the parts depending on $Q/\Lambda_C$, and
retaining the finite parts as those given by the DR
method. Explicitly, the two integrals are given by
\begin{eqnarray}
  &&\int {d^3q\over{(2\pi)^3}} \,
     \,\frac{1}{\omega_+^2\,\omega_-^2}={A(k)\over 4\pi}  \,,\label{eq:ww1r}\\
  &&\int {d^3q\over{(2\pi)^3}} \,
     \,\frac{1}{\omega_+^2\,\omega_-^2}q_i q_j\Rightarrow
     \Big(-\frac{s^2A(k)}{8\pi}-\frac{m_\pi}{8\pi}\Big)\delta_{ij}+
     \Big(\frac{s^2A(k)}{8\pi}-\frac{m_\pi}{8\pi}\Big)\frac{k_i\,k_j}{k^2}\,,\label{eq:ww2r}
\end{eqnarray}
where
\begin{equation}
  A(k)=\frac{1}{2k}\arctan\big(\frac{k}{2m_\pi}\big)\,.\label{eq:funA}
\end{equation}

%\emanuele{Is the argument of the $\arctan$ ok?}

\item {\it Spectral function regularization method.}\ 
Pion loop integrals appearing in the two-pion exchange contributions
discussed in the previous subsection
can be generally expressed using a dispersive
representation. Writing the momentum-space potentials in the general
form $V = \sum_i O_i W_i (k)$ with $O_i$ being spin-isospin-momentum
operators and $W_i$ the corresponding structure functions that depend
only on the  momentum transfer $k \equiv |\bs{k} |$,  the unsubtracted
dispersion relations for the functions $W_i (k)$ have the form \cite{Kaiser:1997mw}
\begin{equation}
  \label{spectralf}
  W_i (k) = \frac{2}{\pi} \int_{2 m_\pi}^\infty d \mu \, \mu
  \frac{\rho_i (\mu )}{\mu^2 + k^2} \,,
\end{equation}
where the spectral functions $\rho_i (\mu )$ are given by $\rho_i =
\Im \left( W_i (0^+ - i \mu \right)$. Notice that the spectral
integrals in Eq.~(\ref{spectralf}) do not converge for potentials
derived in chiral EFT since $\rho_i (\mu )$ generally grow with $\mu$,
and must be subtracted the appropriate number of times.  The
subtractions introduce terms which are polynomial in $k^2$ and can be absorbed into the
corresponding contact interactions. It was shown in Ref.~\cite{Epelbaum:2003gr} that
even at fairly large internucleon distances, the potentials receive 
significant contributions from the spectral function in the region of $\mu \gtrsim
\Lambda_\chi$, where the chiral expansion cannot be trusted. It was,
therefore, proposed in that paper to employ an ultraviolet cutoff
$\Lambda$ in the spectral integrals. This can be shown to be
equivalent to introducing a particular ultraviolet cutoff in the loop
integrals over the momentum $q$. Using a sharp cutoff $\Lambda$ in the
spectral integrals over $\mu$ leads to the following modification of
the loop functions $L(k)$ and $A (k)$:
\begin{eqnarray}
  \label{def_LA}
  L^\Lambda(k) &=& \theta (\Lambda - 2 m_\pi ) \, \frac{s}{2 k} \, 
  \ln \frac{\Lambda^2 s^2 + k^2 l^2 + 2 \Lambda k 
  s l}{4 m_\pi^2 ( \Lambda^2 + k^2)}\,, \nonumber \\
  A^\Lambda(k) &=& \theta (\Lambda - 2 m_\pi ) \, \frac{1}{2 k} \, 
  \arctan \frac{k ( \Lambda - 2 m_\pi )}{k^2 + 2 \Lambda m_\pi}\,,
\end{eqnarray}
where we have introduced  $l = \sqrt{\Lambda^2 - 4 m_\pi^2}$.
The resulting approach is referred to as the spectral function
regularization. The
limit of an infinitely large cutoff $\Lambda$ corresponds to the
previously considered case of dimensional regularization with
$L^\infty (k) = L (k) $ and $A^\infty (k) = A (k) $. The spectral
function regularization approach with a finite value of $\Lambda$ was
employed in the PCTC potentials of Ref.~\cite{Epelbaum:2004fk} and the more recent work \cite{Entem:2017gor}, and in the derivation of the N${}^2$LO PVTC potential in Ref.~\cite{deVries:2014vqa}.

\item {\it Local regularization in momentum space.}\ 
The previously introduced spectral function regularization approach
has an unpleasant feature of inducing \emph{long-range}
finite-$\Lambda$ artifacts as can be seen by expanding the functions 
$L^\Lambda (k)$ and $A^\Lambda (k)$ in inverse powers of $\Lambda$. 
This feature may affect the applicability of chiral EFT  for softer cutoff
choices. Recently, local regulators in coordinate \cite{Epelbaum:2014efa,Epelbaum:2014sza} and
momentum space \cite{Reinert:2017usi} were introduced, which do not affect the
analytic structure of the pion-exchange interactions and thus maintain
the long-range part of the nuclear force.  The approach of
Ref.~\cite{Reinert:2017usi} amounts to replacing the static propagators of pions exchanged
between different nucleons via
\begin{equation}
  \label{1pilocal}
  \frac{1}{q^2 + m_\pi^2 } \quad \longrightarrow \quad
  \frac{1}{q^2 + m_\pi^2 } \; \exp \left( - \frac{q^2 +
    m_\pi^2}{\Lambda^2} \right)\,,
\end{equation}
with $q \equiv |\bs{q} |$. 
Such a regulator obviously does not induce any long-range artifacts 
at any order in the $1/\Lambda$-expansion. This regularization
approach can be easily implemented 
for two-pion exchange NN potentials with no need to recalculate the
various loop integrals. Using the feature that the regulator does not
affect long-range interactions, it is easy to show that the
regularization of a generic two-pion exchange contribution simply amounts to
introducing a specific cutoff in the dispersive representation (modulo
short-range interactions), namely \cite{Reinert:2017usi}
\begin{equation}
  \label{2pilocal}
  \frac{2}{\pi} \int_{2 m_\pi}^\infty d \mu \, \mu
  \frac{\rho_i (\mu )}{\mu^2 + k^2}  \quad \longrightarrow \quad
  \frac{2}{\pi} \int_{2 m_\pi}^\infty d \mu \, \mu
  \frac{\rho_i (\mu )}{\mu^2 + k^2} \; \exp \left( - \frac{\mu^2 + k^2}{2 \Lambda^2} \right)\,.
\end{equation}
In Ref.~\cite{Reinert:2017usi}, the regularized two-pion exchange contributions were
defined using the requirement (i.e.~a convention) that the corresponding potentials in
coordinate space and derivatives thereof vanish at the origin. This
is achieved by adding to the right-hand side of Eq.~(\ref{2pilocal}) a
specific combination of (locally regularized) contact interactions
allowed by the power counting. For more details and explicit
expressions see Ref.~\cite{Reinert:2017usi}. This local regularization scheme has not been used 
for PVTC or PVTV nuclear potentials.

\end{itemize}

\subsubsection{The regularized PVTC potential}\label{sec:potkspace}
Once the loop integrals have been manipulated as discussed previously,
we can now write the PVTC potential up to N$^2$LO derived from $\chi$EFT.
In the following, some of the LEC's have been further redefined to absorb
the singular parts coming from the loop integrals. If one has chosen
to regularize the loop integral using the SFR method, then the
functions $L(k)$ and $A(k)$ below have to be substituted with
$L^{\Lambda}(k)$ and $A^{\Lambda}(k)$, the spectral
regularized functions, see Eq.~(\ref{def_LA}). In summary,
\begin{eqnarray}
  V_{PVTC}&=& V^{(-1)}_{PVTC}({\rm OPE})+V_{PVTC}^{(1)}({\rm CT})
  +V_{PVTC}^{(1)}({\rm TPE}) +  V_{PVTC}^{(2)}({\rm TPE})\,, \label{eq:PVpot}
\end{eqnarray}
where 
\begin{eqnarray}
  V^{(-1)}_{PVTC}({\rm OPE})&=&  \frac{g_A h^1_{\pi}}{2\sqrt{2}f_{\pi}} \left(\bmta_1\times\bmta_2\right)_z
        \frac{i\bs{k}\cdot(\bs{\sigma}_1+\bs{\sigma}_2)}
         {\omega_k^2} \,,\label{eq:PVpotrOPE}  \\
  V_{PVTC}^{(1)}({\rm CT})&=& \frac{1}{\Lambda_{\chi}^2 f_\pi} [
 C_1  i (\bmsi_1\times\bmsi_2)\cdot\bmk   
 +C_2 (\bmta_1\cdot\bmta_2)i(\bmsi_1\times\bmsi_2)\cdot\bmk   
    \nonumber\\
 &&+C_3
    (\bmta_1\times\bmta_2)_z i(\bmsi_1+\bmsi_2)\cdot\bmk   
+C_4 
  (\tau_{1z}+\tau_{2z}) i(\bmsi_1 \times\bmsi_2)\cdot\bmk
    \nonumber\\
 &&+C_5
    {\mathcal I}_{ab}\tau_{1a}\tau_{2b} i (\bmsi_1\times\bmsi_2)\cdot\bmk]
    \,, \label{eq:PVpotrCT}\\
    V_{PVTC}^{(1)}({\rm TPE}) &=&
  -{g_A h^1_\pi\over 2\sqrt{2} f_\pi} \frac{1}{\Lambda_{\chi}^2}
   {     (\bmta_1\times\bmta_2)_z  } 
          i\bmk\cdot(\bmsi_1+\bmsi_2)  L(k) \nonumber \\
 && -{g_A^3 h^1_\pi\over 2\sqrt{2}f_\pi} {1\over\Lambda_{\chi}^2} 
     \biggl[4(\tau_{1z}+\tau_{2z})\; i\bmk\cdot(\bmsi_1\times\bmsi_2)\;
     L(k)\nonumber \\
  && +(\bmta_1\times\bmta_2)_z i\bmk\cdot(\bmsi_1+\bmsi_2)\; 
     \Bigl(H(k)-3L(k)\Bigr)\biggr]\,, \label{eq:PVpotrTPE}\\
   V^{(2)}_{PVTC}({\rm TPE}) &=& - \frac{\mathit{c}_4h_{\pi}^1g_A}{\sqrt{2}f_{\pi}}
   \frac{\pi}{\Lambda_{\chi}^2}
   i\bmk \cdot(\sone \times \stwo)(\tau_{1z}+\tau_{2z}) s^2A(k) \nonumber\\
   &+&\frac{g_A^2}{2 f_{\pi}^2}\frac{\pi}{\Lambda_{\chi}^2}
   \Big\{\Big[\frac{3g_Ah^0_V}{4}+\frac{g_Ah^0_V}{2}\bmta_1\cdot\bmta_2+
   \Big(\frac{g_Ah^1_V}{4}-h^1_A\Big)(\tau_{1z}+\tau_{2z})\nonumber \\ 
  &&-\Big(h^2_A+\frac{g_Ah^2_V}{3}\Big)
    {\mathcal I}_{ab}\tau_{1b}\tau_{2b}\Big]i\bmk\cdot(\sone\times\stwo) \nonumber \\
   &&-\frac{g_Ah^1_V}{2}(\bmta_1 \times \bmta_2)_zi\bmk\cdot(\sone+\stwo)
   \Big(1-\frac{2m_{\pi}^2}{s^2}\Big)\Big\}s^2A(k)
   \,. \label{eq:PVpotrLEC}
\end{eqnarray}
The NLO term $ V_{PVTC}^{(1)}({\rm TPE})$ derives from the regularized
parts of $ V_{PVTC}^{(1)}(e)$ and $ V_{PVTC}^{(1)}(f,g)$, while the N$^2$LO term
$ V_{PVTC}^{(2)}({\rm TPE})$ from  $ V_{PVTC}^{(1)}(j)$, $ V_{PVTC}^{(1)}(k)$,
and $V_{PVTC}^{(1)}(l)$. 
Let us note that we have in total 11 LECs that must be determined from
the experimental data: one in the LO term, five in the subleading
order and five at N$^2$LO. This potential is
the same as the one derived using the MUT in Ref.~\cite{deVries:2015pza}.

Finally, the potential to be used in calculation of PVTC observables
has to be regularized for large values of $\bmp$, $\bmp'$. The
frequently used
procedure is to multiply by a cutoff function containing a parameter
$\Lambda_C$
\begin{equation}
  V_{PVTC}(\bmp,\bmp') \rightarrow f_{\Lambda_C}(\bmp,\bmp')
    V_{PVTC}(\bmp,\bmp')\,.
\end{equation}
Usual choices for $f_{\Lambda_C}$ are~\cite{Epelbaum:2004fk}
\begin{equation}
  \label{nonlocalcutoff}
  f_{\Lambda_C}(\bmp,\bmp')= \exp\left[
    -\biggl({p\over\Lambda_C}\biggr)^n-\biggl({p'\over \Lambda_C}\biggr)^n\right]\,,
\end{equation}
where usually $n=6$, adopted for example in Ref.~\cite{deVries:2015pza}, or
\begin{equation}
   \label{localcutoff}
  f_{\Lambda_C}(\bmp,\bmp')= \exp\left[
    -\biggl({|\bmp-\bmp'|\over\Lambda_C}\biggr)^4\right]
  \,,
\end{equation}
adopted in Ref.~\cite{Viviani:2014zha}. The value of the cutoff
$\Lambda_C$ is chosen to be around $400$--$600$ MeV, and consistent with
the analogous parameter used to regularize the PCTC potential.

  The currently most accurate and precise PCTC NN potentials of
  Ref.~\cite{Reinert:2017usi} employ the local momentum-space regularization approach
  for pion-exchange contributions as described in section
  \ref{sec:regular} in combination with a nonlocal Gaussian regulator given in
  Eq.~(\ref{nonlocalcutoff}) with $n=2$ and $\Lambda_C = \Lambda$  for
 contact interactions ($\Lambda$ is the cutoff used
 in the local regulator in Eqs.~(\ref{1pilocal}), (\ref{2pilocal})). The superior performance of the 
 momentum-space regulator in Eq.~(\ref{2pilocal}) as compared with both the
 spectral-function regularization and a local multiplicative
 regularization as defined in Eq.~(\ref{localcutoff}) manifests itself in exponentially
 small distortions  at large distances as visualized
 in Fig.~5 of \cite{Reinert:2017usi}. 

Last but not least, we emphasize that using \emph{different} regulators when
calculating loop integrals in the nuclear potentials/currents and 
solving the Schr\"odinger equation to compute observables is generally
incorrect.  This issue
becomes relevant at the chiral order, at which one encounters the 
first loop
contributions to the 3N potentials and to the NN exchange current
operators (i.e.~at order $Q^4$ or N$^3$LO in the PCTC sector)
\cite{Krebs:2019uvm,Epelbaum:2019jbv}, which is
beyond the accuracy of the calculations described in this review
article. For more details and a discussion of a possible solution to
this problem see Ref.~\cite{Epelbaum:2019kcf}.

\subsubsection{Relevant PCTC and PVTC electromagnetic currents}
\label{sec:currents}

Electromagnetic currents can be calculated in the $\chi$EFT
expansion. For our purposes we require currents for the longitudinal
asymmetry in radiative neutron capture on a proton target at thermal
energies. As we deal with a real outgoing photon, the LO PCTC current
is induced by the nucleon magnetic moment. At NLO there are
contributions from the convection currents and one-pion-exchange
currents proportional to $g_A^2$. At NLO the relevant currents become 
\begin{eqnarray}
  \bmJ_{PCTC} &=& \sum_{j=1}^A \frac{e}{4M}\left\{ -\left[ (1+\kappa_0) +  (1+\kappa_1) \tau_{jz} \right]
  i (\bmsi_j \times \bmq) + (1+\tau_{jz})(\bmp_j +
  \bmp_j^{\,\prime})\right\}\delta_{\bmp_j -\bmp_j^{\,\prime},\bmq}
  \nonumber\\
&& +\frac{e g_A^2}{4f_\pi^2} \sum_{j<k}^A i\left(\vec \tau_j\times \vec \tau_k\right)_z
\left\{ 2\bmk\,
\frac{\bmsi_j\cdot(\bmk+\bmq/2)}{(\bmk +\bmq/2)^2+m_\pi^2} \,
\frac{\bmsi_k\cdot(\bmk -\bmq/2)}{(\bmk -\bmq/2)^2+m_\pi^2}  
\right.
\nonumber\\
&& \qquad\quad \left. -\bmsi_j\,
\frac{\bmsi_k\cdot(\bmk-\bmq/2)}{(\bmk -\bmq/2)^2+m_\pi^2}-
\bmsi_k\,
\frac{\bmsi_j\cdot(\bmk+\bmq/2)}{(\bmk +\bmq/2)^2+m_\pi^2} 
\right\}\,, \label{eq:PCcurrent}
\end{eqnarray}
where $\kappa_0 = -0.12$  and $\mu_v=3.71$ are the isoscalar  and isovector
anomalous nucleon magnetic moments. $\bmp_j$ and $\bmp^{\,\prime}_j$
denote the incoming and outgoing momenta of nucleon $j$ interacting
with a photon of outgoing momentum $\bmq$. The intermediate
pions carry momenta $\bmk+\bmq/2= \bmp_j-\bmp^{\,\prime}_j$ or
$\bmk-\bmq/2= \bmp^{\,\prime}_k- \bmp_k$. Ref. \cite{deVries:2015pza} used these
currents in combination with N${}^3$LO $\chi$EFT potentials from
Ref.~\cite{Epelbaum:2014sza} to calculate the total $ n p\rightarrow d \gamma$ capture
cross section. Using just the LO currents gives a cross section of $305\pm 4$ mb,
which grows to $319\pm 5$ at NLO. The remaining $4\%$ discrepancy
to the experimental cross section $334.2 \pm 0.5$, indicates that N$^2$LO currents
should probably be included. 

A consistent calculation of PVTC observables as the photon asymmetry
in the $\vec np\rightarrow d\gamma$ radiative capture  also requires
the inclusion of PVTC currents. There is no one-body current in this
case, as the anapole moment vanishes for on-shell photons \cite{Maekawa:2000bd}. As such,
the leading PVTC currents arises from one-pion-exchange currents
\begin{eqnarray}
  \bmJ_{PVTC} &=& \frac{eg_A h^1_\pi}{2\sqrt{2}f_\pi}\sum_{j<k}^A \left(\vec \tau_j\cdot\vec \tau_k -
  \tau_{jz} \tau_{kz} \right) \bigg\{2 \bmk \frac{\bmsi_j\cdot(\bmk+\bmq/2)+
 \bmsi_k\cdot(\bmk-\bmq/2)}{[(\bmk +\bmq/2)^2+m_\pi^2][(\bmk -\bmq/2)^2+m_\pi^2]}\nonumber\\
    && -\frac{\bmsi_j}{(\bmk -\bmq/2)^2+m_\pi^2}- \frac{\bmsi_k}{(\bmk +\bmq/2)^2+m_\pi^2} 
  \bigg\}\,,
\end{eqnarray}
where we stress the dependence on the PVTC pion-nucleon LEC $h^1_\pi$. Higher-order PVTC currents have not been developed.

\subsection{The PVTV potential up to order $Q$}
\label{sec:pvtvpot}
In this section, we discuss the derivation of the PVTV NN and 3N
potentials at N$^2$LO. The final expressions are given in terms of a sum
of diagrams, which can be obtained either using the
MUT~\cite{Epelbaum:1998ka,Epelbaum:1999dj,Epelbaum:2008ga},
standard dimensional regularization \cite{Maekawa:2011vs} or the 
TOPT method~\cite{Gnech:2019dod}. In the following,
we briefly report the derivation of the PVTV potential in the framework
of TOPT approach.

\begin{figure}[t]
  \begin{center}
    \includegraphics[scale=.7]{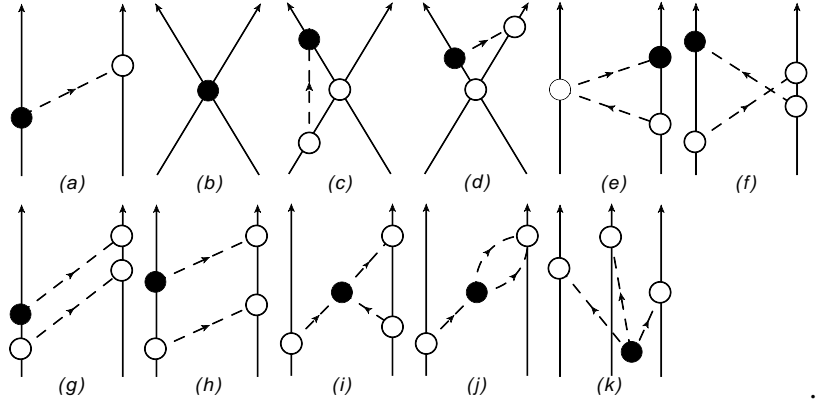}.
    \caption{   \label{fig:diagNN}
      Time-ordered diagrams contributing to the PVTV
      potential (only a single time ordering is shown).
      Nucleons and pions are denoted by solid and dashed lines,
      respectively. The open (solid) circle represents a PCTC (PVTV)
      vertex.} 
  \end{center}
\end{figure}

The TOPT diagrams that give contribution to the NN PVTV potential up to N$^2$LO (order $Q^1$) are shown in
Fig.~\ref{fig:diagNN}. We do not consider diagrams which give contributions only to the
renormalization of the LECs. In this section we write the final expression of
the NN PVTV potential $ V_{PVTV}$ by having already taken into account
the singular parts coming from loops. Note that for the PVTV potential the LO term is of order
$Q^{-1}$ as for the PVTC case. However, now there will be terms of order $Q^0$, which
will be denoted as NLO terms, etc. We have
\begin{eqnarray}
    V_{PVTV}(\bmp,\bmp')&=&
    V^{(-1)}_{PVTV}({\rm OPE}) +
    V^{(1)}_{PVTV}({\rm CT})  +
    V^{(1)}_{PVTV}({\rm TPE}) \nonumber \\
    && + V^{(0)}_{PVTV}({\rm 3\pi})+ V^{(1)}_{PVTV}({\rm 3\pi})
    \,,\label{eq:dec}
\end{eqnarray}
namely coming from  OPE diagrams at LO, TPE at N$^2$LO, three-pion vertices (3$\pi$) at NLO 
and at N$^2$LO, and contact contributions (CT). 
From now on we define $\bar g_0^*=\bar g_0+\bar g_2/3$.
In this case, we report here the final form of the potential, namely, 
the LECs appearing in the expressions below are the physical ones,
having reabsorbed the various infinities generated by loops and
diagrams like those shown in panels (1)--(3) of Fig.\ref{fig:otherdiag}.
\begin{itemize}
  \item {\it One pion exchange diagram.}   
    The OPE term, depicted in diagram (a) of Fig.~\ref{fig:diagNN}, gives a contribution
    at LO, namely of order $Q^{-1}$, coming from the NR expansion  of the vertices
\bgroup
\arraycolsep=0.7pt
\begin{eqnarray}
  V^{(-1)}_{PVTV}({\rm OPE})
  &=&\frac{{g}_A\bar{g}_0^*}{2{f}_\pi}(\tone\cdot\ttwo)\frac{i\bmk\cdot
    (\sone-\stwo)}{\omk^2}+
  \frac{{g}_A\bar{g}_2}{6{f}_{\pi}}
  (3\tau_{1z}\tau_{2z}-\tone\cdot\ttwo)\frac{i\bmk\cdot
    (\sone-\stwo)}{\omk^2}  \nonumber\\
  &+&
  \frac{{g}_A\bar{g}_1}{4{f}_\pi}\Big[(\tau_{1z}+\tau_{2z})
    \frac{i\bmk\cdot(\sone-\stwo)}{\omega_k^2}
    +(\tau_{1z}-\tau_{2z})\frac{i \bmk\cdot(\sone+\stwo)}{\omega_k^2}\Big]\,,
  \label{eq:isotope}
\end{eqnarray}
\egroup
where there are an isoscalar, an isovector and an isotensor components.
Contributions coming from the $1/M$ expansion are considered
to be suppressed at least by four orders with respect to $V^{(-1)}_{PVTV}({\rm OPE})$.

\item {\it Contact term diagrams.}
The potential $V^{(1)}_{PVTV}({\rm CT})$, derived from the NN
contact diagrams (b) of Fig.~\ref{fig:diagNN}, reads
\begin{eqnarray}
  V^{(1)}_{PVTV}({\rm CT})&=&\frac{1}{\Lambda_\chi^2f_\pi}\big\{{\bar C}_1\ i\bmk\cdot
      \left(\sone-\stwo\right)+{\bar C}_2\ i\bmk\cdot
      \left(\sone-\stwo\right)\tone\cdot\ttwo\nonumber\\
      &+&{{\bar C}_3\over 2}\ 
      \big[i\bmk\cdot\left(\sone-\stwo\right)\left(\tau_{1z}+\tau_{2z}\right)
        +i\bmk\cdot\left(\sone+\stwo\right)
        \left(\tau_{1z}-\tau_{2z}\right)\big]\nonumber\\
      &+&{{\bar C}_4\over 2}\ 
      \big[i\bmk\cdot\left(\sone-\stwo\right)\left(\tau_{1z}+\tau_{2z}\right)
        -i\bmk\cdot\left(\sone+\stwo\right)
        \left(\tau_{1z}-\tau_{2z}\right)\big]\nonumber\\
      &+&{\bar C}_5\ i\bmk\cdot
      \left(\sone-\stwo\right)\left(3\tau_{1z}\tau_{2z}-\tone\cdot\ttwo\right)
      \big\}\label{eq:ct}\,.
\end{eqnarray}
Notice that the above LECs ${\bar C}_1$, ${\bar C}_2$, ${\bar C}_3$, ${\bar
C}_4$ and ${\bar C}_5$ have been redefined to absorb 
various singular terms coming from the TPE and 3$\pi$ diagrams.
It is possible to write ten operators which can enter
$V_{PVTV}^{(1)}({\rm CT})$ at order $Q$ but only five of them are
independent as discussed in Chapter~\ref{sec:pionless}.
In this work we have chosen to write the operators in terms of $\bmk$,
so that the $r$-space version of $V_{PVTV}^{(1)}({\rm CT})$
will assume a simple local form with no gradients.

\item {\it Contact terms with an OPE.}
Diagrams like (c) and (d) of Fig.~\ref{fig:diagNN} vanish directly due to
the integration over the loop momentum.

\item {\it Two pions exchange diagrams}.
  The TPE term comes from the not singular contributions of panels (e)-(h) of
  Fig.~\ref{fig:diagNN}. This term has no isovector component,
as shown for the first time in~\cite{Bsaisou:2012rg}. It reads
\begin{eqnarray}
  V_{PVTV}^{(1)}(\rm TPE)&=&\frac{g_A{\bar g}_0^*}{ f_\pi\Lx^2}
   \tone\cdot\ttwo \ i\bmk\cdot(\sone-\stwo)\ L(k)
  +\frac{ g_A^3 {\bar g}_0^*}{ f_\pi\Lx^2}\ \tone\cdot\ttwo\  i\bmk\cdot(\sone-\stwo)
  \ (H(k)-3L(k))\nonumber\\
  &-&\frac{ g_A {\bar g}_2}{3 f_\pi\Lx^2}
   (3\tau_{1z}\tau_{2z}-\tone\cdot\ttwo)
   \ i\bmk\cdot(\sone-\stwo)\ L(k)\nonumber\\
   &-&\frac{ g_A^3 {\bar g}_2}{3 f_\pi\Lx^2}(3\tau_{1z}\tau_{2z}-\tone\cdot\ttwo)
   i\bmk\cdot(\sone-\stwo)\ 
   (H(k)-3L(k))\,,
\end{eqnarray}
where the loop functions $L(k)$ and $H(k)$ are defined in Eq.~(\ref{eq:funk}).

\item {\it Diagrams with three pion vertices} 
The $3\pi$-exchange term gives a NLO contribution through the diagram (i) of
Fig.~\ref{fig:diagNN},
\bgroup
\arraycolsep=1.0pt
\begin{align}
  V_{PVTV}^{(0)}(3\pi)&=&-\frac{5 g_A^3{\bar\Delta} M}{4 f_\pi\Lx^2}\pi
  \Big[(\tau_{1z}+\tau_{2z})
    \frac{i\bmk\cdot(\sone-\stwo)}{\omega_k^2}
    +
    (\tau_{1z}-\tau_{2z})\frac{i \bmk\cdot(\sone+\stwo)}{\omega_k^2}\Big]
  \nonumber\\
  &&\qquad\times\Big(\big(1-\frac{2 m_\pi^2}{s^2}\big)s^2A(k)+m_\pi\Big)\label{eq:3pnlo}\,,
\end{align}
\egroup
where $A(k)$ is given in Eq.~(\ref{eq:funA}).
Additional contributions coming from diagram (i) deriving from the
$1/M$ expansion of the energy denominators and vertex functions are
here neglected since we count them as N$^3$LO. 

The diagram in panel (j) of Fig.~\ref{fig:diagNN} contributes to
$V_{PVTV}^{(3\pi)}$ at N$^2$LO,
\bgroup
\arraycolsep=1.0pt
\begin{align}
  &V_{PVTV}^{(1)}(3\pi)=\frac{5 g_A{\bar\Delta} Mc_1}{2 f_\pi\Lx^2}
      \Big[(\tau_{1z}+\tau_{2z})i\bmk\cdot(\sone-\stwo)\nonumber\\
        &\qquad+(\tau_{1z}-\tau_{2z})i \bmk\cdot(\sone+\stwo)\Big]
       \ 4\frac{m_\pi^2}{\omega_k^2}L(k)\nonumber\\
       &\qquad-\frac{5 g_A{\bar\Delta} Mc_2}{6 f_\pi\Lx^2}
       \Big[(\tau_{1z}+\tau_{2z})i\bmk\cdot(\sone-\stwo)\nonumber\\
         &\qquad+(\tau_{1z}-\tau_{2z}) i \bmk\cdot(\sone+\stwo)\Big]
       \Big(2L(k)+6\frac{m_\pi^2}{\omega_k^2}L(k)\Big)\nonumber\\
       &\qquad-\frac{5 g_A{\bar\Delta} Mc_3}{4 f_\pi\Lx^2}
       \Big[(\tau_{1z}+\tau_{2z})i\bmk\cdot(\sone-\stwo)\nonumber\\
         &\qquad+(\tau_{1z}-\tau_{2z})i \bmk\cdot(\sone+\stwo)\Big]
       \Big(3L(k)+5\frac{m_\pi^2}{\omega_k^2}L(k)\Big)\label{eq:3pc3}\,.
\end{align}
\egroup
Note in Eq.~(\ref{eq:3pc3}) the presence of the
$c_1$, $c_2$ and $c_3$ LECs, which belong to the PCTC Lagrangian given
in Eq.~(\ref{eq:lagpctc}). In Eqs.~(\ref{eq:3pnlo})
and~(\ref{eq:3pc3}), ${\bar\Delta}$ is a
renormalized LEC.
    
The $3\pi$ PVTV vertex gives rise to a three body
interaction through the diagram $(k)$ in Fig.~\ref{fig:diagNN}.
The lowest contribution appears at NLO while at
N$^2$LO the various time orderings cancel out~\cite{Gnech:2019dod}.  
The final expression for the NLO of the 3N PVTV potential is,
\begin{eqnarray}
  V_{PVTV}^{(0)}(3\rm{N})&=&\frac{{\bar\Delta} g_A^3 M}{4 f_\pi^3}(\tone\cdot\ttwo\ \tau_{3z}+
  \tone\cdot\ttre\ \tau_{2z}+\ttwo\cdot\ttre\ \tau_{1z})\nonumber\\
  &&\times\frac{(i\bmk_1\cdot\sone)\ (i\bmk_2\cdot\stwo)\ (i\bmk_3\cdot\stre)}
                       {\omega^2_{k_1}\omega^2_{k_2}\omega^2_{k_3}}\,,
                       \label{eq:NNNpot}
\end{eqnarray}
where $\bmk_i=\bmp_i'-\bmp$. This expression is in agreement with that
reported in Ref.~\cite{deVries:2012ab,Bsaisou:2014oka}.

\end{itemize}

\subsubsection{The PVTV current}
\label{sec:pvtvc}

The PVTV current up to now has been considered coming from the
    LO one-body contribution
\begin{equation}
  \bmJ_{PVTV} = -\sum_{j=1}^A \left[ d_p {1+\tau_{jz}\over 2}+ d_n {1-\tau_{jz}\over 2} \right]
  i (\bmsi_j \cdot \bmq) \,, \label{eq:TVcurrent}
\end{equation}
where $d_p$ ($d_n$) is the proton (neutron) EDM. 
In the nuclear physics applications, it is customary to
consider $d_p$ and $d_n$ as unknown parameters, although they in principle
can be estimated in terms of the LECs entering the $\chi$EFT, as we
have seen in Sect.~\ref{sec:chiedm}. The complete derivation of PVTV two-body
currents has not been completed. Partial results have been given in Refs.~\cite{deVries:2011an,Bsaisou:2012rg}.

\section{PVTC and PVTV potentials in pionless EFT}
\label{sec:pionless}

In this section, we specifically focus on
the contact few-nucleon interactions which enter the potentials in both
chiral and pionless EFT formulations. We also discuss the expected hierarchy of
the corresponding LECs as suggested by the large-$N_c$ analysis.

\subsection{Effective Lagrangians}

At distances much larger than the range of the interactions mediated
by pions, the latter degrees of freedom can be integrated out of the
effective theory, and the relevant effective Lagrangian can be written in
terms of nucleon fields only, that interact through contact vertices. 

At leading order these vertices involve one spatial derivative of
fields, responsible for parity violation. Time derivatives can be
eliminated by recursively using the equations of motion order by order
in the low-energy expansion. This reflects our freedom in choosing the
nucleon interpolating field, and amounts to a definite choice of the
off-shell behavior of amplitudes. The theory can be formulated in
terms of non-relativistic nucleon fields represented by
two-component Pauli spinors $N_r(x)$. The relativistic $1/M$ corrections,
which can in principle be worked out (see
e.g.~\cite{Girlanda:2010ya}) will be of no interest
here. Relativistic covariance requires that the interactions depend on
the relative momenta only (momentum-dependent ``drift'' corrections,
which vanish in the center-of-mass frame of two nucleon systems, are
part of the above mentioned relativistic corrections). Thus, gradients
of nucleon fields in two-nucleon contact operators may only enter in
the combinations 
\begin{equation}
  \bmna (N_r^\dagger O_1 N_r) N_r^\dagger O_2 N_r, \quad [ (N_r^\dagger i
    \bmnalr O_1 N_r) N_r^\dagger O_2 N_r -  N_r^\dagger
    O_1 N_r (N_r^\dagger i \bmnalr O_2 N_r) ]\,, 
\end{equation}
where $(a i \bmnalr b) \equiv a (i \bmna
b) - ( i\bmna a) b$ and the factor $i$, meant to ensure the
hermiticity, makes it odd under time-reversal.

Since the underlying mechanism of parity violation in the SM
may induce $\Delta I =0,1,2$ transitions (at least to order
$G_F^2$), the effective Lagrangian will contain contact operators
which transform as isoscalars or the neutral components of isovector
and isotensors. In the two-nucleon case all these flavor structures
are real, and therefore unaffected by the time-reversal operation,
except for $({\vec \tau_1} \times {\vec \tau_2})_z$, which changes
sign. 
  
  \subsection{PVTC Lagrangian}
  Following the general considerations outlined above, there are ten
  possible structures entering the two-nucleon contact Lagrangian in
  the PVTC case, 
  \begin{equation}
    \begin{array}{ll}
      \Delta I=0 & \left\{ \begin{array}{l}
        O_1^{PVTC} = \bmna \times (N_r^\dagger
        \bmsi N_r) \cdot  N_r^\dagger \bmsi N_r\,, \\ 
                O_2^{PVTC} = \bmna \times (N_r^\dagger
                \bmsi \tau^a N_r) \cdot  N_r^\dagger \bmsi
                \tau^aN_r\,, \\ 
                O_{1'}^{PVTC} = (N_r^\dagger i
                \bmnalr \cdot \bmsi N_r)
                N_r^\dagger  N_r - N_r^\dagger \bmsi N_r \cdot
                (N_r^\dagger i \bmnalr N_r)\,,  \\ 
                O_{2'}^{PVTC} = (N_r^\dagger i
                \bmnalr \cdot \bmsi
                \tau^a N_r)   N_r^\dagger  \tau^aN_r - N_r^\dagger
                \bmsi \tau^aN_r \cdot (N_r^\dagger i
                    \bmnalr \tau^aN_r)\,,  \\ 
      \end{array} \right. \\
      \Delta I = 1 & \left\{ \begin{array}{l}
        O_3^{PVTC} = \epsilon^{ab3} \bmna \cdot
        (N_r^\dagger \bmsi \tau^a N_r)   N_r^\dagger \tau^b N_r\,, \\ 
        O_4^{PVTC} = \bmna \times (N_r^\dagger
        \bmsi \tau^3 N_r) \cdot  N_r^\dagger \bmsi N_r\,, \\ 
        O_{3'}^{PVTC} =  (N_r^\dagger i
        \bmnalr \cdot \bmsi \tau^3 N_r)
        N_r^\dagger  N_r -   N_r^\dagger \bmsi  \tau^3  N_r \cdot
        (N_r^\dagger i \bmnalr  N_r)\,,  \\ 
        O_{4'}^{PVTC} =  (N_r^\dagger i
        \bmnalr \cdot \bmsi  N_r)
        N_r^\dagger \tau^3 N_r -   N_r^\dagger \bmsi    N_r \cdot
        (N_r^\dagger i \bmnalr \tau^3 N_r)\,, \\ 
      \end{array} \right. \\
      \Delta I=2 & \left\{ \begin{array}{l}
       O_5^{PVTC} = {\mathcal I}_{ab} \bmna \times
       (N_r^\dagger \bmsi \tau^a N_r) \cdot  N_r^\dagger \bmsi
       \tau^bN_r\,, \\ 
        O_{5'}^{PVTC} =  {\mathcal I}_{ab} \left[ (N_r^\dagger i
          \bmnalr \cdot \bmsi \tau^a N_r)
          N_r^\dagger \tau^b N_r -   N_r^\dagger \bmsi  \tau^a  N_r
          \cdot  (N_r^\dagger i \bmnalr \tau^b N_r)
          \right]. \\ 
\end{array} \right.
      \end{array}
    \end{equation}
The Fermi statistics of nucleon fields, together with Fierz's
reshuffling of spin-isospin indices allow to establish linear
relations between primed and unprimed operators, 
  \begin{equation} \label{eq:pvtcfierz}
    \begin{array}{l}
      O_{1'}^{PVTC}= \frac{1}{2} \left( O_1^{PVTC} + O_2^{PVTC} \right)\,,\\
      O_{2'}^{PVTC}= \frac{1}{2} \left( 3 O_1^{PVTC} - O_2^{PVTC} \right)\,,\\
      O_{3'}^{PVTC}=  O_3^{PVTC} + O_4^{PVTC} \,,\\
      O_{4'}^{PVTC}=  -O_3^{PVTC} + O_4^{PVTC} \,,\\
      O_{5'}^{PVTC}=  O_5^{PVTC} \,,
    \end{array}
  \end{equation}
thus reducing the number of independent operators to five, so that the effective Lagrangian can be written as
  \begin{equation} \label{eq:pvtclagr}
    {\mathcal L}_{PVTC,NN}^{(1)} = \frac{1}{\Lambda_\chi^2 f_\pi} \Bigl[ {1\over 2}C_1 O_1^{PVTC}+
      {1\over 2}C_2 O_2^{PVTC} + C_3 O_3^{PVTC} +  C_4 O_4^{PVTC}
      +{1\over 2}  C_5 O_5^{PVTC}\Bigr]
      \,,
  \end{equation}
where $C_i$ are LECs. This Lagrangian is identical to that reported in Eq.~(\ref{eq:LPVNN1}).
From this Lagrangian, one can derive the
potential given in Eq.~(\ref{eq:potctpv}).

The five LECs are in a one-to-one correspondence with the possible S-P
transitions in two-nucleon systems \cite{Danilov:1965hc}, namely
$^1$S$_0$-$^3$P$_0$ ($\Delta I=0,1,2$), $^3$S$_1$-$^1$P$_1$ ($\Delta
I=0$) and $^3$S$_1$-$^3$P$_1$ ($\Delta I=1$). 
This may be shown explicitly by using the spin-isospin projection
operators \cite{Danilov:1965hc,Kaplan:1998sz,Savage:1998rx,Phillips:2008hn} 
\begin{equation}
    P_{0,0} = \frac{1}{\sqrt{8}} \sigma_2 \tau_2\,, \quad P_{0,a} =
    \frac{1}{\sqrt{8}} \sigma_2 \tau_2\tau_a \,, \quad P_{i,0} =
    \frac{1}{\sqrt{8}} \sigma_2 \sigma_i \tau_2\,, \quad P_{i,a} =
    \frac{1}{\sqrt{8}} \sigma_2 \sigma_i \tau_2\tau_a \,, 
\end{equation}
normalized according to
\begin{equation}
   {\mathrm{Tr}} P_{\mu,\alpha} P^\dagger_{\nu,\beta} = \frac{1}{2}
   \delta_{\mu\nu} \delta_{\alpha\beta}, \quad \mu(\nu)=0,i(j)\,, \quad
   \alpha(\beta)=0,a(b)\,,  
\end{equation}
  such that the operator $(N_r^T P_{\mu,\alpha} N_r)^\dagger$ creates a
  correctly normalized two-nucleon state with the appropriate
  spin-isospin quantum numbers. 
  The relevant operators \cite{Phillips:2008hn}
    \begin{equation}
      \begin{array}{l}
        O^{(^1{\mathrm{S}}_0-^3{\mathrm{P}}_0)}_{\Delta I=0} =(N_r^T
        \sigma^2 \tau^2 \tau^a N_r)^\dagger (N_r^T  i
              \bmnalr \cdot \sigma^2
              \bmsi \tau^2 \tau^a N_r) + {\mathrm{h.c.}}\,,\\ 
        O^{(^1{\mathrm{S}}_0-^3{\mathrm{P}}_0)}_{\Delta I=1} =-i
        \epsilon^{ab3} (N_r^T \sigma^2 \tau^2 \tau^a N_r)^\dagger (N_r^T  i
                \bmnalr \cdot \sigma^2
                \bmsi \tau^2 \tau^b N_r) + {\mathrm{h.c.}}\,,\\ 
        O^{(^1{\mathrm{S}}_0-^3{\mathrm{P}}_0)}_{\Delta I=2}
        ={\mathcal I}_{ab} (N_r^T \sigma^2 \tau^2 \tau^a N_r)^\dagger (N_r^T
        i \bmnalr \cdot \sigma^2 \bmsi
        \tau^2 \tau^b N_r) + {\mathrm{h.c.}}\,,\\ 
        O^{(^3{\mathrm{S}}_1-^1{\mathrm{P}}_1)}_{\Delta I=0} =(N_r^T
        \sigma^2 \bmsi \tau^2  N_r)^\dagger \cdot (N_r^T  i
              \bmnalr  \sigma^2  \tau^2  N_r) +
              {\mathrm{h.c.}}\,,\\ 
        O^{(^3{\mathrm{S}}_1-^3{\mathrm{P}}_1)}_{\Delta I=1} = (N_r^T
        \sigma^2 \bmsi \tau^2  N_r)^\dagger \cdot (N_r^T  
              \bmnalr \times \sigma^2
              \bmsi \tau^2 \tau^3 N_r) + {\mathrm{h.c.}}\,, 
      \end{array}\label{eq:ghhop}
    \end{equation}
are related to the original basis via Fierz's transformations as follows,
    \begin{equation}
      \begin{array}{l}
        O^{(^1{\mathrm{S}}_0-^3{\mathrm{P}}_0)}_{\Delta I=0} = 3 O_1^{PVTC} + O_2^{PVTC}\,,\\
        O^{(^1{\mathrm{S}}_0-^3{\mathrm{P}}_0)}_{\Delta I=1} =4 O_4^{PVTC}\,,\\
        O^{(^1{\mathrm{S}}_0-^3{\mathrm{P}}_0)}_{\Delta I=2} =-2 O_5^{PVTC}\,,\\
        O^{(^3{\mathrm{S}}_1-^1{\mathrm{P}}_1)}_{\Delta I=0} =- O_1^{PVTC} +  O_2^{PVTC}\,,\\
        O^{(^3{\mathrm{S}}_1-^3{\mathrm{P}}_1)}_{\Delta I=1} =-4 O_3^{PVTC} \,,
      \end{array}
      \end{equation}
whence one can read the relation between the  partial-waves projected
LECs and the $C_i$. The potential derived from the operators given in
Eq.~(\ref{eq:ghhop}) has been often used in studies of PVTC
observables. It is given explicitly as~\cite{Haxton:2013aca,Gardner:2017xyl}
\begin{eqnarray}
   V_{PVTC}^{(1)}(GHH)&=& \frac{1}{2 M m_\rho^2} \biggl\{
   \Lambda^{(^1{\mathrm{S}}_0-^3{\mathrm{P}}_0)}_{\Delta I=0}
   \Bigl[2(\bmsi_1-\bmsi_2)\cdot\bmK + i
     (\bmsi_1\times\bmsi_2)\cdot\bmk\Bigr]\nonumber\\   
   &+&\Lambda^{(^3{\mathrm{S}}_1-^1{\mathrm{P}}_1)}_{\Delta I=0}    \Bigl[2(\bmsi_1-\bmsi_2)\cdot\bmK - i
     (\bmsi_1\times\bmsi_2)\cdot\bmk\Bigr]\nonumber\\   
 &+&\Lambda^{(^1{\mathrm{S}}_0-^3{\mathrm{P}}_0)}_{\Delta I=1}
   (\tau_{1z}+\tau_{2z}) 2 (\bmsi_1-\bmsi_2)\cdot\bmK
   \nonumber\\
  &+&\Lambda^{(^3{\mathrm{S}}_1-^3{\mathrm{P}}_1)}_{\Delta I=1}  (\tau_{1z}-\tau_{2z}) 2 (\bmsi_1 +\bmsi_2)\cdot\bmK
    \nonumber\\
    &+&\Lambda^{(^1{\mathrm{S}}_0-^3{\mathrm{P}}_0)}_{\Delta I=2}
        {I}^{ab}\tau_{1a} \tau_{2b} {2\over\sqrt{6}}  (\bmsi_1-\bmsi_2)\cdot\bmK\biggr\}
    \,, \label{eq:ghhpot}
\end{eqnarray}
where the five LECs $\Lambda^{(\ldots)}_{\Delta T}$ are in one-to-one
correspondence with $C_{1-5}$. Explicitly
\begin{eqnarray}
  \Lambda^{(^1{\mathrm{S}}_0-^3{\mathrm{P}}_0)}_{\Delta I=0} &= &
         {\kappa\over 2} (C_1+C_2)\,,\nonumber\\
  \Lambda^{(^3{\mathrm{S}}_1-^1{\mathrm{P}}_1)}_{\Delta I=0} &= &
    {\kappa\over 2} (3C_2-C_1)\,,\nonumber\\
  \Lambda^{(^1{\mathrm{S}}_0-^3{\mathrm{P}}_0)}_{\Delta I=1} &=&
  \kappa C_4\,, \label{eq:conv} \\
  \Lambda^{(^1{\mathrm{S}}_1-^3{\mathrm{P}}_1)}_{\Delta I=1} &=&
  \kappa C_3\,, \nonumber \\
  \Lambda^{(^1{\mathrm{S}}_0-^3{\mathrm{P}}_0)}_{\Delta I=2}  &=&
  \sqrt{6} \kappa C_5\,, \nonumber 
 \end{eqnarray} 
where $\kappa=  2M m_\rho^2/f_\pi\Lambda_\chi^2$. 

\subsection{PVTV Lagrangian}
        The $T$-odd sector is very similar (see also Ref.~\cite{Vanasse:2019fzl}): one starts with a list of
        ten redundant operators, 
\begin{equation}
    \begin{array}{ll}
      \Delta I=0 & \left\{ \begin{array}{l}
        O_1^{PVTV} = \bmna \cdot (N_r^\dagger
        \bmsi N_r)   N_r^\dagger  N_r\,, \\ 
                O_2^{PVTV} = \bmna \cdot
                (N_r^\dagger \bmsi \tau^a N_r)   N_r^\dagger
                \tau^aN_r\,, \\ 
                O_{1'}^{PVTV} = (N_r^\dagger i
                \bmnalr \times \bmsi N_r)
                \cdot N_r^\dagger \bmsi  N_r \,,  \\ 
                O_{2'}^{PVTV} = (N_r^\dagger i
                \bmnalr \times \bmsi
                \tau^a N_r)   \cdot N_r^\dagger \bmsi \tau^a N_r \,,
                \\ 
      \end{array} \right. \\
      \Delta I = 1 & \left\{ \begin{array}{l}
        O_3^{PVTV} = \bmna \cdot (N_r^\dagger
        \bmsi \tau^3 N_r) N_r^\dagger  N_r\,, \\ 
        O_4^{PVTV} = \bmna \cdot (N_r^\dagger
        \bmsi  N_r) N_r^\dagger \tau^3 N_r\,, \\ 
        O_{3'}^{PVTV} =(N_r^\dagger i
        \bmnalr \times \bmsi \tau^3 N_r) \cdot
        N_r^\dagger \bmsi  N_r +(N_r^\dagger i
        \bmnalr \times \bmsi  N_r) \cdot
        N_r^\dagger \bmsi \tau^3 N_r \,,  \\ 
        O_{4'}^{PVTV} =\epsilon^{ab3} \left[
          (N_r^\dagger i \bmnalr \cdot\bmsi
          \tau^a N_r)    N_r^\dagger \tau^b  N_r +(N_r^\dagger i
              \bmnalr  \tau^a N_r)    \cdot
              N_r^\dagger  \bmsi\tau^b N_r \right] \,,  \\ 
      \end{array} \right. \\
      \Delta I=2 & \left\{ \begin{array}{l}
       O_5^{PVTV} = {\mathcal I}_{ab} \bmna 
       \cdot (N_r^\dagger \bmsi \tau^a N_r)  N_r^\dagger  \tau^bN_r\,,
       \\ 
        O_{5'}^{PVTV} =  {\mathcal I}_{ab}
        (N_r^\dagger i \bmnalr \times \bmsi
        \tau^a N_r)   \cdot N_r^\dagger \bmsi \tau^b N_r \,, \\ 
\end{array} \right.
      \end{array}
\end{equation}
and uses Fierz's identities to establish the linear relations,
  \begin{eqnarray} \label{eq:pvtvfierz}
      O_{1'}^{PVTV} &=& -     O_{1}^{PVTV}-     O_{2}^{PVTV}\,,\nonumber\\
      O_{2'}^{PVTV} &=& -   3  O_{1}^{PVTV}+     O_{2}^{PVTV}\,,\nonumber\\
      O_{3'}^{PVTV} &=& -   2 O_{3}^{PVTV}-2    O_{4}^{PVTV}\,,\\
      O_{4'}^{PVTV} &=& -   2 O_{3}^{PVTV}+2    O_{4}^{PVTV}\,,\nonumber\\
      O_{5'}^{PVTV} &=& -   2 O_{5}^{PVTV}\,,\nonumber
  \end{eqnarray}
so that the Lagrangian only depends on five LECs,
\begin{equation}
  {\mathcal L}_{PVTV,NN}^{(1)} = {1\over \Lambda_\chi^2 f_\pi}\sum_{i=1}^5  \bar C_i O_i^{PVTV},
\end{equation}
from which one can derive the potential given in Eq.~({\ref{eq:ct}).
  
%\emanuele{There seems to be a sign and a factor of 2 between Eq. 112 and Eq. 31. Should stick to the definition that gives the potential in Eq. 93?}  
  
The five S-P transition operators only differ from the $T$-even case
by a factor of $i$, 
      \begin{equation}
      \begin{array}{l}
        \bar O^{(^1{\mathrm{S}}_0-^3{\mathrm{P}}_0)}_{\Delta I=0}
        =(N_r^T \sigma^2 \tau^2 \tau^a N_r)^\dagger (N_r^T
        \bmnalr \cdot \sigma^2 \bmsi
        \tau^2 \tau^a N_r) + {\mathrm{h.c.}}\,,\\ 
        \bar O^{(^1{\mathrm{S}}_0-^3{\mathrm{P}}_0)}_{\Delta I=1} =
        \epsilon^{ab3} (N_r^T \sigma^2 \tau^2 \tau^a N_r)^\dagger (N_r^T  i
                \bmnalr \cdot \sigma^2
                \bmsi \tau^2 \tau^b N_r) + {\mathrm{h.c.}}\,,\\ 
        \bar O^{(^1{\mathrm{S}}_0-^3{\mathrm{P}}_0)}_{\Delta I=2}
        ={\mathcal I}_{ab} (N_r^T \sigma^2 \tau^2 \tau^a N_r)^\dagger (N_r^T
        \bmnalr \cdot \sigma^2 \bmsi
        \tau^2 \tau^b N_r) + {\mathrm{h.c.}}\,,\\ 
        \bar O^{(^3{\mathrm{S}}_1-^1{\mathrm{P}}_1)}_{\Delta I=0}
        =(N_r^T \sigma^2 \bmsi \tau^2  N_r)^\dagger \cdot (N_r^T
        \bmnalr  \sigma^2  \tau^2  N_r) +
        {\mathrm{h.c.}}\,,\\ 
        \bar O^{(^3{\mathrm{S}}_1-^3{\mathrm{P}}_1)}_{\Delta I=1} =
        (N_r^T \sigma^2 \bmsi \tau^2  N_r)^\dagger \cdot (N_r^T  i
             \bmnalr \times \sigma^2
             \bmsi \tau^2 \tau^3 N_r) + {\mathrm{h.c.}}\,, 
      \end{array}
    \end{equation}
    related to the original basis as follows,
    \begin{eqnarray}
        \bar O^{(^1{\mathrm{S}}_0-^3{\mathrm{P}}_0)}_{\Delta I=0} &=& 6
        O_1^{PVTV} +
        2 O_2^{PVTV}\,,\\ 
        \bar O^{(^1{\mathrm{S}}_0-^3{\mathrm{P}}_0)}_{\Delta I=1} &=& -4
        O_3^{PVTV} -4
        O_4^{PVTV}\,,\\ 
        \bar O^{(^1{\mathrm{S}}_0-^3{\mathrm{P}}_0)}_{\Delta I=2} &=&-4
        O_5^{PVTV}\,,\\ 
        \bar O^{(^3{\mathrm{S}}_1-^1{\mathrm{P}}_1)}_{\Delta I=0}
        &=& 2  O_1^{PVTV} - 2
        O_2^{PVTV}\,,\\ 
        \bar O^{(^3{\mathrm{S}}_1-^3{\mathrm{P}}_1)}_{\Delta I=1} &=&4
        O_3^{PVTV} -4 O_4^{PVTV}\,. 
      \end{eqnarray}

\subsection{Constraints from the large-$N_c$ limit}
        In 't Hooft combined large-$N_c$ and small coupling limit,
        with $g^2_s N_c$ fixed \cite{tHooft:1973alw}, QCD considerably
        simplifies, while maintaining many of the features of the
        actual theory, becoming a theory of stable hadrons. The
        baryons emerge as dense systems of many quarks, subjected to a
        mean field potential \cite{Witten:1979kh}. 
        Nucleon-nucleon interactions exhibit in this limit a
        spin-flavor symmetry
        \cite{Dashen:1994qi,Kaplan:1995yg,Kaplan:1996rk}. Indeed, due
        to the fact that nucleons carry definite spin and isospin of
        $O(1)$, interactions inducing a change in either spin or
        isospin are suppressed relative to the dominant $O(N_c)$ ones,
        that are either spin-isospin independent ($\sim {\bf 1}$) or
        dependent on both ($\sim \sigma \tau$). The large-$N_c$
        counting of momenta follows from the observation that the
        nucleon-nucleon scattering amplitude is in this limit a sum of
        meson exchange poles, each one depending only on the relative
        momentum transfer. The average relative momenta can only
        appear as  relativistic corrections, which are suppressed by
        inverse powers of $M \sim O(N_c)$. 

                Apparently the resulting scaling laws do not conform
                with the operator identities (\ref{eq:pvtcfierz})
                and~(\ref{eq:pvtvfierz}) and seem to imply a dependence
                on the choice of operator basis. However, one can
                start with the redundant set of operators, pertinent
                to a theory of distinguishable nucleons, since the
                large-$N_c$ arguments outlined above are completely
                general and do not rely on the statistics of the
                interacting baryons (the only assumption is that they
                both carry spin and isospin of $O(1)$).  
 As a result  one obtains the large-$N_c$ scaling of the LECs in the PVTV contact potential,
        \begin{equation}
          \begin{array}{l}
            C_2 \sim C_5 \sim O(1)\,,\\
            C_3 \sim C_4\sim C_{3'} \sim O(1/N_c)\,,\\
            C_1\sim C_{1'}\sim
            C_{2'}\sim C_{5'} \sim
            O(1/N_c^2)\,,\\ 
            C_{4'}\sim O(1/N_c^3)\,,
\end{array}
          \end{equation}
        and in the PVTV one,
        \begin{equation}
          \begin{array}{l}
            \bar C_3 \sim O(1)\,,\\
            \bar C_1\sim
            \bar C_2\sim
            \bar C_{2'} \sim
            \bar C_5 \sim
            \bar C_{5'} \sim O(1/N_c)\,,\\ 
            \bar C_4\sim
            \bar C_{3'}\sim
            \bar C_{4'} \sim O(1/N_c^2)\,,\\ 
            \bar C_{1'}\sim O(1/N_c^3).
          \end{array}
        \end{equation}
        Therefore we have only two leading LECs in the PVTC potential
        ($C_2$ and $C_5$ corresponding to
        $\Delta I=0,2$ respectively) and only one in the PVTV
        potential ($\bar C_3$ with $\Delta I=1$)
        \cite{Schindler:2015nga,Samart:2016ufg}. This
        % fact
        largely
        increases the predictive power for low-energy hadronic parity
        violation, and allows to put more severe constraints on the
        forthcoming  experimental results. Notice however that the
        above results are obtained by simply projecting the Hartree
        Hamiltonian in the nucleon-nucleon sector. A consistent 
        treatment would require to consider the induced effect on NN
        contact vertices of  $\Delta$ exchanges, since the latter are
        enhanced, in the large-$N_c$ limit, due to the degeneracy
        between nucleon and delta masses implied by the spin-flavor
        symmetry. 
        
Moreover, for the PVTV case, this picture is obscured by the fact that
the magnitude of the five contact LECs depends
% very much
strongly on the
particular type of the CP-violating source at the quark level. For
example,  the QCD $\bar\theta$ term conserves isospin symmetry such that
$\bar C_{3,4,5}$ are suppressed by powers of $\varepsilon \epsilon_{m_\pi}$
compared to $\bar C_{1,2}$ (see Table 1). Despite the possible $1/N_c$
suppression of $\bar C_{1,2}$ compared to $\bar C_3$ the former are still
expected to dominate.

\section{One-meson exchange models}
\label{sec:obep}
In the past, a simple and rather efficient description of the strong PCTC NN interaction
was obtained in terms of a sum of
single meson exchanges~\cite{Nagels:1975fb,Nagels:1976xq}.
These models started to be popular since the discovery of various
meson resonances during the sixties. The potentials were generally  constructed
taking into account the exchanges of pions ($J^P=0^-$, $m_\pi=138$ MeV),
$\eta$-mesons ($J^P=0^-$, $m_\eta=550$ MeV), and
$\rho$- and $\omega$-mesons ($J^P=1^-$, $m_{\rho,\omega}=770,780$ MeV),
but clearly, the number of mesons to be included is somewhat arbitrary.
This picture has been extended also to describe PVTC and PVTV interactions,
simply considering single meson exchanges where one of the vertex is strong
and PCTC, while the other violates P and conserves T or violates both P and T.  Then, all the dynamics of such interactions is contained in a number of
PVTC and PVTC nucleon-nucleon-meson (NNM) coupling constants.

\begin{table}[bth]
  \begin{center}
\begin{tabular}{l|c|c|c|c|c|c|c|c|c}
\hline
  \hline
  &&&&&&&&&  \\[-10pt]
&$\overline{N}N$ 
& $\overline{N}i\, \gamma_5N$
& $\overline{N} \gamma_\mu N$
& $\overline{N} \gamma_\mu \gamma_5 N$
& $\overline{N} \sigma_{\mu \nu} N $
& $\pi_a$
& $\rho_a$
& $\eta$
& $\omega$ \\
\hline
H & + & + & + & + & + & + & + & + & +   \\
\hline
P & + & -- & + & -- & + & -- & + & -- & +  \\
\hline
C & + & + & -- & + & -- & $(-)^{a+1}$ & $-(-)^{a+1}$ & +  & --  \\
\hline
\hline
\end{tabular}
\caption{ \label{tab:sign1}
  Transformation properties of fermion bilinears
  with different elements of the Clifford algebra
  and various meson fields
  under hermitian conjugation (H), parity (P), and
  charge conjugation (C). Note that the pion and rho-meson fields
  are isospin triplets, $a=1,2,3$.}
\end{center}
\end{table}

One starts by writing the Lagrangian
consistent of Yukawa-like NNM vertices,  invariant under the proper Lorentz
transformations, and either conserving or violating the discrete P, C, T
symmetries. The building blocks of the Lagrangian are therefore nucleon
bilinears multiplied by a meson field arranged so that Lorentz symmetry is satisfied.
For the construction of the PCTC Lagrangian, one usually includes only
isospin-conserving terms, however, for the PVTC and PVTV Lagrangians also
isospin-violating terms have to be included as the underlying operators at the quark level are not necessarily isospin symmetric.
A summary of the transformation properties of nucleon bilinears
with different elements of the Clifford algebra and the
various meson fields under hermitian conjugation (H), parity P, and
charge conjugation C are reported in Table~\ref{tab:sign1}.

Using these properties it is not difficult to write the
Lagrangians. For example, the strong $\mathcal L_{\rm PCTC}$
Lagrangian constructed with these mesons is given by (here we list
only isospin -conserving terms) 
\begin{eqnarray}
\mathcal L_{\rm PCTC}&=& g_{\pi}\bar{N} i\gamma_5 \vec\tau\cdot\vec\pi N
              +g_{\eta }\bar{N}i\gamma_5\eta N \nonumber \\
              &- &  g_{\rho }\bar{N}\left(\gamma^\mu-i\frac{\chi_V}{2 M}\sigma^{\mu\nu} q_\nu\right)
              \vec\tau\cdot \vec\rho_{\mu} N
               \nonumber \\
               &- &  g_{\omega }\bar{N}\left(\gamma^\mu-i\frac{\chi_S}{2M}\sigma^{\mu\nu} q_\nu\right)\omega_\mu N\,,
\end{eqnarray}
where $q^\mu$ is the meson momentum~\footnote{More appropriately, these Lagrangian terms should
  be written in terms of four-gradients. For example
  \begin{equation}
    \bar{N}i\frac{\chi_V}{2 M}\sigma^{\mu\nu} q_\nu \vec\tau\cdot \vec\rho_{\mu} N \rightarrow
    -\bar{N}\frac{\chi_V}{2 M}\Bigl[ \partial_\nu  \, , \, \sigma^{\mu\nu} \vec\tau\cdot \vec\rho_{\mu}\Bigr] N \,.
    \nonumber
  \end{equation}
where $[,]$ denotes the commutator.}, $\pi_a$, $\rho^\mu_a$, $\eta$ and $\omega^\mu$
are meson fields and $g_\pi$, $\ldots$
PCTC coupling constants. Above,  $\chi_V$ and $\chi_S$ are the ratios of
the tensor to vector coupling constant for $\rho$ and $\omega$,
respectively. Assuming  vector-meson dominance~\cite{Sakurai:1960ju}, they can be related
to the iso-vector and iso-scalar magnetic moments of a nucleon
($\chi_V=3.70$ and $\chi_S=-0.12$). Note 
that the pion and rho-meson are isospin triplets, therefore the
fields have the isospin index $a=1,\ldots,3$. Moreover, the rho- and
omega-mesons have spin 1, and their fields correspondingly are vector
fields with index $\mu=0,\ldots,3$.

Let us now consider the PVTC Lagrangian constructed in terms of the same
mesons. In this case one has to take into account
Barton's theorem~\cite{Barton:1961eg}, which asserts that exchange of neutral
and spinless mesons between on-shell nucleons is forbidden by CP invariance,
and therefore they cannot enter in a PVTC Lagrangian. 
Therefore only $\pi^\pm,\rho$ and $\omega$ vertices need to be considered and
the form of the PVTC effective Lagrangian is~\cite{Haeberli:1995uz}
\begin{eqnarray}
  \mathcal L_{\rm PVTC}&=&{h^1_\pi\over \sqrt{2}}\bar{N}(\vec\pi\times\vec\tau)_3N\nonumber\\
  &+&\bar{N}\left(h_\rho^0\vec\tau\cdot(\vec\rho)^\mu +h_\rho^1\rho_3^\mu
  +{h_\rho^2\over 2\sqrt{6}}(3\tau_3\rho_3^\mu -\vec\tau\cdot(\vec\rho)^\mu)\right)
  \gamma_\mu\gamma_5N\nonumber\\
  &+&\bar{N}(h_\omega^0\omega^\mu+h_\omega^1\tau_3\omega^\mu
  )\gamma_\mu\gamma_5N
  -h_\rho^{'1}\bar{N}(\vec\tau\times(\vec\rho)^\mu)_3{\sigma_{\mu\nu}q^\nu\over
  2M}\gamma_5N\,,
  \end{eqnarray}
where $h^1_\pi$, $\ldots$ are PVTC coupling constants to be determined.
As discussed also in Section~\ref{sec:chieft}, where we focused in particular on
the pion-nucleon PVTC constant $h^1_\pi$, attempts to estimate the
magnitude of these couplings from the fundamental theory were reported in several
papers~\cite{Michel:1964zz,Donoghue:1992dd,
  McKellar:1967mxj,Fischbach:1968zz,Tadic:1969xx,Kummer:1968ra,Mckellar:1973rr}.
In particular, in the DDH paper~\cite{Desplanques:1979hn}, the authors presented
{\it reasonable ranges} inside of which these parameters
were extremely likely to be found, together with a set of ``best values''
(see Table~\ref{tab:ddh}). Clearly, these values have to be considered as
educated guesses in view of all the uncertainties of their evaluation.
Of the seven unknown weak couplings $h^1_\pi,h_\rho^0, \ldots$, there are
estimates that indicate that $h_\rho^{'1}$ is quite small~\cite{Holstein:1981cg}
and this term was generally omitted, leaving PVTC observables to be
described in terms of six constants.

\begin{table}
\begin{center}
\begin{tabular}{|c|c|c|}
\hline
\quad   & DDH~\cite{Desplanques:1979hn} & DDH~\cite{Desplanques:1979hn} \\
Coupling & Reasonable Range & ``Best" Value \\ \hline
$h^1_\pi$ & $0\rightarrow 30$ &12\\
$h_\rho^0$& $30\rightarrow -81$&-30\\
$h_\rho^1$& $-1\rightarrow 0$& -0.5\\
$h_\rho^2$& $-20\rightarrow -29$&-25\\
$h_\omega^0$&$15\rightarrow -27$&-5\\
$h_\omega^1$&$-5\rightarrow -2$&-3\\ \hline
\end{tabular}
\caption{\label{tab:ddh}
  Weak NNM couplings as estimated in Ref.~\protect\cite{Desplanques:1979hn}.  All
numbers are quoted in units of the value $3.8\times 10^{-8}$.}
\end{center}
\end{table}
In the same manner, we can write the PVTV Lagrangian
composed of NNM vertices~\cite{Herczeg:1987gp,Liu:2004tq}
\begin{eqnarray}
\mathcal L_{\rm PVTV}
&=&\bar{N}[\bar{g}_\pi^{0} \vec \tau \cdot \vec \pi+\bar{g}_\pi^{1}\pi_3
           +\bar{g}_\pi^{2}(3\tau_3 \pi_3-\vec \tau \cdot \vec \pi)]N\\
& + &\bar{N}[\bar{g}^{0}_\eta\eta+\bar{g}^{1}_\eta \tau_3 \eta] N\\
& + &\bar{N}\frac{1}{2M}[\bar{g}_\rho^{0}\vec \tau \cdot (\vec \rho)^\mu
                         +\bar{g}^{1}_\rho \rho_3^\mu
                         +\bar{g}^{2}_\rho(3\tau_z \rho_3^\mu-\vec \tau\cdot(\vec\rho)^\mu )]
                         \sigma_{\mu\nu}q^\nu\gamma_5 N \\
& + &\bar{N}\frac{1}{2M}[\bar{g}^{0}_\omega\omega_\mu
                         +\bar{g}^{1}_\omega \tau_z \omega_\mu]
                         \sigma^{\mu\nu}q_\nu\gamma_5 N\,,
\end{eqnarray}
where $\bar{g}^{i}_\alpha$, $i=0,1,2$, are PVTV
meson-nucleon coupling constants. In this case, there were no attempts to
obtain the values of these coupling constants from the fundamental theory, as also
the magnitude of the parameters entering the underlying theory is unknown.

From these Lagrangians, the PVTC and PVTV interactions are obtained as a sum of
single-meson exchange diagrams. Regarding PVTC, below we report the
potential in the form obtained by DDH~\cite{Desplanques:1979hn}
\begin{eqnarray}\label{eq:DDHpot}
  V_{\mathrm{PVTC}}&=&
  - \frac{g_{\pi}h^1_\pi}{ 2\sqrt{2} M} i(\vec \tau_1\times \vec \tau_2)_z
  \frac{(\bmsi_1+\bmsi_2)\cdot \bmk }{k^2+ m_\pi^2}\nonumber\\  
 &&-\frac{ g_\rho}{M}
    \left[  \vec \tau_1\cdot \vec\tau_2  \,h_\rho^0 + \frac{(
      \tau_{1z}+\tau_{2z})}{2}  h_\rho^1
      +\frac{3 \tau_{1z} \tau_{2z}-\vec \tau_1\cdot \vec
        \tau_2}{2\sqrt 6} h_\rho^2\right] \nonumber\\
 && \quad \times \left[ {2(\bmsi_1 - \bmsi_2)\cdot\bmK + (1+\chi_V) i
        (\bmsi_1 \times \bmsi_2)\cdot \bmk \over k^2+m_\rho^2} \right]
    \nonumber\\
 &&-\frac{ g_\omega}{M}\left[h_\omega^0+ \frac{(\tau_{1z}+
        \tau_{2z})}{2}  h_\omega^1 \right]
   \left[ {2(\bmsi_1 - \bmsi_2)\cdot\bmK + (1+\chi_S) i
        (\bmsi_1 \times \bmsi_2)\cdot \bmk \over k^2+m_\omega^2} \right]
    \nonumber\\
    &&+ \left[\frac{g_\rho h_\rho^1}{M} {(
        \tau_{1z}-\tau_{2z})(\bmsi_1 + \bmsi_2)\cdot\bmK
        \over k^2+m_\rho^2}\right]
    - \left[\frac{g_\omega h_\omega^1}{M} {(
	\tau_{1z}-\tau_{2z})(\bmsi_1 + \bmsi_2)\cdot\bmK
        \over k^2+m_\omega^2}\right]\nonumber\\
 && -\left[ \frac{g_\rho h_\rho^{1\,\prime}}{2 M} { i(\vec \tau_1 \times
        \vec \tau_2)_z (\bmsi_1 + \bmsi_2)\cdot \bmk
        \over k^2+m_\rho^2}\right]\,,    
\end{eqnarray}
where $\bmk$ and $\bmK$ are defined in Eq.~(\ref{eq:Kk}). Often the
potential is regularized for large values of $k$, modifying the
meson propagators so that
$1/( k^2+m_x^2)\rightarrow f_{\Lambda_x}(k^2)/( k^2+m_x^2)$,
where $x=\pi$, $\rho$, and $\omega$. For example, in Ref.~\cite{Schiavilla:2008ic}
the following regularization was chosen
\begin{equation}\label{eq:DDHcutoff}
{1\over k^2+m_x^2} \rightarrow {1\over k^2+m_x^2} \left(\frac{\Lambda_x^2 - m_x^2}{\Lambda^2_x+k^2} \right)^2\,,
\end{equation}
The parameters $\Lambda_\pi$, $\Lambda_\rho$, and $\Lambda_\omega$
were chosen to have values $1.72$ GeV, $1.31$ GeV, and $1.50$ GeV, respectively~\cite{Schiavilla:2008ic}.
However, the cutoff functions $f_{\Lambda_x}(k^2)$ were not always applied and
also their form can vary.

Several PVTC observables have been studied using the DDH potential, with the aim to identify
the values of the six or seven coupling constants, see for example
Refs.~\cite{RamseyMusolf:2006dz,Schindler:2013yua,Haxton:2013aca}. Up to now the lack
of accurate experimental values has prevented the completion of this task.

The PVTV potential was derived in Ref.~\cite{Haxton:1983dq,Gudkov:1992yc,Towner:1994qe,Liu:2004tq}. The momentum space version reads
\begin{eqnarray}
\label{eq:pot}
V_{\mathrm{PVTV}}&=& + \frac{g_{\pi}}{ 2 M}
   \left[  \bar g^0_\pi \vec \tau_1\cdot \vec\tau_2  + \bar g^1_\pi
   \frac{(\tau_{1z}+\tau_{2z})}{2}  
      +\bar g^2_\pi (3 \tau_{1z} \tau_{2z}-\vec \tau_1\cdot \vec
        \tau_2)\right]   \frac{i(\bmsi_1-\bmsi_2)\cdot \bmk }{m_\pi^2+k^2}\nonumber\\  
 &&-\frac{ g_\rho}{2M}
   \left[  \bar g^0_\rho \vec \tau_1\cdot \vec\tau_2
     + \bar g^1_\rho \frac{(\tau_{1z}+\tau_{2z})}{2}
     +\bar g^2_\rho (3 \tau_{1z} \tau_{2z}-\vec \tau_1\cdot \vec \tau_2) \right]
   \frac{i(\bmsi_1-\bmsi_2)\cdot \bmk }{m_\rho^2+k^2} \nonumber\\
    && + \frac{ g_\eta}{2M}\left[\bar g_\eta^0+ \bar g^1_\eta
      \frac{(\tau_{1z}+\tau_{2z})}{2} \right]
   \frac{i(\bmsi_1-\bmsi_2)\cdot \bmk }{m_\eta^2+k^2} \nonumber\\
    && -\frac{ g_\omega}{2M}
   \left[  \bar g^0_\omega + \bar g^1_\omega \frac{(\tau_{1z}+\tau_{2z})}{2}\right]
   \frac{i(\bmsi_1-\bmsi_2)\cdot \bmk }{m_\omega^2+k^2} \nonumber\\
      &&+  \left[\frac{g_\pi \bar g_\pi^1}{4M}  {(
        \tau_{1z}-\tau_{2z}) i(\bmsi_1 + \bmsi_2)\cdot\bmk
        \over k^2+m_\pi^2}\right]
     + \left[\frac{g_\rho \bar g_\rho^1}{4M} {(
        \tau_{1z}-\tau_{2z}) i(\bmsi_1 + \bmsi_2)\cdot\bmk
        \over k^2+m_\rho^2}\right]\nonumber\\
   &&- \left[\frac{g_\eta \bar g_\eta^1}{4M} {(
        \tau_{1z}-\tau_{2z}) i (\bmsi_1 + \bmsi_2)\cdot\bmk
        \over k^2+m_\eta^2}\right]
    - \left[\frac{g_\omega \bar g_\omega^1}{4M} {(
	\tau_{1z}-\tau_{2z}) i(\bmsi_1 + \bmsi_2)\cdot\bmk
        \over k^2+m_\omega^2}\right]\,.
\end{eqnarray}
Also in this case, cut off functions can be applied in order to regularize the large $k$ behavior of
$V_{\mathrm{PVTV}}$. It is worthwhile to stress that the PVTV
  meson-exchange potential involves significantly more parameters than
  the LO PVTV chiral potential which depends in principle only on 4 LECs $\bar
  g_{0,1}$ and $\bar C_{1,2}$, with $\bar g_2$, $\bar \Delta$, and
  $\bar C_{3,4,5}$ appearing at subleading orders. While the
  meson-exchange potential can be mapped onto the short-distance $\bar
  C_i$ operators, the dynamics from the 3-pion $\bar \Delta$
  interaction is not captured in this way.

\section{Selected results for various PVTC and PVTV observables}
\label{sec:res}
In this section we present a selection of results obtained with the 
chiral EFT potentials and currents described in
Sect.~\ref{sec:chieft} for various PVTC and PVTV 
observables. We will discuss first in the next three subsections the parity violation in 
the radiative neutron capture on the proton,  then the longitudinal asymmetry
in $\vec{p}p$ scattering, and subsequently
the $\vec{n}$-$p$ and $\vec{n}$-$d$ spin rotations. Finally, in the last
subsection, we present some results for the EDM of light nuclei.
% The aim
Our motivation
to include these results in the review is mainly
to establish benchmarks  to help future applications. We include also
a ``minimum'' analysis how the current experimental data 
constrain some of the values of the LECs entering the $\chi$EFT interactions.

Results obtained using the pionless EFT can be found, for example,
in Refs.~\cite{Schindler:2013yua,Haxton:2013aca,Gardner:2017xyl}.
The meson-exchange potentials (in particular the DDH model) were used to analyze the results
of several experiments of PVTC observables also in medium and heavy nuclei. For a summary of
the obtained results, see, for example, Refs.~\cite{Haeberli:1995uz,RamseyMusolf:2006dz,Gardner:2017xyl}.
Calculations of the EDM of light nuclei using the meson exchange potential were performed in
Refs.~\cite{Liu:2004tq,Song:2012yh, Yamanaka:2016umw}.

\subsection{Parity violation in radiative neutron capture on the
  proton}
\label{sec:resnpdgamma}
The radiative neutron capture on the proton $\vec n p\rightarrow d \gamma$,
where $d$ denotes the deuteron and $\vec n$ a longitudinally polarized neutron,
represents a very interesting process to study 
PVTC effects in nuclear physics. The longitudinal analyzing power for
this process is defined as 
\begin{eqnarray}\label{LAP}
  A_\gamma(\theta) & = &\frac{d\sigma_+(\theta) - d \sigma_-(\theta)}{d\sigma_+(\theta)
    + d \sigma_-(\theta)} = a_\gamma\,\cos \theta\,,
\end{eqnarray}
where $d\sigma_\pm(\theta)$ is the differential cross section for positive/negative
helicity neutrons, and $\theta$ is defined as the angle between the neutron spin
and the outgoing photon momentum. $a_\gamma$ has been targeted by several
experiments in the last decades. A first nonzero signal was reported last year
for incoming neutrons of thermal energies~\cite{Blyth:2018aon},
\begin{equation}\label{agammaEXP}
    a_\gamma = (-3.0\pm1.4\pm0.2)\cdot 10^{-8}\,.
\end{equation}
albeit only two standard deviations away from a null measurement. 

The theoretical asymmetry is given by
\begin{equation}
  a_\gamma=\frac{ -\sqrt{2}\, {\rm Re}\, \Big[ M_1^*(^1{\rm S}_0) E_1(^3{\rm S}_1) +
                                             E_1^*(^1{\rm S}_0) M_1(^3{\rm S}_1)\Big]
  +  {\rm Re}\,\Big[E_1^*(^3{\rm S}_1) M_1(^3{\rm S}_1)\Big] }
  {|M_1(^1{\rm S}_0)|^2 + |E_1(^1{\rm S}_0)|^2 + |M_1(^3{\rm S}_1)|^2 +  |E_1(^3{\rm S}_1)|^2} \,,
  \label{agammaTHEO}
\end{equation}
where $X_\ell({}^{2S+1}\mathrm{S}_J)$ are reduced matrix elements (RMEs) either of electric ($X=E$)
or magnetic ($X=M$) type, of multipolarity $\ell$, and describing the EM transition
from the $n-p$ system in the scattering state ${}^{2S+1}\mathrm{S}_J$~\cite{Schiavilla:2004wn}.

Compared to the PVTC longitudinal analyzing power in proton-proton scattering discussed later,
$a_\gamma$ comes with a big advantage. The initial neutron-proton state can be in a ${}^3S_1$ state,
such that the process is sensitive to ${}^3S_1 \leftrightarrow {}^3 P_1$ transitions and thus
depends on the LO PVTC NN potential. In chiral EFT the LO potential only depends
on the LEC $h^1_\pi$, such that measurements of $a_\gamma$ provide a unique chance to
pin down the value of this LEC, something which is much more difficult in
proton-proton scattering, where the contribution of the LO potential vanishes.
The disadvantage is that $\vec n p\rightarrow d \gamma$
is an electromagnetic process and therefore depends on P-conserving
and P-violating electromagnetic currents.

As can be seen from Eq.~(\ref{agammaTHEO}), nonzero $a_\gamma$ requires an
interference between electric and magnetic
dipole currents. As such, just including the leading magnetic moment current
in the presence of the LO PVTC NN potential leads to a vanishing result
and NLO currents are necessary. There are then three relevant contributions
that consist of interference between the isovector nucleon magnetic moment and
\begin{enumerate}
\item the one-body convection current in combination with the  PVTC NN potential,
\item the two-body PCTC currents in combination with the  PVTC NN potential,
\item the two-body PVTC currents.
\end{enumerate}
Each of these contributions is sizeable:
$a_\gamma^1 = (-0.27 \pm 0.03)h^1_\pi$, $a_\gamma^2 =  (-0.53 \pm 0.02)h^1_\pi$,
and $a_\gamma^3 = (0.72 \pm 0.03)h^1_\pi$ where the theoretical error bands
are obtained from cut-off variations in the strong NN potential and do
not reflect uncertainties from higher-order
contributions~\cite{deVries:2015pza}. While these
uncertainties are small on the individual contributions, they lead to a
sizeable uncertainty on the total analyzing power~\cite{deVries:2015pza}
\begin{equation}\label{agamma}
a_\gamma = a_\gamma^1 +a_\gamma^2 + a_\gamma^3 = (-0.11 \pm 0.05)h^1_\pi\,.
\end{equation}
The cancellations between the different contributions are related to gauge invariance
\cite{Hyun:2001yg,Schiavilla:2004wn,deVries:2015pza}
and this explains the relatively large total theoretical uncertainty.
While the electromagnetic
currents given above are explicitly gauge invariant as they result from the gauge-invariant
$\chi$EFT Lagrangian, explicit gauge invariance is lost due to applied regulator when solving
the NN scattering and bound-state equations. Future calculations can probably reduce the
uncertainty by using regulators that do not violate explicit gauge invariance, but such schemes
have not been applied to PVTC processes. Alternatively, it is possible to apply the Siegert
theorem to relate part of the electric dipole currents to the one-body charge density.
Ref.~\cite{Schiavilla:2002uc} applied the Siegert theorem in combination with phenomenological strong potentials
to calculate $a_\gamma$ finding a result in good agreement with the central value in Eq.~\eqref{agamma}.
Such calculations however do not include an uncertainty estimate, for instance from missing
transverse currents that are not included when applying the Siegert theorem. In this light,
Eq.~\eqref{agamma} can be interpreted as a conservative result. It would be interesting to
redo the calculation of $a_\gamma$ in an updated framework to reduce the theoretical uncertainty. 

The contribution to $a_\gamma$ of the short range components of the potential is
considered to be negligible. For example, using the meson-exchange model, the calculations
have shown that $a_\gamma$ is essentially unaffected by short-range
contributions~\cite{Desplanques:1975vle,McKellar:1975iy,Desplanques:1980pa,Schiavilla:2004wn},
represented in this case by $\rho$ and $\omega$ exchanges. Within $\chi$EFT, a resonance saturation estimate of the short-distance LECs 
contributing to the asymmetry led to short-distance contributions to $a_\gamma$ of roughly $5 \cdot 10^{-9}$ and thus very small \cite{deVries:2015pza}. 
 Therefore, 
considering the theoretical expression given in Eq.~(\ref{agammaTHEO}) and the experimental
value given in Eq.~(\ref{agammaEXP}), we obtain an estimate for the LEC $h^1_\pi$
\begin{equation}
  h^1_\pi = (2.7\pm 1.8)\times 10^{-7}\,.\label{h1piest}
\end{equation}
Note that the large experimental error and the large theoretical uncertainty only
allow to establish the positive sign and that the magnitude of this LEC is
consistent with the Lattice QCD preliminary evaluation reported in
Eq.~(\ref{eq:lattice})~\cite{Wasem:2011zz}.

\subsection{Parity violation in $\vec{p}p$ scattering}
\label{sec:respp}
PVTC effects in proton-proton scattering can be studied by looking at the
longitudinal analyzing power $A_z(E,\theta)$ defined as,
\begin{equation}
  A_z(E,\theta)=\frac{\sigma_{+}(\theta ,E)-\sigma_{-}(\theta ,E)}
  {\sigma_{+}(\theta ,E)+\sigma_{-}(\theta ,E)} \,,\label{eq:azpptheta}
\end{equation}
where $\theta$ is the scattering angle and $E$ the energy of the
protons in the laboratory frame,
and $\sigma_{+}(\theta ,E)$($\sigma_{-}(\theta ,E)$) the cross section when the
polarization of the incoming proton is parallel (anti-parallel) to the
beam direction.
Actually the experiments detect the particles scattered
in angular range $\left[ \theta_1, \theta_2 \right]$
and  the measured quantity is an ``average" of the
asymmetry over the total cross-section in this range, explicitly
\begin{equation}
    \overline{A}_z(E)=\frac{\int_{\theta_1\leq\theta\leq\theta_2} d\cos \theta
      \;A_z(\theta,E)\sigma(\theta,E)} 
   {\int_{\theta_1\leq\theta\leq\theta_2}d\cos \theta \;\sigma(\theta,E)}\,,
   \label{eq:azpp}
\end{equation}
where
\begin{equation}
  \sigma(\theta,E)=\frac{1}{2}\left(\sigma_{+}(\theta,E)+\sigma_{-}(\theta,E)\right)
\end{equation}
is the unpolarized differential cross-section for the process.
There exist three  measurements of the angle-averaged
$\vec{p}p$ longitudinal asymmetry $\overline{A}_z(E)$,
see Eq.~(\ref{eq:azpp}), obtained at different laboratory energies
$E$~\cite{Eversheim:1991tg,Kistryn:1987tq,Berdoz:2002sn}. 
The measurements and the angle ranges are reported in Table~\ref{tab:anglerange}.

The isospin state of two proton system 
is $|pp\rangle \equiv |T=1, T_z=1\rangle$, therefore the LO contribution
that comes from the OPE  vanishes and the LEC $h^1_\pi$ will
contribute to the observable only via the TPE box diagrams
that appear at NLO and N$^2$LO. Taking into account the isospin selection
rules, the longitudinal asymmetry can be written as
\begin{equation}
  \overline{A}_z =  h^1_\pi\, a^{(pp)}_0+ C \, a^{(pp)}_1 + \tilde{h}\, a^{(pp)}_2\,,
  \label{eq:aaa} 
\end{equation}
where the first two terms are NLO contributions and the third term enters at N${}^2$LO. We defined
\begin{equation}
  C=C_1+C_2 + 2\, (C_4+C_5)\,,\label{eq:lecc}
\end{equation}
\begin{equation}
  \tilde{h}=\frac{5g_A}{4}h^0_V+2\Big(\frac{g_A}{4}h^1_V
  -h^1_A\Big)-2\Big(\frac{g_A}{3}h^2_V+h^2_A\Big) \,, \label{eq:lech}
\end{equation}
and $a^{(pp)}_0$, $a^{(pp)}_1$, $a^{(pp)}_2$ are numerical coefficients independent
on the LEC values (but they depend on the energy). 
The values of the coefficients $a^{(pp)}_0$, $a^{(pp)}_1$, and $a^{(pp)}_2$
calculated with the $\chi$EFT N$^2$LO PVTC potential described in
Sect.~\ref{sec:pvtcpot} and the N$^4$LO  PCTC potential derived
in Ref.~\cite{Entem:2017gor} are reported in Table~\ref{tab:pp-eft-coef}. 
The only coefficient which receives contributions from both the NLO and
N$^2$LO potentials is $a^{(pp)}_0$. In the table, we report separately
the two contributions and also the total contribution, given simply as
\begin{equation}
  a^{(pp)}_0({\rm TOT}) = a^{(pp)}_0({\rm NLO}) + a^{(pp)}_0({\rm N^2LO}) \,,
  \qquad  a^{(pp)}_0({\rm N^2LO})= c_4 a^{(pp)}_0(4)\,.\label{eq:a0pp}
\end{equation}
The value of $a^{(pp)}_0({\rm N^2LO})$ has been obtained assuming a
value $c_4=3.56$ GeV$^{-1}$~\cite{Hoferichter:2015tha}. This
correction to $a^{(pp)}_0$ is of the order of $\sim50\%$ with respect
to the NLO value, somewhat larger than expected. This is related by
the unnaturally large value of the $\pi NN $ LEC $c_4$ appearing in
the PCTC Lagrangian~(\ref{eq:lagpctc}). This value has been obtained  
from the Roy-Steiner analysis of $\pi N$ scattering data
at N$^2$LO performed in Ref.~\cite{Hoferichter:2015tha}. 

\begin{table}
\begin{center}
\begin{tabular}{lcc}
\hline 
$E \, ({\rm MeV})$ & $\overline{A}_z$ $(10^{-7})$& $(\theta_1,\theta_2)$ \\ 
\hline
13.6 & $-0.97\pm 0.20$ & $(20^\circ, 78^\circ)$ \\ 
45 & $-1.53\pm 0.21$ & $(23^\circ,52^\circ)$ \\ 
221 & $+0.84\pm 0.34$ &$(5^\circ,90^\circ)$ \\ 
\hline 
\end{tabular} 
\caption{Values of $\overline{A}_z$ and angle ranges for the three measurements
  of the $\vec p p$ longitudinal analyzing
  power~\protect\cite{Eversheim:1991tg,Kistryn:1987tq,Berdoz:2002sn}.  
  }\label{tab:anglerange}
\end{center}
\end{table}
\begin{table}[t]
\begin{center}
\begin{tabular}{lccccc}
\hline
$E$ [MeV]   & $a^{(pp)}_0$(NLO) &  $a^{(pp)}_0$(N$^2$LO) & $a^{(pp)}_0$(TOT) & $a^{(pp)}_1$  & $a^{(pp)}_2$       \\
\hline
$13.6$      &  $\m 0.289$ & $\m 0.160$ & $\m 0.449$  & $  -0.044$ & $  -0.215$\\
$45$        &  $\m 0.595$ & $\m 0.355$ & $\m 0.950$  & $  -0.084$ & $  -0.475$\\
$221$       &  $  -0.281$ & $-  0.187$ & $  -0.468$  & $\m 0.036$ & $\m 0.251$\\
\hline
\end{tabular}
\caption{ \label{tab:pp-eft-coef}
  Values of the coefficients $a^{(pp)}_i$ calculated with the $\chi$EFT N$^2$LO PVTC potential described in
  Sect.~\ref{sec:pvtcpot} and the N$^4$LO  PCTC potential derived in Ref.~\cite{Entem:2017gor}
  at the three energies corresponding to the experimental data
  points. The PVTC potential has been regularized as in
  Eq.~(\ref{localcutoff}) adopting the value $\Lambda_C=500$ MeV for
  the cutoff parameter. The PCTC potential has been regularized with
  the same value of the cutoff parameter.
  For the coefficient $a^{(pp)}_0$
   we give separately the contributions of the NLO and N$^2$LO terms
   only and then their sum, see Eq.~(\ref{eq:a0pp}).}
\end{center}
\end{table}

Unfortunately, of the performed measurements, the two at the lowest energy
do not give independent information. In fact, the observable
$\overline{A}_z$ at low energy scales as $\sqrt{E}$, since its energy
dependence in this energy range is driven solely by that of the S-wave
(strong interaction) phase shift~\cite{Carlson:2001ma}. Because of
this scaling, it is not possible to fit from these data all three LECs
$h^1_\pi$, $C$, and $\tilde h$ at the same time. If we fix the 
value $h^1_\pi=2.7\times10^{-7}$ from the central value as extracted
from the $\vec n p\rightarrow d \gamma$ observable, see
Eq.~(\ref{h1piest}), then we can perform a  $\chi^2$
analysis of the three data points listed in Table~\ref{tab:anglerange}
in order to fix the values of $C$ and $\tilde h$.
    Note that this value of $h^1_\pi$ was obtained from the
     $\vec n p\rightarrow d \gamma$ calculation performed in
    Ref.~\cite{deVries:2015pza} using a different PCTC potential than
    that one used compute the $a^{(pp)}_i$ coefficients . However,
    since the  $\vec n p\rightarrow d \gamma$  experiment 
    depends mainly on the peripheral regions of the process, the value of
    $a_\gamma$ is not very sensitive to the PCTC interaction (see also
    the calculations reported in Ref.~\cite{Desplanques:2000ej}).
    
First of all, if we restrict ourselves to an NLO analysis, using
$h^1_\pi=2.7\times10^{-7}$ we would obtain
$C = (49 \pm 2) \cdot 10^{-7}$. 
If we take into account also the N${}^2$LO LEC, we report  in
Fig.~\ref{fig:ellipse}
the $C$ and $\tilde h$ values for which $\chi^2\le 2$,
which form an elliptic region. As can be seen, there appears to be 
a strong correlation between $C$ and $\tilde h$ and 
the range of allowed values of the LECs is rather large
$5\times 10^{-7}<C< 67\times 10^{-7}$ and
$-1.5\times 10^{-7}<\tilde h< 2.5\times 10^{-7}$.
Note that the ellipse is rather narrow and
almost coincides with a straight line.
See also Refs.~\cite{deVries:2014vqa,Viviani:2014zha} for a similar
analysis performed at NLO for the LECs $h^1_\pi$ and $C$ only.

\begin{figure}[t]
  \begin{center}
   \includegraphics[width=12cm,clip]{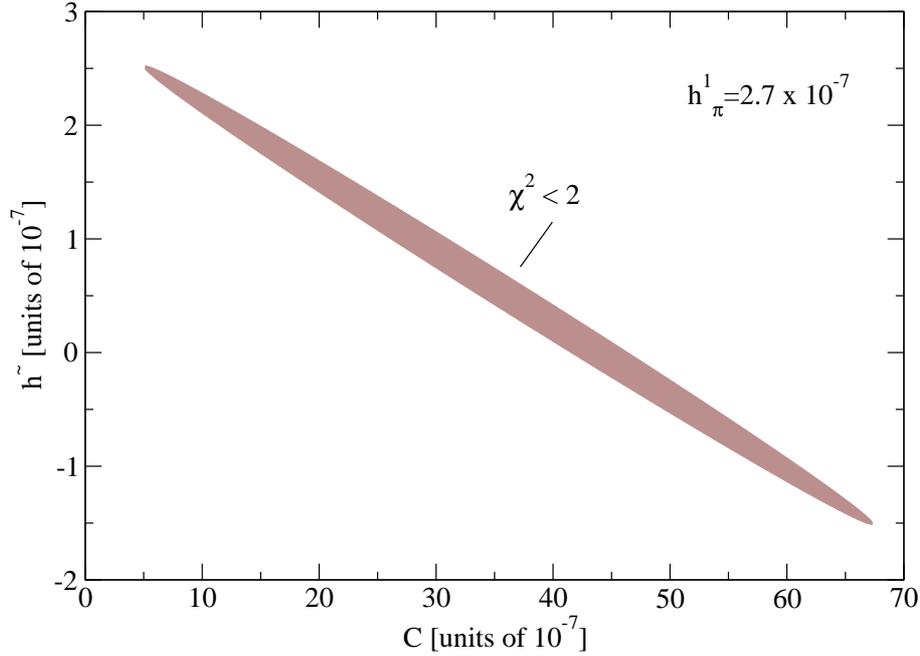}
   \caption{   \label{fig:ellipse}
     Region of $C$ and $\tilde h$ values
     for which  $\chi^2\le 2$ for the $\vec p\,$-$p$
     longitudinal asymmetry. The calculation is based on the coefficients
     $a^{(pp)}_0$, $a^{(pp)}_1$, and $a^{(pp)}_2$ reported in Table~\ref{tab:pp-eft-coef} assuming the value
     $h^1_\pi=2.7\times 10^{-7}$.
   }
   \end{center}
\end{figure}

The previous discussion did not take into account the large uncertainty
of the $h^1_\pi$ coupling constant after the fit of the $\vec n\,$-$p$
radiative capture asymmetry. In Table~\ref{tab:lecs}, we report representative
values of $C$ and $\tilde h$ giving the minimum value of $\chi^2$ corresponding
to range of values for $h^1_\pi$ as given in Eq.~(\ref{h1piest}).
In the fourth column we report values for $C$ if we neglect
  the N${}^2$LO contributions (setting $\tilde h =0$). We conclude that
  the combination of the $\vec pp$ and $\vec n p \rightarrow d \gamma$
  asymmetries allows for a rough extraction of the LO and NLO LECs $
  h^1_\pi$ and $C$, but is insufficient to also pinpoint the
  N${}^2$LO LEC $\tilde h$. The uncertainty of the extractions of
  $h^1_\pi$ and $C$ is dominated by theoretical and experimental
  uncertainties related to the PVTC asymmetry in the radiative neutron
  capture process.

\begin{table}[t]
\begin{center}
\begin{tabular}{cccc}
\hline
$ \quad h_\pi^1\quad$  &  $\quad C\quad$&  $\quad \tilde h \qquad$  &  $\quad C(\tilde h=0)\quad$\\
\hline
 $0.9$ & $27.7$ & $0.11$  & $28\pm 2$ \\ 
 $2.7$ & $34.5$ & $0.97$  & $49\pm 2$ \\  
 $4.5$ & $41.2$ & $1.84$  & $69\pm 3$ \\
\hline
\end{tabular}
\end{center}
\caption{\label{tab:lecs}
  Values for $C$ and $\tilde h$ corresponding to different values of $h^1_\pi$
  (all LECs are given in units of $10^{-7}$) giving the minimum value of the $\chi^2$ in the fit
  of the three experimental  $\vec p\,$-$p$ data points.
  For example, for $h^1_\pi=2.7\times 10^{-7}$, the $C$, $\tilde h$ values are those
  lying in the center of the elliptical contour shown in
  Fig.~\protect\ref{fig:ellipse}. The fourth column corresponds to an
  analysis where we ignore the N${}^2$LO contributions and thus set
  $\tilde h=0$.} 
   \end{table}

\subsection{The $\vec{n}$-$p$ and $\vec{n}$-$d$ spin rotation}
\label{sec:resnp}

The spin rotation of neutron traversing a slab of matter
in a plane transverse to the beam direction induced by the PVTC
potential is given by
\bgroup
\arraycolsep=1.0pt
\begin{eqnarray}
  \frac{{\rm d}\phi^{(nX)}}{{\rm d}z} &=& \frac{2\pi\rho}{(2S_X+1)\,
    v_{\rm rel}} {\rm Re} 
  \! \sum_{m_n m_X} \epsilon_{m_n}
 \!^{(-)}\langle p\hat{\bf z}; m_n , m_X | V_{PVTC} | 
  p\hat{\bf z};m_n , m_X\rangle^{(+)}\,,
  \label{eq:nphi}
\end{eqnarray}
\egroup
where $\rho$ is the density of hydrogen or deuterium nuclei for $X = p$ or $d$,
$\mid\! p \hat{\bf z};m_n,m_X\rangle^{(\pm)}$
are the $n$-$X$ scattering states with outgoing-wave $(+)$ and
incoming-wave $(-)$ boundary conditions and relative momentum
${\bf p}=p\, \hat{\bf z}$ taken along the spin-quantization axis
(the $\hat{\bf z}$-axis), $S_X$ is the $X$ spin, and $v_{\rm rel}=p/\mu$ is
the magnitude of the relative velocity, $\mu$ being the $n$-$X$
reduced mass.  The expression above is averaged over the spin
projections $m_X$; however, the phase
factor $\epsilon_{m_n}= (-)^{1/2-m_n}$ is $\pm 1$ depending
on whether the neutron has $m_n=\pm 1/2$. 
We consider the $n$-$p$ and $n$-$d$ spin rotations for vanishing
incident neutron energy (measurements of this observable are
performed using ultracold neutron beams). In the following,
we assume $\rho=0.4\times 10^{23}$ cm${}^{-3}$.
The rotation angle depends linearly on the PVTC LECs,
as higher-order weak corrections are negligible. We write
\begin{eqnarray}
  \frac{{\rm d}\phi^{(nX)}}{{\rm d}z}&=& h^1_\pi \, a^{(nX)}_0+ C_1 \, a^{(nX)}_1
   + C_2\,  a^{(nX)}_2 + \, C_3\,  a^{(nX)}_3+ C_4 \, a^{(nX)}_4+ C_5\,  a^{(nX)}_5 \nonumber \\
   &&\qquad +
   h^0_V \, b^{(nX)}_1 + h^1_V \, b^{(nX)}_2 +h^2_V \, b^{(nX)}_3 +h^1_A \, b^{(nX)}_4 + h^2_A \, b^{(nX)}_5
   \,, \label{eq:nphii}
\end{eqnarray}
where the $a^{(nX)}_i$ for $i=0, \dots ,5$ and $b^{(nX)}_i$ for $i=1, \dots ,5$
are numerical coefficients.
The coefficient $a^{(nX)}_0$ receives contributions from different chiral orders,
in particular 
\begin{equation}
    a^{(nX)}_0=a^{(nX)}_0(\text{LO})+a^{(nX)}_0(\text{NLO})+a^{(nX)}_0(\text{N$^2$LO})\,.
\end{equation}
The values of these coefficients for the $n$-$p$ case and the cut-off
value $\Lambda=500$ MeV are listed in Table~\ref{tab:nrotz-eft-coef}.
From that Table, it is possible to  appreciate the chiral convergence for the
coefficients $a^{(np)}_0$. The NLO correction is $\sim10\%$ of the
LO result. In this case, the N$^2$LO contribution vanishes
since the LEC $h^1_\pi$ in $V^{(2)}_{PVTC}({\rm TPE})$ multiplies
the operator $(\tau_{1z}+\tau_{2z})$. The $\vec n$-$p$
spin rotation is sensitive to all the LECs except for the LECs $C_4$ and $h^1_A$
multiplying again the isospin term $(\tau_{1z}+\tau_{2z})$;
in particular, there is a large sensitivity to $C_5$ and $h_A^2$,
which multiply the isotensor terms of the PVTC potential.

\begin{table}
\begin{center}
\begin{tabular}{lc|lc|lc}
\hline
$a^{(np)}_0(\text{LO})$      &  $1.227$ & $a^{(np)}_1$  &  $\m0.257$ & $b^{(np)}_1$  &  $\m1.653$  \\
$a^{(np)}_0(\text{NLO})$     &  $0.137$ & $a^{(np)}_2$  &  $\m0.178$ & $b^{(np)}_2$  &  $ -0.181$  \\
$a^{(np)}_0(\text{N$^2$LO})$ &  $0.000$ & $a^{(np)}_3$  &  $\m0.106$ & $b^{(np)}_3$  &  $\m1.882$  \\
$a^{(np)}_0(\text{TOT})$     &  $1.364$ & $a^{(np)}_4$  &  $\m0.000$ & $b^{(np)}_4$  &  $\m0.000$  \\
                             &          & $a^{(np)}_5$  &  $ -0.949$ & $b^{(np)}_5$  &  $\m4.456$  \\
\hline
\end{tabular}
\caption{ \label{tab:nrotz-eft-coef}
   Values of the coefficients entering the expression of the $\vec n$ -$p$ spin
   rotation in units of Rad m${}^{-1}$
   calculated for the $\chi$EFT N$^2$LO PVTC potential described in
   Sect.~\ref{sec:pvtcpot} and the N$^4$LO  PCTC potential
   derived in Ref.~\cite{Entem:2017gor}
   at vanishing neutron beam energy. The PVTC potential has been regularized as in
   Eq.~(\ref{localcutoff}) adopting the value $\Lambda_C=500$ MeV for
   the cutoff parameter. The PCTC potential has been regularized with
   the same value of the cutoff parameter.
   For $a^{(np)}_0$ we give explicitly the contribution of the different orders,
   the sum of the three contributions is given in fourth row.}
\end{center}
\end{table}

Regarding the $\vec{n}$-$d$ spin rotation, the coefficients,
as reported in Table~\ref{tab:nrotz-eft-coef-2}, are calculated
by using only the NLO PVTC potential.
We note the large sensitivity to
$h^1_\pi$ (this fact is well
known~\cite{Schiavilla:2008ic,Song:2010sz}), and to 
the LEC's $C_2$ and $C_3$.

\begin{table}[t]
\begin{center}
\begin{tabular}{lc}
\hline
$a^{(nd)}_0$  & $\m2.179$  \\
$a^{(nd)}_1$  & $ -0.010$  \\
$a^{(nd)}_2$  & $ -0.160$  \\
$a^{(nd)}_3$  & $\m0.191$  \\
$a^{(nd)}_4$  & $\m0.064$  \\
$a^{(nd)}_5$  & $\m0.000$  \\
\hline
\end{tabular}
\caption{ \label{tab:nrotz-eft-coef-2}
   The same as in Table~\ref{tab:nrotz-eft-coef} 
   but for the $\vec n$ -$d$ spin rotation and using
   the $\chi$EFT NLO PVTC potential and the N$^3$LO  PCTC potential 
   derived in Ref.~\cite{Machleidt:2011zz}.
   }
\end{center}
\end{table}

At present there are no measurements of these quantities, however
their experimental knowledge could be very useful in isolating certain combinations of LECs.

\subsection{EDM of light nuclei}
\label{sec:resedm}

The EDM operator $\EDM$ is composed by two parts,
\begin{equation}
  \EDM=\EDM_{\rm PCTC}+\EDM_{\rm PVTV}.
\end{equation}
$\EDM_{\rm PCTC}$ is the electric dipole operator derived from the
current  $\bmJ_{PCTC}$ given in Eq.~(\ref{eq:PCcurrent}), after using
the long wavelength approximation and the continuity equation~\cite{Walecka:1995mi},
explicitly
\begin{equation}
  \EDM_{\rm PCTC}=e\sum_i\frac{1+\tau_z(i)}{2}\boldsymbol{r}_i\,,
  \label{eq:dpc}
\end{equation}
where $e>0$ is the electric unit charge, $\tau_z(i)$ and $\bmr_i$ are the
$z$ component of the isospin and the position of the i-th particle.
This operator implicitly takes into account also the main part of the two-body
PCTC currents. The $\EDM_{\rm PVTV}$ contribution comes from the
PVTV current at LO given in Eq.~(\ref{eq:TVcurrent}) and it reads
\begin{equation}
  \EDM_{\rm PVTV}=\frac{1}{2}\sum_i\left[(d_p+d_n)+(d_p-d_n)\tau_z(i)
    \right]\bmsi_i\,,
  \label{eq:dtrv}
\end{equation}
where $d_p$ and $d_n$ are the EDM of proton and neutron, respectively and
$\bmsi_i$ is the spin operator which act on the i-th particle.
As discussed in Sect.~\ref{sec:pvtvc} and in Refs.~\cite{deVries:2011an,Bsaisou:2012rg}
the $\EDM_{\rm PVTV}$ should also include contributions from transition currents at
N$^2$LO. These are not considered in this review.

The EDM of an $A$ nucleus can be expressed as
\begin{eqnarray}
  d^A & = & \bra \psi^A_{+} | \hat{D}_{\rm PVTV} | \psi^A_{+} \ket
  +2\, \bra \psi^A_{+} | \hat{D}_{\rm PCTC} | \psi^A_{-} \ket
  \nonumber\\
  & \equiv & d_{\rm PVTV}^A+e\; d_{\rm PCTC}^A\,,
\end{eqnarray}
where $|\psi^A_{+}\ket$ $(|\psi^A_{-}\ket)$
is defined to be the even-parity (odd-parity) component of the wave function.
In general, due to the smallness of the LECs, the EDM  depends linearly on the PVTV LECs
\begin{eqnarray}
  d^A_{\rm PVTV}&=&d_pa_p+d_na_n\\
  d^A_{\rm PCTC}&=&\bar g_0 a_0+\bar g_1 a_1+\bar g_2 a_2\nonumber\\
  &+&\bar C_1 A_1+\bar C_2A_2+\bar C_3A_3+\bar C_4A_4+\bar C_5A_5+\bar \Delta a_\Delta\,,
  \label{eq:da}
\end{eqnarray}
where the $a_i$ for $i=0,1,2$, $A_i$ for $i=1,\dots,5$, $a_\Delta$,  and
$a_p$, $a_n$ are coefficients independent on the LEC values
(all coefficients except $a_p$ and $a_n$ have the unit of a length).
For the deuteron, $d^{2}_{\rm PVTV}$ is dominated by one-body
components, proportional to the neutron and proton EDM. 
The coefficients $a_p$ and $a_n$ multiplying the intrinsic neutron and
proton EDM, as already pointed out first in Ref.~\cite{Yamanaka:2015qfa}
and then in Ref.~\cite{Bsaisou:2014zwa}, are given by,
\begin{equation}
  a_n=a_p=\left(1-\frac{3}{2}P_D\right)\,,
\end{equation}
where $P_D$ is the percentage of D-wave present in the deuteron wave function.
$d^2_{\rm PCTC}$, in the case of the deuteron, receives contribution only from the LECs
 $\bar g_1$, $\bar\Delta$, $\bar C_3$ and $\bar C_4$.
The coefficients calculated with the $\chi$EFT N$^2$LO PVTV potential described in
Sect.~\ref{sec:pvtvpot} and the N$^4$LO  PCTC potential
derived in Ref.~\cite{Entem:2017gor} are reported in Table~\ref{tab:2hdpc}.
The cutoff for both the PCTC and PVTV potentials has been chosen to be
$\Lambda_C=500$ MeV. 
The coefficients $a_1$, $A_3$ and $A_4$ agree well with the power
counting expectation in Eq. \eqref{dANDA}. The slight suppression of
$a_1$ compared with the naive estimate $a_1 \sim 1$ is in very good
agreement with the perturbative pion power counting
\cite{deVries:2011re}. The LO perturbative pion calculation of $a_1$
agrees with the value in Table \ref{tab:2hdpc} 
at the 20\% level~\cite{deVries:2011re}. Results obtained in chiral
EFT with N$^2$LO PCTC potentials \cite{Bsaisou:2014zwa}, and with
``hybrid'' approaches \cite{deVries:2011an,Yamanaka:2015qfa} based on 
chiral PVTV and phenomenological PCTC potentials, also agree well
with the results reported in Table \ref{tab:2hdpc}.
The contribution of the three-pion coupling $a_{\Delta}$ is a bit more
problematic. We find in this case that the contribution of the N$^2$LO
term is of the order of  $\sim60\%$ of the NLO term. We will discuss
the issue of these large N$^2$LO corrections more in detail below.

\begin{table}
\begin{center}
\begin{tabular}{lc}
    \hline
    $a_n(a_p)$ &  $\m0.939$ \\
    $a_1$ [fm]     &  $\m0.200$  \\
    $A_3$ [fm]     &  $\m0.013$  \\
    $A_4$ [fm]     &  $ -0.013$  \\
    $a_\Delta({\rm NLO})$ [fm] &  $ -0.894$  \\
    $a_\Delta({\rm N^2LO})$ [fm] &  $ +0.590$  \\
    $a_\Delta({\rm TOT})$ [fm] &  $ -0.304$  \\
    \hline
\end{tabular}
\caption{ 
  Values of the coefficients entering the expression
  of the deuteron EDM calculated 
  for the $\chi$EFT N$^2$LO PVTV potential described in
  Sect.~\ref{sec:pvtvpot} and the N$^4$LO  PCTC potential
  derived in Ref.~\cite{Entem:2017gor}.
  The PVTC potential has been regularized as in
  Eq.~(\ref{localcutoff}) adopting the value $\Lambda_C=500$ MeV for
  the cutoff parameter. The PCTC potential has been regularized with
  the same value of the cutoff parameter.
  For $a_\Delta$ we give explicitly the contribution of the different orders,
  the sum of the two contributions is given in the last row.
  }\label{tab:2hdpc}
\end{center}
\end{table}

Depending on the source of CP violation at the quark level, the
    deuteron EDM can be dominated by different LECs. For sources such
    as quark chromo-EDMs and four-quark operators $\Xi$, for which
    $\bar g_1$ is induced without any chiral suppression, the
    pion-exchange contribution proportional to $\bar g_1$ is expected
    to dominate the deuteron EDM. For sources such as quark EDMs or
    the Weinberg operator, however, the deuteron EDM is well
    approximated by the sum of the nucleon EDMs. For the $\theta$-term,
    the pion-exchange contributions are expected to be minor as
    well. Given measurements of the deuteron and nucleon EDMs, one can,
    therefore, identify the underlying source of CP violation
    \cite{Lebedev:2004va,deVries:2011re}.

As regarding the ${}^3$H and ${}^3$He EDMs, the results are summarized in
Table~\ref{tab:3htot1}. The coefficients $a_0$ and $a_1$ are again a
bit smaller than the $\mathcal O(1)$ expectation.  Note that
the value for $a_0$ reported in Table~\ref{tab:2hdpc} is approximately
50\% smaller than that reported in Ref.~\cite{Bsaisou:2014zwa}. This
difference can be traced back to the contribution of the TPE, which
was not included in that work. Performing the
calculations at LO, namely including only the OPE term, the $a_0$
coefficient results to agree with that reported in Ref.~\cite{Bsaisou:2014zwa}.
The values of the numerical
coefficients are mostly equal in modulus between 
$\tri$ and $\hel$ except $a_p$ and $a_n$. The coefficients associated
to isovector terms have the same sign while all the others are opposite. 
Again the contribution of the N$^2$LO potential term to $a_\Delta$ is
significant, about $60\%$. This issue is discussed below.

\begin{table}[t]
\begin{center}
\begin{tabular}{lcc}
  \hline
   &  ${}^3$H  & ${}^3$He \\
  \hline
$a_n$      &$ -0.033$ & $\m0.908$ \\
$a_p$      &$\m0.909$ & $ -0.033$ \\
$a_0$ [fm]      &$ -0.053$ & $\m0.054$ \\
$a_1$ [fm]     &$\m0.158$ & $\m0.158$ \\
$a_2$ [fm]     &$ -0.119$ & $\m0.119$ \\
$A_1$ [fm]     &$\m0.006$ & $ -0.006$ \\
$A_2$ [fm]     &$ -0.010$ & $\m0.010$ \\
$A_3$ [fm]     &$ -0.008$ & $ -0.008$ \\
$A_4$ [fm]     &$\m0.013$ & $\m0.013$ \\
$A_5$ [fm]     &$ -0.022$ & $\m0.022$ \\
$a_\Delta({\rm NLO})$   [fm] &$ -0.941$ & $ -0.929$ \\
$a_\Delta({\rm N^2LO})$ [fm] &$ +0.598$ & $ +0.591$ \\
$a_\Delta({\rm TOT})$   [fm] &$ -0.343$ & $ -0.339$ \\
\hline
\end{tabular}
\caption{
  The same as in Table~\ref{tab:2hdpc} but for the $\tri$ and $\hel$ EDM.
  } \label{tab:3htot1}
\end{center}
\end{table}

Let us now consider in more detail the issue of the NLO and N$^2$LO contributions to
$a_\Delta$. We have seen that in all cases the N$^2$LO correction to
$a_\Delta$ is of the order of 60\%, a bit larger than
expected. Explicitly, the coefficient $a_\Delta$ can 
be written as~\cite{Gnech:2019dod} 
\begin{eqnarray}
  a_\Delta&=&a_\Delta({\rm NLO})+a_\Delta({\rm N^2LO})\,,\label{eq:adelta}\\
  a_\Delta({\rm NLO})&=& a_\Delta(0)+ a_\Delta({\rm 3N})\,, \label{eq:adelta2}\\
  a_\Delta({\rm N^2LO}) &=& c_1a_\Delta(1)+c_2a_\Delta(2)+c_3a_\Delta(3)
  \,,\label{eq:adelta3}
\end{eqnarray}
where $a_\Delta(0)$ comes from the NLO potential
$V_{PVTV}^{(0)}(3\pi)$ given in Eq.~(\ref{eq:3pnlo}) and 
$a_\Delta(3N)$ from the 3N potential given in Eq.~(\ref{eq:NNNpot}). 
The N${}^2$LO terms  come from
$V_{PVTV}^{(1)}(3\pi)$,  where the LECs $c_1$, $c_2$ and $c_3$ appear.
To calculate the values reported in
Tables~\ref{tab:2hdpc} and~\ref{tab:3htot1}, the following values 
were adopted: $c_1=-1.10$ GeV$^{-1}$, $c_2=+3.57$ GeV$^{-1}$,
and $c_3=-5.54$ GeV$^{-1}$ as reported in
Refs.~\cite{Hoferichter:2015tha,Hoferichter:2015hva}.
The large N$^2$LO corrections are caused by the 
large values of these LECs.\footnote{Notice that the values of the
  LECs $c_i$ obtained from the pion-nucleon amplitude at NLO, which
  would be appropriate for $V_{PVTV}^{(1)}$, are considerably smaller
  in magnitude. We, however, decided to adopt the larger values
  to be consistent with the employed PCTC potential.} For more detail, see
Ref.~\cite{Gnech:2019dod}. For the trinucleon systems, the values
of $a_\Delta({\rm 3N})$ give a correction to $a_\Delta({\rm NLO})$
of the order of $\sim25\%$, which is in line with the chiral
perturbation theory prediction because these contributions appear at
the same order. 

\begin{table}[t]
\begin{center}
  \begin{tabular}{lccc}
    \hline
     &  ${}^2$H &  $\tri$ & $\hel$ \\    
    \hline
    $a_\Delta(0) $ [fm] &  $ -0.894$ & $ -0.751$ & $ -0.749$ \\
    $a_\Delta({\rm 3N})$ [fm] &  $-$       & $ -0.190$ & $ -0.180$ \\
    \hline
    $a_\Delta(1) $ [fm GeV] &  $\m0.120$ & $\m0.098$ & $\m0.098$ \\
    $a_\Delta(2) $ [fm GeV] &  $ -0.119$ & $ -0.110$ & $ -0.109$ \\
    $a_\Delta(3) $ [fm GeV] &  $ -0.207$ & $ -0.198$ & $ -0.196$ \\
    \hline
\end{tabular}
\caption{ \label{tab:2had}
  Values of the various components coefficients $a_\Delta$
  as given in Eq.~(\ref{eq:adelta})
  in units of $e$ fm for the different nuclei. The coefficients have
  been evaluated using the N$^4$LO PC potential
  derived in Ref.~\cite{Entem:2017gor}
  and using a cutoff parameter of value $\Lambda_C=500$ MeV.
  } 
\end{center}
\end{table}

Similarly to the deuteron EDM, the trinucleon EDMs can be
    dominated by different terms. As the isoscalar interaction
    proportional to $\bar g_0$ and $\bar C_{1,2}$ now gives a sizable
    contribution, the trinucleon EDMs are noticeably different from
    the nucleon EDMs for the QCD $\theta$-term, the quark chromo-EDMs,
    the four-quark operators $\Xi$, and potentially the Weinberg
    operator and the four-quark operators $\Sigma$. These EDMs
    therefore provide complementary information to the deuteron and
    nucleon EDMs. Combined measurements of all these EDMs would allow
    one to unravel various BSM models of new CP violation
    \cite{Dekens:2014jka}.

\section{Conclusions and perspectives}
\label{sec:conc}

In this paper we have discussed the current status of the PVTC and PVTV
nuclear interactions using the traditional approach based on
phenomenological boson exchange models and utilizing the modern
frameworks of pionless and chiral EFT. 
The study of PVTC signals in nuclei is interesting since
it derives from the nonleptonic weak interactions between quarks.
Furthermore, a solid understanding of the manifestation of PVTC
interactions at the nuclear level would give us confidence in the
analysis of the more exotic PVTV case and other BSM nuclear
observables. In fact, PVTV observables provide very
valuable information since they are sensitive to interactions
originating from the $\theta$-term in the SM and even to more exotic
mechanisms appearing in BSM theories. 

As discussed in this review, the theoretical understanding of the PVTC and PVTV
interactions is already rather advanced. Interactions in $\chi$EFT have
been developed up to N$^2$LO. The convergence of the $\chi$EFT appears
to be problematic only for the contributions proportional to the
$\pi\pi NN$ LECs $c_i$, due to the large values of those coefficients
as it results from $\pi N$ scattering data~\cite{Hoferichter:2015tha}.
Given that the LECs $c_{2,3,4}$ are largely driven by the
$\Delta$(1232) \cite{Siemens:2016jwj}, one may expect a better
convergence in the formulation of chiral EFT that includes the
$\Delta$ as an explicit degree of freedom. Furthermore, large $N_c$
analysis may help in reducing the number of contact LECs. Also Lattice
QCD calculations start to give valuable
information~\cite{Wasem:2011zz,Guo:2018aiq}. 

We have also reported the results of the theoretical calculations of several
observables performed using the potentials derived within the $\chi$EFT framework.
The PVTC observables considered include  {\it i)} the longitudinal asymmetry in $\vec{n}$-$p$
radiative capture, {\it ii)} the longitudinal asymmetry in proton-proton elastic scattering,
and {\it iii)} the spin rotation of a neutron beam passing through a hydrogen and
deuterium gas. As an example of PVTV observable, we have studied the EDMs of
some light nuclei. The main motivation to study these observables is that for
such light systems, the theoretical analysis can be carried out
without invoking 
any uncontrolled approximations. Thus, the comparison with the experimental data
can be performed unambiguously. 
The analyses of PVTC and PVTV observables using meson
    exchange models can be found in other
    review articles~\cite{Haeberli:1995uz,RamseyMusolf:2006dz,Gardner:2017xyl}
    and are not reported here.

As discussed previously, there exists a first measurement of the parameter $a_\gamma$ of the
radiative neutron capture on the proton $\vec n p\rightarrow d \gamma$.
The large error derives from the smallness of this parameter which makes
this measurement very awkward~\cite{Blyth:2018aon}.
This observable is directly connected to the LO pion-nucleon PVTC coupling
constant $a_\gamma \sim h^1_\pi$. However, as we have seen, the theoretical estimate
of the proportionality coefficient has been obtained with a relatively large theoretical
uncertainty due to  sizeable cancellations between different contributions.
Therefore, to infer information from this observable, it will be
necessary to make progress in both the experimental and theoretical analyses.

Other important information is brought forth by the three measurements at
different energies of the  $\vec p$-$p$ longitudinal asymmetry. This observable is
sensitive to $h^1_\pi$ via the TPE component of the PVTC potential and 
also to other LECs. In fact, owing to the isospin quantum numbers
$T=1$, $T_z=1$ of the 
$p$-$p$ system, the LO contribution vanishes. Moreover, at
NLO (N${}^2$LO), this observable 
depends on two (three) combinations of the LECs. Unfortunately,
only two of the performed measurements give independent
information. These two data have not been obtained with enough accuracy,
so the constraints to the (combinations of) LECs which can be obtained are not so
stringent~\cite{deVries:2014vqa,Viviani:2014zha}, as discussed in Subsect.~\ref{sec:respp}.
For this observable the wave functions are easily obtained, however the
vanishing of the LO contribution makes the $\chi$PT convergence more uncertain. 
On the other hand, it would be very useful to have more accurate
experimental measurements.

Regarding the spin rotation observables, no experiments
to
measure the $\vec n$-$p$ and $\vec n$-$d$ spin rotation angles, which could
provide useful information on some of the contact term LECs, are
planned at present.
The experimental detection of a non-vanishing $\vec n$-$p$ spin rotation would be
rather important for two reasons: {\it i)} the theoretical treatment of the
two-nucleon system does not present any difficulty numerically, while {\it ii)} this
observable is sensitive to the LO term and therefore the chiral expansion
of the potential is well under control, as discussed in
Subect.~\ref{sec:resnp}. Regarding the $\vec n$-$d$ spin 
rotation, the same is not completely true since, as discussed in
Sect.~\ref{sec:pvtcpot} one has to include also the PVTC 3N
interaction terms which start to appear at N${}^2$LO. This is an
interesting extension of $\chi$EFT which will be considered in  
future. From the experimental point of view, we note that
there is an experiment in progress trying to 
measure the $\vec n$-${}^4$He spin rotation at NIST~\cite{Bass:2009fs}.
Some years ago there was a measure of the longitudinal asymmetry in
$\vec p$-${}^4$He scattering, 
but this experiment has been performed at a rather high energy of the
proton beam (46 MeV)~\cite{Lang:1985jv}
and this makes the theoretical treatment very difficult and
impossible without some approximations.

Also a measurement of the $\vec n$-${}^3$He longitudinal asymmetry at
the SNS facility is in progress,
and the experimental value should be published soon. For this $A=4$ system
it is possible to perform accurate calculations of the wave functions~\cite{Viviani:2014zha}, and therefore
this observable may give valuable information in the near future.

To have the possibility to pin down all the LECs (6 LECs at NLO and 5 more
at N${}^2$LO) more experimental information will be necessary in any case. In particular,
an interesting possibility would be to measure PVTC observables
in the $A=3$ system, such as the longitudinal asymmetry of $\vec p$-$ d$ elastic scattering
and the photon asymmetry in $\vec n$-$ d$ radiative capture. For both reactions,
the theoretical treatment would be straightforward, once
the PVTC 3N force has been taken into account.

Regarding the PVTV observables, the measurement of EDMs of particles is the most
promising observable for studying CP violation beyond CKM mixing matrix effects. 
Currently,  there are proposals for the direct measurement of EDMs of
electrons, single nucleons and light nuclei in dedicated storage rings
~\cite{Orlov:2006su,Semertzidis:2011qv,Rathmann:2013rqa,Guidoboni:2016bdn,Abusaif:2019gry}. This new
approach plans to reach an accuracy of $\sim 10^{-16}\ e$ fm,  improving
the sensitivity in particular in the hadronic sector. Any measurement of a
non-vanishing EDM of this magnitude would provide evidence of
PVTV beyond CKM effects~\cite{Pospelov:2005pr,Czarnecki:1997bu,Mannel:2012qk,Mannel:2012hb}.
However, a single measurement will be insufficient to
identify the source of PVTV, only the availability of the measurement of EDM of various
light nuclei such as ${}^2$H, ${}^3$H, and ${}^3$He can impose
constrains on all the LECs. {Other light nuclear EDMs have been
  discussed in Refs.~\cite{Yamanaka:2016umw,Yamanaka:2019vec}.  
EDMs of heavy diamagnetic systems provide very important 
information as well, but the systems are too large for chiral EFT
calculations.

Other observables sensitive to PVTV effects are the transmission of polarized neutrons
through a polarized target~\cite{Kabir:1981tp,Stodolsky:1981vn}, in particular
for heavy nuclei the PVTV effects can be enhanced by factors as large as
$10^6$~\cite{Bunakov:1982is,Gudkov:1991qg}, see also Ref.~\cite{Bowman:2014fca}.
In order to exploit this enhancement, some experiments are being planned,
such as the NOPTREX experiment at RIKEN~\cite{Shimizu:2017zan,Gudkov:2017vqn}. 
Also polarized nucleon -- polarized deuteron scattering has been proposed
as a way to detect PVTV signals~\cite{Song:2010sz,Uzikov:2016lsc}.
Finally, searching for a large PVTV violations in polarized $\beta$-decay of
${}^8$Li via measurements of the triple vector correlation is under
consideration~\cite{Murata:2017ifq}. Clearly, it would be important
to be able to detect a nonzero PVTV signal in all these experiments to be
able to pin down the values of all the LECs.

From the theoretical point of view, calculations of the EDM of ${}^2$H,
${}^3$H and ${}^3$He can be performed very accurately,
also taking into account the contributions of the PVTV 3N force.
The robustness of the calculation has been checked by evaluating
the EDMs of the nuclei using different chiral orders in the PCTC potential.
The discrepancy between the use of the N$^2$LO and the N$^4$LO PCTC potential
has been found to be approximately $5\%$~\cite{Gnech:2019dod}.

Currently, the only missing ingredient is
the two-body PVTV N$^2$LO currents~\cite{deVries:2011an,Bsaisou:2012rg}.
Once this problem will be solved, one will achieve a fully consistent calculation
of the EDM of light nuclei up to N$^2$LO. There are also plans to 
perform theoretical studies of PVTV observables
in  $\vec{n}$-$\vec{p}$ and $\vec{n}$-$\vec{d}$ scattering
in order to have independent and complementary information of PVTV
effects.

The PVTV $\chi$EFT interaction developed in the previous sections depends
on 11 coupling constants that should be determined by comparing with
experimental data. As already pointed out by  many authors~\cite{Mereghetti:2010tp,deVries:2012ab,Bsaisou:2014oka}
and discussed in Subsect.~\ref{sec:chitvsources}, 
the LECs $\bar g_2$, $\bar C_3$, $\bar C_4$ and $\bar C_5$
are suppressed for all CP-violation sources. However, for certain sources,
the suppression is not too severe. For example, in Ref.~\cite{Bsaisou:2014zwa},
an analysis of the nuclear EDM in the minimal left-right scenario is
presented in which the Lagrangian terms with LECs
$\bar C_3$ and $\bar C_4$ appear at N$^2$LO. In any case, since the CP-violation
sources are not known, the only way to determine them is to fit all 
possible LECs and then compare the results with predictions
for various scenarios. 

Most of the observables discussed so far were obtained (or they are planned to be
studied) at low energies, where also the pionless EFT framework is valid. 
The advantage of this framework is related to the fact
that the resulting potentials depends on five LECs. Then, assuming the
validity of
the large $N_c$ analysis~\cite{Schindler:2015nga,Vanasse:2018buq}, the number
of dominant LECs could be further reduced. This new paradigm is advocated
for the PVTC case in Ref.~\cite{Gardner:2017xyl}.  For this case, only two LECs
are expected to be dominant, the other three demoted to be subleading.
Unfortunately, the photon asymmetry of $\vec n p\rightarrow d \gamma$  depends on 
the subleading LECs (this could explain its relative smallness) and therefore cannot be used
to give information on the two leading LECs. Moreover, only one of the
measured $\vec p$-$ p$ longitudinal asymmetry at low energy may be used to test if
this hierarchy is realized in Nature (the other measurement is taken at too high energy
to be used in the pionless EFT framework).  The other observable which could give valuable
information could be $\vec n$-${}^3$He longitudinal  asymmetry, for
which experimental data should appear soon. However, no
theoretical calculations of this observable performed in the framework
of pionless EFT are available at present.
Additional information could be obtained by calculations
of these LECs using Lattice QCD, presently in progress.
Regarding the PVTV observables in pionless EFT, here the large $N_c$ analysis
predicts that only one of the LECs should be dominant, while the other four being
suppressed. However, this picture is partially obscured by the fact that
the magnitude of the five contact LECs would depend very much on the
particular type of the CP-violating source. 

In conclusions, the study of PVTC and PVTV observables is an active area of research that provide 
important tests of the SM and hopefully future evidence for BSM physics.
%they could in the future give important tests of the SM and information of BSM theories.

\section*{Conflict of Interest Statement}
%All financial, commercial or other relationships that might be
%perceived by the academic community as representing a potential
%conflict of interest must be disclosed. If no such relationship
%exists, authors will be asked to confirm the following statement:  

The Authors declare that the research was conducted in the absence of
any commercial or financial relationships that could be construed as a
potential conflict of interest. 

\section*{Author Contributions}
All Authors contributed to this article at about the same level.

\section*{Funding}
JdV was supported by the RHIC Physics Fellow Program of
the RIKEN BNL Research Center.

\noindent EE was supported by BMBF (Grant No. 05P18PCFP1) and by DFG
through funds provided to the Sino-German CRC 110 ``Symmetries and
the Emergence of Structure in QCD'' (Grant No. TRR110).

\noindent EM was supported in part by the LDRD program at
Los Alamos National Laboratory, the DOE topical collaboration on
``Nuclear Theory for Double-Beta Decay and Fundamental Symmetries'',
the US DOE, Office of Science, Office of Nuclear Physics, under award
numbers DE-AC52-06NA25396.

\noindent LG, AG, and MV were supported by INFN through the
National Initiative ``FBS''.

%\section*{Acknowledgments}

\bibliography{TVreview}{}

\begin{thebibliography}{239}
\providecommand{\natexlab}[1]{#1}
\expandafter\ifx\csname urlstyle\endcsname\relax
  \providecommand{\doi}[1]{doi:\discretionary{}{}{}#1}\else
  \providecommand{\doi}{doi:\discretionary{}{}{}\begingroup
  \urlstyle{rm}\Url}\fi
\providecommand{\selectlanguage}[1]{\relax}
\providecommand{\bibAnnoteFile}[1]{%
  \IfFileExists{#1}{\begin{quotation}\noindent\textsc{Key:} #1\\
  \textsc{Annotation:}\ \input{#1}\end{quotation}}{}}
\providecommand{\bibAnnote}[2]{%
  \begin{quotation}\noindent\textsc{Key:} #1\\
  \textsc{Annotation:}\ #2\end{quotation}}

\bibitem[{Machleidt(2017)}]{Machleidt:2017vls}
Machleidt, R. (2017).
\newblock {Historical perspective and future prospects for nuclear
  interactions}.
\newblock \emph{Int. J. Mod. Phys.} E26, 1730005.
\newblock \doi{10.1142/S0218301317300053}
\bibAnnoteFile{Machleidt:2017vls}

\bibitem[{Weinberg(1990)}]{Weinberg:1990rz}
Weinberg, S. (1990).
\newblock {Nuclear forces from chiral Lagrangians}.
\newblock \emph{Phys. Lett.} B251, 288--292.
\newblock \doi{10.1016/0370-2693(90)90938-3}
\bibAnnoteFile{Weinberg:1990rz}

\bibitem[{Ordonez et~al.(1996)Ordonez, Ray, and van Kolck}]{Ordonez:1995rz}
Ordonez, C., Ray, L., and van Kolck, U. (1996).
\newblock {The Two nucleon potential from chiral Lagrangians}.
\newblock \emph{Phys. Rev.} C53, 2086--2105.
\newblock \doi{10.1103/PhysRevC.53.2086}
\bibAnnoteFile{Ordonez:1995rz}

\bibitem[{Epelbaum et~al.(2009)Epelbaum, Hammer, and
  Mei{\ss}ner}]{Epelbaum:2008ga}
Epelbaum, E., Hammer, H.-W., and Mei{\ss}ner, U.-G. (2009).
\newblock {Modern Theory of Nuclear Forces}.
\newblock \emph{Rev. Mod. Phys.} 81, 1773--1825.
\newblock \doi{10.1103/RevModPhys.81.1773}
\bibAnnoteFile{Epelbaum:2008ga}

\bibitem[{Machleidt and Entem(2011)}]{Machleidt:2011zz}
Machleidt, R. and Entem, D.~R. (2011).
\newblock {Chiral effective field theory and nuclear forces}.
\newblock \emph{Phys. Rept.} 503, 1--75.
\newblock \doi{10.1016/j.physrep.2011.02.001}
\bibAnnoteFile{Machleidt:2011zz}

\bibitem[{Weinberg(1966)}]{Weinberg:1966kf}
Weinberg, S. (1966).
\newblock {Pion scattering lengths}.
\newblock \emph{Phys. Rev. Lett.} 17, 616--621.
\newblock \doi{10.1103/PhysRevLett.17.616}
\bibAnnoteFile{Weinberg:1966kf}

\bibitem[{Weinberg(1968)}]{Weinberg:1968de}
Weinberg, S. (1968).
\newblock {Nonlinear realizations of chiral symmetry}.
\newblock \emph{Phys. Rev.} 166, 1568--1577.
\newblock \doi{10.1103/PhysRev.166.1568}
\bibAnnoteFile{Weinberg:1968de}

\bibitem[{Weinberg(1979)}]{Weinberg:1978kz}
Weinberg, S. (1979).
\newblock {Phenomenological Lagrangians}.
\newblock \emph{Physica} A96, 327--340.
\newblock \doi{10.1016/0378-4371(79)90223-1}
\bibAnnoteFile{Weinberg:1978kz}

\bibitem[{Coleman et~al.(1969)Coleman, Wess, and Zumino}]{Coleman:1969sm}
Coleman, S.~R., Wess, J., and Zumino, B. (1969).
\newblock {Structure of phenomenological Lagrangians. 1.}
\newblock \emph{Phys. Rev.} 177, 2239--2247.
\newblock \doi{10.1103/PhysRev.177.2239}
\bibAnnoteFile{Coleman:1969sm}

\bibitem[{Callan et~al.(1969)Callan, Coleman, Wess, and Zumino}]{Callan:1969sn}
Callan, C.~G., Jr., Coleman, S.~R., Wess, J., and Zumino, B. (1969).
\newblock {Structure of phenomenological Lagrangians. 2.}
\newblock \emph{Phys. Rev.} 177, 2247--2250.
\newblock \doi{10.1103/PhysRev.177.2247}
\bibAnnoteFile{Callan:1969sn}

\bibitem[{Gasser and Leutwyler(1984)}]{Gasser:1983yg}
Gasser, J. and Leutwyler, H. (1984).
\newblock {Chiral Perturbation Theory to One Loop}.
\newblock \emph{Annals Phys.} 158, 142.
\newblock \doi{10.1016/0003-4916(84)90242-2}
\bibAnnoteFile{Gasser:1983yg}

\bibitem[{Bernard et~al.(1995)Bernard, Kaiser, and
  Mei{\ss}ner}]{Bernard:1995dp}
Bernard, V., Kaiser, N., and Mei{\ss}ner, U.-G. (1995).
\newblock {Chiral dynamics in nucleons and nuclei}.
\newblock \emph{Int. J. Mod. Phys.} E4, 193--346.
\newblock \doi{10.1142/S0218301395000092}
\bibAnnoteFile{Bernard:1995dp}

\bibitem[{Bernard(2008)}]{Bernard:2007zu}
Bernard, V. (2008).
\newblock {Chiral Perturbation Theory and Baryon Properties}.
\newblock \emph{Prog. Part. Nucl. Phys.} 60, 82--160.
\newblock \doi{10.1016/j.ppnp.2007.07.001}
\bibAnnoteFile{Bernard:2007zu}

\bibitem[{Bijnens and Ecker(2014)}]{Bijnens:2014lea}
Bijnens, J. and Ecker, G. (2014).
\newblock {Mesonic low-energy constants}.
\newblock \emph{Ann. Rev. Nucl. Part. Sci.} 64, 149--174.
\newblock \doi{10.1146/annurev-nucl-102313-025528}
\bibAnnoteFile{Bijnens:2014lea}

\bibitem[{Epelbaum et~al.(2017)Epelbaum, Gegelia, and
  Mei{\ss}ner}]{Epelbaum:2017byx}
Epelbaum, E., Gegelia, J., and Mei{\ss}ner, U.-G. (2017).
\newblock {Wilsonian renormalization group versus subtractive renormalization
  in effective field theories for nucleon–nucleon scattering}.
\newblock \emph{Nucl. Phys.} B925, 161--185.
\newblock \doi{10.1016/j.nuclphysb.2017.10.008}
\bibAnnoteFile{Epelbaum:2017byx}

\bibitem[{Entem et~al.(2015)Entem, Kaiser, Machleidt, and
  Nosyk}]{Entem:2014msa}
Entem, D.~R., Kaiser, N., Machleidt, R., and Nosyk, Y. (2015).
\newblock {Peripheral nucleon-nucleon scattering at fifth order of chiral
  perturbation theory}.
\newblock \emph{Phys. Rev.} C91, 014002.
\newblock \doi{10.1103/PhysRevC.91.014002}
\bibAnnoteFile{Entem:2014msa}

\bibitem[{Epelbaum et~al.(2015{\natexlab{a}})Epelbaum, Krebs, and
  Mei{\ss}ner}]{Epelbaum:2014sza}
Epelbaum, E., Krebs, H., and Mei{\ss}ner, U.-G. (2015{\natexlab{a}}).
\newblock {Precision nucleon-nucleon potential at fifth order in the chiral
  expansion}.
\newblock \emph{Phys. Rev. Lett.} 115, 122301.
\newblock \doi{10.1103/PhysRevLett.115.122301}
\bibAnnoteFile{Epelbaum:2014sza}

\bibitem[{Entem et~al.(2017)Entem, Machleidt, and Nosyk}]{Entem:2017gor}
Entem, D.~R., Machleidt, R., and Nosyk, Y. (2017).
\newblock {High-quality two-nucleon potentials up to fifth order of the chiral
  expansion}.
\newblock \emph{Phys. Rev.} C96, 024004.
\newblock \doi{10.1103/PhysRevC.96.024004}
\bibAnnoteFile{Entem:2017gor}

\bibitem[{Reinert et~al.(2018)Reinert, Krebs, and Epelbaum}]{Reinert:2017usi}
Reinert, P., Krebs, H., and Epelbaum, E. (2018).
\newblock {Semilocal momentum-space regularized chiral two-nucleon potentials
  up to fifth order}.
\newblock \emph{Eur. Phys. J.} A54, 86.
\newblock \doi{10.1140/epja/i2018-12516-4}
\bibAnnoteFile{Reinert:2017usi}

\bibitem[{Kaplan et~al.(1998)Kaplan, Savage, and Wise}]{Kaplan:1998tg}
Kaplan, D.~B., Savage, M.~J., and Wise, M.~B. (1998).
\newblock {A New expansion for nucleon-nucleon interactions}.
\newblock \emph{Phys. Lett.} B424, 390--396.
\newblock \doi{10.1016/S0370-2693(98)00210-X}
\bibAnnoteFile{Kaplan:1998tg}

\bibitem[{Nogga et~al.(2005)Nogga, Timmermans, and van Kolck}]{Nogga:2005hy}
Nogga, A., Timmermans, R. G.~E., and van Kolck, U. (2005).
\newblock {Renormalization of one-pion exchange and power counting}.
\newblock \emph{Phys. Rev.} C72, 054006.
\newblock \doi{10.1103/PhysRevC.72.054006}
\bibAnnoteFile{Nogga:2005hy}

\bibitem[{Birse(2006)}]{Birse:2005um}
Birse, M.~C. (2006).
\newblock {Power counting with one-pion exchange}.
\newblock \emph{Phys. Rev.} C74, 014003.
\newblock \doi{10.1103/PhysRevC.74.014003}
\bibAnnoteFile{Birse:2005um}

\bibitem[{Valderrama(2011)}]{Valderrama:2009ei}
Valderrama, M.~P. (2011).
\newblock {Perturbative renormalizability of chiral two pion exchange in
  nucleon-nucleon scattering}.
\newblock \emph{Phys. Rev.} C83, 024003.
\newblock \doi{10.1103/PhysRevC.83.024003}
\bibAnnoteFile{Valderrama:2009ei}

\bibitem[{Long and Yang(2012)}]{Long:2012ve}
Long, B. and Yang, C.~J. (2012).
\newblock {Short-range nuclear forces in singlet channels}.
\newblock \emph{Phys. Rev.} C86, 024001.
\newblock \doi{10.1103/PhysRevC.86.024001}
\bibAnnoteFile{Long:2012ve}

\bibitem[{Epelbaum and Gegelia(2012)}]{Epelbaum:2012ua}
Epelbaum, E. and Gegelia, J. (2012).
\newblock {Weinberg$^\prime$s approach to nucleon–nucleon scattering
  revisited}.
\newblock \emph{Phys. Lett.} B716, 338--344.
\newblock \doi{10.1016/j.physletb.2012.08.025}
\bibAnnoteFile{Epelbaum:2012ua}

\bibitem[{Lepage(1997)}]{Lepage:1997cs}
Lepage, G.~P. (1997).
\newblock {How to renormalize the Schrodinger equation}.
\newblock In \emph{{Nuclear physics. Proceedings, 8th Jorge Andre Swieca Summer
  School, Sao Jose dos Campos, Campos do Jordao, Brazil, January 26-February 7,
  1997}}. 135--180
\bibAnnoteFile{Lepage:1997cs}

\bibitem[{Epelbaum and Mei{\ss}ner(2013)}]{Epelbaum:2006pt}
Epelbaum, E. and Mei{\ss}ner, U.-G. (2013).
\newblock {On the Renormalization of the One-Pion Exchange Potential and the
  Consistency of Weinberg`s Power Counting}.
\newblock \emph{Few Body Syst.} 54, 2175--2190.
\newblock \doi{10.1007/s00601-012-0492-1}
\bibAnnoteFile{Epelbaum:2006pt}

\bibitem[{Epelbaum and Gegelia(2009)}]{Epelbaum:2009sd}
Epelbaum, E. and Gegelia, J. (2009).
\newblock {Regularization, renormalization and 'peratization' in effective
  field theory for two nucleons}.
\newblock \emph{Eur. Phys. J.} A41, 341--354.
\newblock \doi{10.1140/epja/i2009-10833-3}
\bibAnnoteFile{Epelbaum:2009sd}

\bibitem[{Valderrama(2016)}]{Valderrama:2016koj}
Valderrama, M.~P. (2016).
\newblock {Power Counting and Wilsonian Renormalization in Nuclear Effective
  Field Theory}.
\newblock \emph{Int. J. Mod. Phys.} E25, 1641007.
\newblock \doi{10.1142/S021830131641007X}
\bibAnnoteFile{Valderrama:2016koj}

\bibitem[{Epelbaum et~al.(2018)Epelbaum, Gasparyan, Gegelia, and
  Mei{\ss}ner}]{Epelbaum:2018zli}
Epelbaum, E., Gasparyan, A.~M., Gegelia, J., and Mei{\ss}ner, U.-G. (2018).
\newblock {How (not) to renormalize integral equations with singular potentials
  in effective field theory}.
\newblock \emph{Eur. Phys. J.} A54, 186.
\newblock \doi{10.1140/epja/i2018-12632-1}
\bibAnnoteFile{Epelbaum:2018zli}

\bibitem[{Hammer et~al.(2019)Hammer, K{\"o}nig, and van Kolck}]{Hammer:2019poc}
Hammer, H.~W., K{\"o}nig, S., and van Kolck, U. (2019).
\newblock {Nuclear effective field theory: status and perspectives,
  ArXiv:1906.12122}
\bibAnnoteFile{Hammer:2019poc}

\bibitem[{Bedaque and van Kolck(2002)}]{Bedaque:2002mn}
Bedaque, P.~F. and van Kolck, U. (2002).
\newblock {Effective field theory for few nucleon systems}.
\newblock \emph{Ann. Rev. Nucl. Part. Sci.} 52, 339--396.
\newblock \doi{10.1146/annurev.nucl.52.050102.090637}
\bibAnnoteFile{Bedaque:2002mn}

\bibitem[{Hammer and Platter(2010)}]{Hammer:2010kp}
Hammer, H.-W. and Platter, L. (2010).
\newblock {Efimov States in Nuclear and Particle Physics}.
\newblock \emph{Ann. Rev. Nucl. Part. Sci.} 60, 207--236.
\newblock \doi{10.1146/annurev.nucl.012809.104439}
\bibAnnoteFile{Hammer:2010kp}

\bibitem[{Ramsey-Musolf and Page(2006)}]{RamseyMusolf:2006dz}
Ramsey-Musolf, M.~J. and Page, S.~A. (2006).
\newblock {Hadronic parity violation: A New view through the looking glass}.
\newblock \emph{Ann. Rev. Nucl. Part. Sci.} 56, 1--52.
\newblock \doi{10.1146/annurev.nucl.54.070103.181255}
\bibAnnoteFile{RamseyMusolf:2006dz}

\bibitem[{Hertzog and Ramsey-Musolf(2013)}]{Hertzog:2012zd}
Hertzog, D. and Ramsey-Musolf, M.~J. (2013).
\newblock {Parity- and Time-Reversal Tests in Nuclear Physics}.
\newblock In \emph{100 Years of Subatomic Physics}, eds. E.~M. Henley and S.~D.
  Ellis. 155--170.
\newblock \doi{10.1142/9789814425810_0006}
\bibAnnoteFile{Hertzog:2012zd}

\bibitem[{Schindler and Springer(2013)}]{Schindler:2013yua}
Schindler, M.~R. and Springer, R.~P. (2013).
\newblock {The Theory of Parity Violation in Few-Nucleon Systems}.
\newblock \emph{Prog. Part. Nucl. Phys.} 72, 1--43.
\newblock \doi{10.1016/j.ppnp.2013.05.002}
\bibAnnoteFile{Schindler:2013yua}

\bibitem[{Haxton and Holstein(2013)}]{Haxton:2013aca}
Haxton, W.~C. and Holstein, B.~R. (2013).
\newblock {Hadronic Parity Violation}.
\newblock \emph{Prog. Part. Nucl. Phys.} 71, 185--203.
\newblock \doi{10.1016/j.ppnp.2013.03.009}
\bibAnnoteFile{Haxton:2013aca}

\bibitem[{de~Vries and Mei{\ss}ner(2016)}]{deVries:2015gea}
de~Vries, J. and Mei{\ss}ner, U.-G. (2016).
\newblock {Violations of discrete space–time symmetries in chiral effective
  field theory}.
\newblock \emph{Int. J. Mod. Phys.} E25, 1641008.
\newblock \doi{10.1142/S0218301316410081}
\bibAnnoteFile{deVries:2015gea}

\bibitem[{Gardner et~al.(2017)Gardner, Haxton, and Holstein}]{Gardner:2017xyl}
Gardner, S., Haxton, W.~C., and Holstein, B.~R. (2017).
\newblock {A New Paradigm for Hadronic Parity Nonconservation and its
  Experimental Implications}.
\newblock \emph{Ann. Rev. Nucl. Part. Sci.} 67, 69--95.
\newblock \doi{10.1146/annurev-nucl-041917-033231}
\bibAnnoteFile{Gardner:2017xyl}

\bibitem[{Zhu et~al.(2005)Zhu, Maekawa, Holstein, Ramsey-Musolf, and van
  Kolck}]{Zhu:2004vw}
Zhu, S.-L., Maekawa, C.~M., Holstein, B.~R., Ramsey-Musolf, M.~J., and van
  Kolck, U. (2005).
\newblock {Nuclear parity-violation in effective field theory}.
\newblock \emph{Nucl. Phys.} A748, 435--498.
\newblock \doi{10.1016/j.nuclphysa.2004.10.032}
\bibAnnoteFile{Zhu:2004vw}

\bibitem[{de~Vries et~al.(2013{\natexlab{a}})de~Vries, Mei{\ss}ner, Epelbaum,
  and Kaiser}]{deVries:2013fxa}
de~Vries, J., Mei{\ss}ner, U.-G., Epelbaum, E., and Kaiser, N.
  (2013{\natexlab{a}}).
\newblock {Parity violation in proton-proton scattering from chiral effective
  field theory}.
\newblock \emph{Eur. Phys. J.} A49, 149.
\newblock \doi{10.1140/epja/i2013-13149-9}
\bibAnnoteFile{deVries:2013fxa}

\bibitem[{Viviani et~al.(2014)Viviani, Baroni, Girlanda, Kievsky, Marcucci, and
  Schiavilla}]{Viviani:2014zha}
Viviani, M., Baroni, A., Girlanda, L., Kievsky, A., Marcucci, L.~E., and
  Schiavilla, R. (2014).
\newblock {Chiral effective field theory analysis of hadronic parity violation
  in few-nucleon systems}.
\newblock \emph{Phys. Rev.} C89, 064004.
\newblock \doi{10.1103/PhysRevC.89.064004}
\bibAnnoteFile{Viviani:2014zha}

\bibitem[{Adelberger and Haxton(1985)}]{Adelberger:1985ik}
Adelberger, E.~G. and Haxton, W.~C. (1985).
\newblock {Parity Violation in the Nucleon-Nucleon Interaction}.
\newblock \emph{Ann. Rev. Nucl. Part. Sci.} 35, 501--558.
\newblock \doi{10.1146/annurev.ns.35.120185.002441}
\bibAnnoteFile{Adelberger:1985ik}

\bibitem[{Girlanda(2008)}]{Girlanda:2008ts}
Girlanda, L. (2008).
\newblock {On a redundancy in the parity-violating 2-nucleon contact
  Lagrangian}.
\newblock \emph{Phys. Rev.} C77, 067001.
\newblock \doi{10.1103/PhysRevC.77.067001}
\bibAnnoteFile{Girlanda:2008ts}

\bibitem[{Danilov(1965)}]{Danilov:1965hc}
Danilov, G.~S. (1965).
\newblock {}.
\newblock \emph{Phys. Lett.} 18, 40
\bibAnnoteFile{Danilov:1965hc}

\bibitem[{de~Vries et~al.(2014)de~Vries, Li, Mei{\ss}ner, Kaiser, Liu, and
  Zhu}]{deVries:2014vqa}
de~Vries, J., Li, N., Mei{\ss}ner, U.-G., Kaiser, N., Liu, X.~H., and Zhu,
  S.~L. (2014).
\newblock {A study of the parity-odd nucleon-nucleon potential}.
\newblock \emph{Eur. Phys. J.} A50, 108.
\newblock \doi{10.1140/epja/i2014-14108-8}
\bibAnnoteFile{deVries:2014vqa}

\bibitem[{Schindler et~al.(2016)Schindler, Springer, and
  Vanasse}]{Schindler:2015nga}
Schindler, M.~R., Springer, R.~P., and Vanasse, J. (2016).
\newblock {Large-$N_c$ limit reduces the number of independent few-body
  parity-violating low-energy constants in pionless effective field theory}.
\newblock \emph{Phys. Rev.} C93, 025502.
\newblock \doi{10.1103/PhysRevC.97.059901, 10.1103/PhysRevC.93.025502}.
\newblock [Erratum: Phys. Rev.C97,no.5,059901(2018)]
\bibAnnoteFile{Schindler:2015nga}

\bibitem[{Phillips et~al.(2015)Phillips, Samart, and Schat}]{Phillips:2014kna}
Phillips, D.~R., Samart, D., and Schat, C. (2015).
\newblock {Parity-Violating Nucleon-Nucleon Force in the 1/$N_c$ Expansion}.
\newblock \emph{Phys. Rev. Lett.} 114, 062301.
\newblock \doi{10.1103/PhysRevLett.114.062301}
\bibAnnoteFile{Phillips:2014kna}

\bibitem[{Vanasse(2019)}]{Vanasse:2018buq}
Vanasse, J. (2019).
\newblock {Parity-violating three-nucleon interactions at low energies and
  large $N_C$}.
\newblock \emph{Phys. Rev.} C99, 054001.
\newblock \doi{10.1103/PhysRevC.99.054001}
\bibAnnoteFile{Vanasse:2018buq}

\bibitem[{Desplanques et~al.(1980)Desplanques, Donoghue, and
  Holstein}]{Desplanques:1979hn}
Desplanques, B., Donoghue, J.~F., and Holstein, B.~R. (1980).
\newblock {Unified Treatment of the Parity Violating Nuclear Force}.
\newblock \emph{Annals Phys.} 124, 449.
\newblock \doi{10.1016/0003-4916(80)90217-1}
\bibAnnoteFile{Desplanques:1979hn}

\bibitem[{'t~Hooft(1976{\natexlab{a}})}]{tHooft:1976rip}
't~Hooft, G. (1976{\natexlab{a}}).
\newblock {Symmetry Breaking Through Bell-Jackiw Anomalies}.
\newblock \emph{Phys. Rev. Lett.} 37, 8--11.
\newblock \doi{10.1103/PhysRevLett.37.8}.
\newblock [,226(1976)]
\bibAnnoteFile{tHooft:1976rip}

\bibitem[{Pospelov and Ritz(2005)}]{Pospelov:2005pr}
Pospelov, M. and Ritz, A. (2005).
\newblock {Electric dipole moments as probes of new physics}.
\newblock \emph{Annals Phys.} 318, 119--169.
\newblock \doi{10.1016/j.aop.2005.04.002}
\bibAnnoteFile{Pospelov:2005pr}

\bibitem[{Sakharov(1967)}]{Sakharov:1967dj}
Sakharov, A.~D. (1967).
\newblock {Violation of CP Invariance, C asymmetry, and baryon asymmetry of the
  universe}.
\newblock \emph{Pisma Zh. Eksp. Teor. Fiz.} 5, 32--35.
\newblock \doi{10.1070/PU1991v034n05ABEH002497}.
\newblock [Usp. Fiz. Nauk161,no.5,61(1991)]
\bibAnnoteFile{Sakharov:1967dj}

\bibitem[{Cohen et~al.(1993)Cohen, Kaplan, and Nelson}]{Cohen:1993nk}
Cohen, A.~G., Kaplan, D.~B., and Nelson, A.~E. (1993).
\newblock {Progress in electroweak baryogenesis}.
\newblock \emph{Ann. Rev. Nucl. Part. Sci.} 43, 27--70.
\newblock \doi{10.1146/annurev.ns.43.120193.000331}
\bibAnnoteFile{Cohen:1993nk}

\bibitem[{Czarnecki and Krause(1997)}]{Czarnecki:1997bu}
Czarnecki, A. and Krause, B. (1997).
\newblock {Neutron electric dipole moment in the standard model: Valence quark
  contributions}.
\newblock \emph{Phys. Rev. Lett.} 78, 4339--4342.
\newblock \doi{10.1103/PhysRevLett.78.4339}
\bibAnnoteFile{Czarnecki:1997bu}

\bibitem[{Mannel and Uraltsev(2012)}]{Mannel:2012qk}
Mannel, T. and Uraltsev, N. (2012).
\newblock {Loop-Less Electric Dipole Moment of the Nucleon in the Standard
  Model}.
\newblock \emph{Phys. Rev.} D85, 096002.
\newblock \doi{10.1103/PhysRevD.85.096002}
\bibAnnoteFile{Mannel:2012qk}

\bibitem[{Mannel and Uraltsev(2013)}]{Mannel:2012hb}
Mannel, T. and Uraltsev, N. (2013).
\newblock {Charm CP Violation and the Electric Dipole Moments from the Charm
  Scale}.
\newblock \emph{JHEP} 03, 064.
\newblock \doi{10.1007/JHEP03(2013)064}
\bibAnnoteFile{Mannel:2012hb}

\bibitem[{Wirzba et~al.(2017)Wirzba, Bsaisou, and Nogga}]{Wirzba:2016saz}
Wirzba, A., Bsaisou, J., and Nogga, A. (2017).
\newblock {Permanent Electric Dipole Moments of Single-, Two-, and
  Three-Nucleon Systems}.
\newblock \emph{Int. J. Mod. Phys.} E26, 1740031.
\newblock \doi{10.1142/S0218301317400316}
\bibAnnoteFile{Wirzba:2016saz}

\bibitem[{Seng(2015)}]{Seng:2014lea}
Seng, C.-Y. (2015).
\newblock {Reexamination of The Standard Model Nucleon Electric Dipole Moment}.
\newblock \emph{Phys. Rev.} C91, 025502.
\newblock \doi{10.1103/PhysRevC.91.025502}
\bibAnnoteFile{Seng:2014lea}

\bibitem[{Baker et~al.(2006)}]{Baker:2006ts}
Baker, C.~A. et~al. (2006).
\newblock {An Improved experimental limit on the electric dipole moment of the
  neutron}.
\newblock \emph{Phys. Rev. Lett.} 97, 131801.
\newblock \doi{10.1103/PhysRevLett.97.131801}
\bibAnnoteFile{Baker:2006ts}

\bibitem[{Pendlebury et~al.(2015)}]{Afach:2015sja}
Pendlebury, J.~M. et~al. (2015).
\newblock {Revised experimental upper limit on the electric dipole moment of
  the neutron}.
\newblock \emph{Phys. Rev.} D92, 092003.
\newblock \doi{10.1103/PhysRevD.92.092003}
\bibAnnoteFile{Afach:2015sja}

\bibitem[{Graner et~al.(2016)Graner, Chen, Lindahl, and
  Heckel}]{Graner:2016ses}
Graner, B., Chen, Y., Lindahl, E.~G., and Heckel, B.~R. (2016).
\newblock {Reduced Limit on the Permanent Electric Dipole Moment of Hg199}.
\newblock \emph{Phys. Rev. Lett.} 116, 161601.
\newblock \doi{10.1103/PhysRevLett.119.119901, 10.1103/PhysRevLett.116.161601}.
\newblock [Erratum: Phys. Rev. Lett.119,no.11,119901(2017)]
\bibAnnoteFile{Graner:2016ses}

\bibitem[{Dmitriev and Sen'kov(2003)}]{Dmitriev:2003sc}
Dmitriev, V.~F. and Sen'kov, R.~A. (2003).
\newblock {Schiff moment of the mercury nucleus and the proton dipole moment}.
\newblock \emph{Phys. Rev. Lett.} 91, 212303.
\newblock \doi{10.1103/PhysRevLett.91.212303}
\bibAnnoteFile{Dmitriev:2003sc}

\bibitem[{Andreev et~al.(2018)}]{Andreev:2018ayy}
Andreev, V. et~al. (2018).
\newblock {Improved limit on the electric dipole moment of the electron}.
\newblock \emph{Nature} 562, 355--360.
\newblock \doi{10.1038/s41586-018-0599-8}
\bibAnnoteFile{Andreev:2018ayy}

\bibitem[{Mereghetti et~al.(2010)Mereghetti, Hockings, and van
  Kolck}]{Mereghetti:2010tp}
Mereghetti, E., Hockings, W.~H., and van Kolck, U. (2010).
\newblock {The Effective Chiral Lagrangian From the Theta Term}.
\newblock \emph{Annals Phys.} 325, 2363--2409.
\newblock \doi{10.1016/j.aop.2010.03.005}
\bibAnnoteFile{Mereghetti:2010tp}

\bibitem[{Bsaisou et~al.(2015{\natexlab{a}})Bsaisou, de~Vries, Hanhart, Liebig,
  Mei{\ss}ner, Minossi et~al.}]{Bsaisou:2014zwa}
Bsaisou, J., de~Vries, J., Hanhart, C., Liebig, S., Mei{\ss}ner, U.-G.,
  Minossi, D., et~al. (2015{\natexlab{a}}).
\newblock {Nuclear Electric Dipole Moments in Chiral Effective Field Theory}.
\newblock \emph{JHEP} 03, 104.
\newblock \doi{10.1007/JHEP03(2015)104, 10.1007/JHEP05(2015)083}.
\newblock [Erratum: JHEP05,083(2015)]
\bibAnnoteFile{Bsaisou:2014zwa}

\bibitem[{Grzadkowski et~al.(2010)Grzadkowski, Iskrzynski, Misiak, and
  Rosiek}]{Grzadkowski:2010es}
Grzadkowski, B., Iskrzynski, M., Misiak, M., and Rosiek, J. (2010).
\newblock {Dimension-Six Terms in the Standard Model Lagrangian}.
\newblock \emph{JHEP} 10, 085.
\newblock \doi{10.1007/JHEP10(2010)085}
\bibAnnoteFile{Grzadkowski:2010es}

\bibitem[{de~Vries et~al.(2013{\natexlab{b}})de~Vries, Mereghetti, Timmermans,
  and van Kolck}]{deVries:2012ab}
de~Vries, J., Mereghetti, E., Timmermans, R. G.~E., and van Kolck, U.
  (2013{\natexlab{b}}).
\newblock {The Effective Chiral Lagrangian From Dimension-Six Parity and
  Time-Reversal Violation}.
\newblock \emph{Annals Phys.} 338, 50--96.
\newblock \doi{10.1016/j.aop.2013.05.022}
\bibAnnoteFile{deVries:2012ab}

\bibitem[{Bsaisou et~al.(2015{\natexlab{b}})Bsaisou, Mei{\ss}ner, Nogga, and
  Wirzba}]{Bsaisou:2014oka}
Bsaisou, J., Mei{\ss}ner, U.-G., Nogga, A., and Wirzba, A.
  (2015{\natexlab{b}}).
\newblock {P- and T-Violating Lagrangians in Chiral Effective Field Theory and
  Nuclear Electric Dipole Moments}.
\newblock \emph{Annals Phys.} 359, 317--370.
\newblock \doi{10.1016/j.aop.2015.04.031}
\bibAnnoteFile{Bsaisou:2014oka}

\bibitem[{de~Vries et~al.(2011{\natexlab{a}})de~Vries, Mereghetti, Timmermans,
  and van Kolck}]{deVries:2011re}
de~Vries, J., Mereghetti, E., Timmermans, R. G.~E., and van Kolck, U.
  (2011{\natexlab{a}}).
\newblock {Parity- and Time-Reversal-Violating Form Factors of the Deuteron}.
\newblock \emph{Phys. Rev. Lett.} 107, 091804.
\newblock \doi{10.1103/PhysRevLett.107.091804}
\bibAnnoteFile{deVries:2011re}

\bibitem[{Dekens et~al.(2014)Dekens, de~Vries, Bsaisou, Bernreuther, Hanhart,
  Mei{\ss}ner et~al.}]{Dekens:2014jka}
Dekens, W., de~Vries, J., Bsaisou, J., Bernreuther, W., Hanhart, C.,
  Mei{\ss}ner, U.-G., et~al. (2014).
\newblock {Unraveling models of CP violation through electric dipole moments of
  light nuclei}.
\newblock \emph{JHEP} 07, 069.
\newblock \doi{10.1007/JHEP07(2014)069}
\bibAnnoteFile{Dekens:2014jka}

\bibitem[{Maekawa et~al.(2011)Maekawa, Mereghetti, de~Vries, and van
  Kolck}]{Maekawa:2011vs}
Maekawa, C.~M., Mereghetti, E., de~Vries, J., and van Kolck, U. (2011).
\newblock {The Time-Reversal- and Parity-Violating Nuclear Potential in Chiral
  Effective Theory}.
\newblock \emph{Nucl. Phys.} A872, 117--160.
\newblock \doi{10.1016/j.nuclphysa.2011.09.020}
\bibAnnoteFile{Maekawa:2011vs}

\bibitem[{Bsaisou et~al.(2013)Bsaisou, Hanhart, Liebig, Mei{\ss}ner, Nogga, and
  Wirzba}]{Bsaisou:2012rg}
Bsaisou, J., Hanhart, C., Liebig, S., Mei{\ss}ner, U.-G., Nogga, A., and
  Wirzba, A. (2013).
\newblock {The electric dipole moment of the deuteron from the QCD
  $\theta$-term}.
\newblock \emph{Eur. Phys. J.} A49, 31.
\newblock \doi{10.1140/epja/i2013-13031-x}
\bibAnnoteFile{Bsaisou:2012rg}

\bibitem[{Epelbaum et~al.(2005)Epelbaum, Glockle, and
  Mei{\ss}ner}]{Epelbaum:2004fk}
Epelbaum, E., Glockle, W., and Mei{\ss}ner, U.-G. (2005).
\newblock {The Two-nucleon system at next-to-next-to-next-to-leading order}.
\newblock \emph{Nucl. Phys.} A747, 362--424.
\newblock \doi{10.1016/j.nuclphysa.2004.09.107}
\bibAnnoteFile{Epelbaum:2004fk}

\bibitem[{Gnech and Viviani(2019)}]{Gnech:2019dod}
Gnech, A. and Viviani, M. (2019).
\newblock {Time Reversal Violation in Light Nuclei, arXiv:1906.09021}
\bibAnnoteFile{Gnech:2019dod}

\bibitem[{Dobaczewski et~al.(2018)Dobaczewski, Engel, Kortelainen, and
  Becker}]{Dobaczewski:2018nim}
Dobaczewski, J., Engel, J., Kortelainen, M., and Becker, P. (2018).
\newblock {Correlating Schiff moments in the light actinides with octupole
  moments}.
\newblock \emph{Phys. Rev. Lett.} 121, 232501.
\newblock \doi{10.1103/PhysRevLett.121.232501}
\bibAnnoteFile{Dobaczewski:2018nim}

\bibitem[{Orlov et~al.(2006)Orlov, Morse, and Semertzidis}]{Orlov:2006su}
Orlov, Y.~F., Morse, W.~M., and Semertzidis, Y.~K. (2006).
\newblock {Resonance method of electric-dipole-moment measurements in storage
  rings}.
\newblock \emph{Phys. Rev. Lett.} 96, 214802.
\newblock \doi{10.1103/PhysRevLett.96.214802}
\bibAnnoteFile{Orlov:2006su}

\bibitem[{Semertzidis(2011)}]{Semertzidis:2011qv}
Semertzidis, Y.~K. (2011).
\newblock {A Storage Ring proton Electric Dipole Moment experiment: most
  sensitive experiment to CP-violation beyond the Standard Model}.
\newblock In \emph{{Particles and fields. Proceedings, Meeting of the Division
  of the American Physical Society, DPF 2011, Providence, USA, August 9-13,
  2011}}
\bibAnnoteFile{Semertzidis:2011qv}

\bibitem[{Lehrach et~al.(2012)Lehrach, Lorentz, Morse, Nikolaev, and
  Rathmann}]{Lehrach:2012eg}
Lehrach, A., Lorentz, B., Morse, W., Nikolaev, N., and Rathmann, F. (2012).
\newblock {Precursor Experiments to Search for Permanent Electric Dipole
  Moments (EDMs) of Protons and Deuterons at COSY, arXiv:1201.5773}
\bibAnnoteFile{Lehrach:2012eg}

\bibitem[{Pretz(2013)}]{Pretz:2013us}
Pretz, J. (2013).
\newblock {Measurement of Permanent Electric Dipole Moments of Charged Hadrons
  in Storage Rings}.
\newblock \emph{Hyperfine Interact.} 214, 111--117.
\newblock \doi{10.1007/s10751-013-0799-4}
\bibAnnoteFile{Pretz:2013us}

\bibitem[{Rathmann et~al.(2013)Rathmann, Saleev, and
  Nikolaev}]{Rathmann:2013rqa}
Rathmann, F., Saleev, A., and Nikolaev, N.~N. (2013).
\newblock {The search for electric dipole moments of light ions in storage
  rings}.
\newblock \emph{J. Phys. Conf. Ser.} 447, 012011.
\newblock \doi{10.1088/1742-6596/447/1/012011}
\bibAnnoteFile{Rathmann:2013rqa}

\bibitem[{Abusaif et~al.(2019)}]{Abusaif:2019gry}
Abusaif, F. et~al. (2019).
\newblock {Storage Ring to Search for Electric Dipole Moments of Charged
  Particles -- Feasibility Study, arXiv:1912.07881}
\bibAnnoteFile{Abusaif:2019gry}

\bibitem[{Wu et~al.(1957)Wu, Ambler, Hayward, Hoppes, and Hudson}]{Wu:1957my}
Wu, C.~S., Ambler, E., Hayward, R.~W., Hoppes, D.~D., and Hudson, R.~P. (1957).
\newblock {Experimental Test of Parity Conservation in Beta Decay}.
\newblock \emph{Phys. Rev.} 105, 1413--1414.
\newblock \doi{10.1103/PhysRev.105.1413}
\bibAnnoteFile{Wu:1957my}

\bibitem[{Prescott et~al.(1978)}]{Prescott:1978tm}
Prescott, C.~Y. et~al. (1978).
\newblock {Parity Nonconservation in Inelastic Electron Scattering}.
\newblock \emph{Phys. Lett.} B77, 347--352.
\newblock \doi{10.1016/0370-2693(78)90722-0}.
\newblock [,6.31(1978)]
\bibAnnoteFile{Prescott:1978tm}

\bibitem[{Androić et~al.(2018)}]{Androic:2018kni}
Androić, D. et~al. (2018).
\newblock {Precision measurement of the weak charge of the proton}.
\newblock \emph{Nature} 557, 207--211.
\newblock \doi{10.1038/s41586-018-0096-0}
\bibAnnoteFile{Androic:2018kni}

\bibitem[{Tiburzi(2012)}]{Tiburzi:2012hx}
Tiburzi, B.~C. (2012).
\newblock {Hadronic Parity Violation at Next-to-Leading Order}.
\newblock \emph{Phys. Rev.} D85, 054020.
\newblock \doi{10.1103/PhysRevD.85.054020}
\bibAnnoteFile{Tiburzi:2012hx}

\bibitem[{Buchmuller and Wyler(1986)}]{Buchmuller:1985jz}
Buchmuller, W. and Wyler, D. (1986).
\newblock {Effective Lagrangian Analysis of New Interactions and Flavor
  Conservation}.
\newblock \emph{Nucl. Phys.} B268, 621--653.
\newblock \doi{10.1016/0550-3213(86)90262-2}
\bibAnnoteFile{Buchmuller:1985jz}

\bibitem[{Kaplan and Savage(1993)}]{Kaplan:1992vj}
Kaplan, D.~B. and Savage, M.~J. (1993).
\newblock {An Analysis of parity violating pion - nucleon couplings}.
\newblock \emph{Nucl. Phys.} A556, 653--671.
\newblock \doi{10.1016/0375-9474(93)90475-D, 10.1016/0375-9474(94)90787-0,
  10.1016/0375-9474(94)90086-8}.
\newblock [Erratum: Nucl. Phys.A580,679(1994)]
\bibAnnoteFile{Kaplan:1992vj}

\bibitem[{Christenson et~al.(1964)Christenson, Cronin, Fitch, and
  Turlay}]{Christenson:1964fg}
Christenson, J.~H., Cronin, J.~W., Fitch, V.~L., and Turlay, R. (1964).
\newblock {Evidence for the $2\pi$ Decay of the $K_2^0$ Meson}.
\newblock \emph{Phys. Rev. Lett.} 13, 138--140.
\newblock \doi{10.1103/PhysRevLett.13.138}
\bibAnnoteFile{Christenson:1964fg}

\bibitem[{Abouzaid et~al.(2011)}]{Abouzaid:2010ny}
Abouzaid, E. et~al. (2011).
\newblock {Precise Measurements of Direct CP Violation, CPT Symmetry, and Other
  Parameters in the Neutral Kaon System}.
\newblock \emph{Phys. Rev.} D83, 092001.
\newblock \doi{10.1103/PhysRevD.83.092001}
\bibAnnoteFile{Abouzaid:2010ny}

\bibitem[{Batley et~al.(2002)}]{Batley:2002gn}
Batley, J.~R. et~al. (2002).
\newblock {A Precision measurement of direct CP violation in the decay of
  neutral kaons into two pions}.
\newblock \emph{Phys. Lett.} B544, 97--112.
\newblock \doi{10.1016/S0370-2693(02)02476-0}
\bibAnnoteFile{Batley:2002gn}

\bibitem[{Aubert et~al.(2001)}]{Aubert:2001nu}
Aubert, B. et~al. (2001).
\newblock {Observation of CP violation in the $B^0$ meson system}.
\newblock \emph{Phys. Rev. Lett.} 87, 091801.
\newblock \doi{10.1103/PhysRevLett.87.091801}
\bibAnnoteFile{Aubert:2001nu}

\bibitem[{Abe et~al.(2001)}]{Abe:2001xe}
Abe, K. et~al. (2001).
\newblock {Observation of large CP violation in the neutral $B$ meson system}.
\newblock \emph{Phys. Rev. Lett.} 87, 091802.
\newblock \doi{10.1103/PhysRevLett.87.091802}
\bibAnnoteFile{Abe:2001xe}

\bibitem[{Aaij et~al.(2019)}]{Aaij:2019kcg}
Aaij, R. et~al. (2019).
\newblock {Observation of $C\!P$ violation in charm decays}.
\newblock \emph{Phys. Rev. Lett.} 122, 211803.
\newblock \doi{10.1103/PhysRevLett.122.211803}
\bibAnnoteFile{Aaij:2019kcg}

\bibitem[{Khriplovich and Zhitnitsky(1982)}]{Khriplovich:1981ca}
Khriplovich, I.~B. and Zhitnitsky, A.~R. (1982).
\newblock {What Is the Value of the Neutron Electric Dipole Moment in the
  {Kobayashi-Maskawa} Model?}
\newblock \emph{Phys. Lett.} 109B, 490--492.
\newblock \doi{10.1016/0370-2693(82)91121-2}
\bibAnnoteFile{Khriplovich:1981ca}

\bibitem[{Pospelov and Khriplovich(1991)}]{Pospelov:1991zt}
Pospelov, M.~E. and Khriplovich, I.~B. (1991).
\newblock {Electric dipole moment of the W boson and the electron in the
  Kobayashi-Maskawa model}.
\newblock \emph{Sov. J. Nucl. Phys.} 53, 638--640.
\newblock [Yad. Fiz.53,1030(1991)]
\bibAnnoteFile{Pospelov:1991zt}

\bibitem[{Booth(1993)}]{Booth:1993af}
Booth, M.~J. (1993).
\newblock {The Electric dipole moment of the W and electron in the Standard
  Model, arXiv:hep-ph/9301293}
\bibAnnoteFile{Booth:1993af}

\bibitem[{Pospelov and Ritz(2014)}]{Pospelov:2013sca}
Pospelov, M. and Ritz, A. (2014).
\newblock {CKM benchmarks for electron electric dipole moment experiments}.
\newblock \emph{Phys. Rev.} D89, 056006.
\newblock \doi{10.1103/PhysRevD.89.056006}
\bibAnnoteFile{Pospelov:2013sca}

\bibitem[{Callan et~al.(1976)Callan, Dashen, and Gross}]{Callan:1976je}
Callan, C.~G., Jr., Dashen, R.~F., and Gross, D.~J. (1976).
\newblock {The Structure of the Gauge Theory Vacuum}.
\newblock \emph{Phys. Lett.} B63, 334--340.
\newblock \doi{10.1016/0370-2693(76)90277-X}.
\newblock [,357(1976)]
\bibAnnoteFile{Callan:1976je}

\bibitem[{'t~Hooft(1976{\natexlab{b}})}]{tHooft:1976snw}
't~Hooft, G. (1976{\natexlab{b}}).
\newblock {Computation of the Quantum Effects Due to a Four-Dimensional
  Pseudoparticle}.
\newblock \emph{Phys. Rev.} D14, 3432--3450.
\newblock \doi{10.1103/PhysRevD.18.2199.3, 10.1103/PhysRevD.14.3432}.
\newblock [,70(1976)]
\bibAnnoteFile{tHooft:1976snw}

\bibitem[{Baluni(1979)}]{Baluni:1978rf}
Baluni, V. (1979).
\newblock {CP Violating Effects in QCD}.
\newblock \emph{Phys. Rev.} D19, 2227--2230.
\newblock \doi{10.1103/PhysRevD.19.2227}
\bibAnnoteFile{Baluni:1978rf}

\bibitem[{Crewther et~al.(1979)Crewther, Di~Vecchia, Veneziano, and
  Witten}]{Crewther:1979pi}
Crewther, R.~J., Di~Vecchia, P., Veneziano, G., and Witten, E. (1979).
\newblock {Chiral Estimate of the Electric Dipole Moment of the Neutron in
  Quantum Chromodynamics}.
\newblock \emph{Phys. Lett.} 88B, 123.
\newblock \doi{10.1016/0370-2693(80)91025-4, 10.1016/0370-2693(79)90128-X}.
\newblock [Erratum: Phys. Lett.91B,487(1980)]
\bibAnnoteFile{Crewther:1979pi}

\bibitem[{Dragos et~al.(2019)Dragos, Luu, Shindler, de~Vries, and
  Yousif}]{Dragos:2019oxn}
Dragos, J., Luu, T., Shindler, A., de~Vries, J., and Yousif, A. (2019).
\newblock {Confirming the Existence of the strong CP Problem in Lattice QCD
  with the Gradient Flow, arXiv:1902.03254}
\bibAnnoteFile{Dragos:2019oxn}

\bibitem[{Gavela et~al.(1994{\natexlab{a}})Gavela, Hernandez, Orloff, and
  Pene}]{Gavela:1993ts}
Gavela, M.~B., Hernandez, P., Orloff, J., and Pene, O. (1994{\natexlab{a}}).
\newblock {Standard model CP violation and baryon asymmetry}.
\newblock \emph{Mod. Phys. Lett.} A9, 795--810.
\newblock \doi{10.1142/S0217732394000629}
\bibAnnoteFile{Gavela:1993ts}

\bibitem[{Gavela et~al.(1994{\natexlab{b}})Gavela, Lozano, Orloff, and
  Pene}]{Gavela:1994ds}
Gavela, M.~B., Lozano, M., Orloff, J., and Pene, O. (1994{\natexlab{b}}).
\newblock {Standard model CP violation and baryon asymmetry. Part 1: Zero
  temperature}.
\newblock \emph{Nucl. Phys.} B430, 345--381.
\newblock \doi{10.1016/0550-3213(94)00409-9}
\bibAnnoteFile{Gavela:1994ds}

\bibitem[{Gavela et~al.(1994{\natexlab{c}})Gavela, Hernandez, Orloff, Pene, and
  Quimbay}]{Gavela:1994dt}
Gavela, M.~B., Hernandez, P., Orloff, J., Pene, O., and Quimbay, C.
  (1994{\natexlab{c}}).
\newblock {Standard model CP violation and baryon asymmetry. Part 2: Finite
  temperature}.
\newblock \emph{Nucl. Phys.} B430, 382--426.
\newblock \doi{10.1016/0550-3213(94)00410-2}
\bibAnnoteFile{Gavela:1994dt}

\bibitem[{Huet and Sather(1995)}]{Huet:1994jb}
Huet, P. and Sather, E. (1995).
\newblock {Electroweak baryogenesis and standard model CP violation}.
\newblock \emph{Phys. Rev.} D51, 379--394.
\newblock \doi{10.1103/PhysRevD.51.379}
\bibAnnoteFile{Huet:1994jb}

\bibitem[{Khriplovich and Lamoreaux(1997)}]{Khriplovich:1997ga}
Khriplovich, I.~B. and Lamoreaux, S.~K. (1997).
\newblock \emph{{CP violation without strangeness: Electric dipole moments of
  particles, atoms, and molecules [Berlin, Germany: Springer (1997)]}}
\bibAnnoteFile{Khriplovich:1997ga}

\bibitem[{Dekens and de~Vries(2013)}]{Dekens:2013zca}
Dekens, W. and de~Vries, J. (2013).
\newblock {Renormalization Group Running of Dimension-Six Sources of Parity and
  Time-Reversal Violation}.
\newblock \emph{JHEP} 05, 149.
\newblock \doi{10.1007/JHEP05(2013)149}
\bibAnnoteFile{Dekens:2013zca}

\bibitem[{Engel et~al.(2013)Engel, Ramsey-Musolf, and van
  Kolck}]{Engel:2013lsa}
Engel, J., Ramsey-Musolf, M.~J., and van Kolck, U. (2013).
\newblock {Electric Dipole Moments of Nucleons, Nuclei, and Atoms: The Standard
  Model and Beyond}.
\newblock \emph{Prog. Part. Nucl. Phys.} 71, 21--74.
\newblock \doi{10.1016/j.ppnp.2013.03.003}
\bibAnnoteFile{Engel:2013lsa}

\bibitem[{Jenkins et~al.(2018)Jenkins, Manohar, and Stoffer}]{Jenkins:2017jig}
Jenkins, E.~E., Manohar, A.~V., and Stoffer, P. (2018).
\newblock {Low-Energy Effective Field Theory below the Electroweak Scale:
  Operators and Matching}.
\newblock \emph{JHEP} 03, 016.
\newblock \doi{10.1007/JHEP03(2018)016}
\bibAnnoteFile{Jenkins:2017jig}

\bibitem[{Mereghetti(2018)}]{Mereghetti:2018oxv}
Mereghetti, E. (2018).
\newblock {Electric dipole moments: a theory overview}.
\newblock In \emph{{13th Conference on the Intersections of Particle and
  Nuclear Physics (CIPANP 2018) Palm Springs, California, USA, May 29-June 3,
  2018}}
\bibAnnoteFile{Mereghetti:2018oxv}

\bibitem[{Weinberg(1989)}]{Weinberg:1989dx}
Weinberg, S. (1989).
\newblock {Larger Higgs Exchange Terms in the Neutron Electric Dipole Moment}.
\newblock \emph{Phys. Rev. Lett.} 63, 2333.
\newblock \doi{10.1103/PhysRevLett.63.2333}
\bibAnnoteFile{Weinberg:1989dx}

\bibitem[{Fuyuto et~al.(2019)Fuyuto, Ramsey-Musolf, and Shen}]{Fuyuto:2018scm}
Fuyuto, K., Ramsey-Musolf, M., and Shen, T. (2019).
\newblock {Electric Dipole Moments from CP-Violating Scalar Leptoquark
  Interactions}.
\newblock \emph{Phys. Lett.} B788, 52--57.
\newblock \doi{10.1016/j.physletb.2018.11.016}
\bibAnnoteFile{Fuyuto:2018scm}

\bibitem[{Dekens et~al.(2019)Dekens, de~Vries, Jung, and Vos}]{Dekens:2018bci}
Dekens, W., de~Vries, J., Jung, M., and Vos, K.~K. (2019).
\newblock {The phenomenology of electric dipole moments in models of scalar
  leptoquarks}.
\newblock \emph{JHEP} 01, 069.
\newblock \doi{10.1007/JHEP01(2019)069}
\bibAnnoteFile{Dekens:2018bci}

\bibitem[{Ng and Tulin(2012)}]{Ng:2011ui}
Ng, J. and Tulin, S. (2012).
\newblock {D versus d: CP Violation in Beta Decay and Electric Dipole Moments}.
\newblock \emph{Phys. Rev.} D85, 033001.
\newblock \doi{10.1103/PhysRevD.85.033001}
\bibAnnoteFile{Ng:2011ui}

\bibitem[{Jenkins and Manohar(1991)}]{Jenkins:1990jv}
Jenkins, E.~E. and Manohar, A.~V. (1991).
\newblock {Baryon chiral perturbation theory using a heavy fermion Lagrangian}.
\newblock \emph{Phys. Lett.} B255, 558--562.
\newblock \doi{10.1016/0370-2693(91)90266-S}
\bibAnnoteFile{Jenkins:1990jv}

\bibitem[{Baroni et~al.(2016)Baroni, Girlanda, Pastore, Schiavilla, and
  Viviani}]{Baroni:2015uza}
Baroni, A., Girlanda, L., Pastore, S., Schiavilla, R., and Viviani, M. (2016).
\newblock {Nuclear Axial Currents in Chiral Effective Field Theory}.
\newblock \emph{Phys. Rev.} C93, 015501.
\newblock \doi{10.1103/PhysRevC.93.049902, 10.1103/PhysRevC.93.015501,
  10.1103/PhysRevC.95.059901}.
\newblock [Erratum: Phys. Rev.C95,no.5,059901(2017)]
\bibAnnoteFile{Baroni:2015uza}

\bibitem[{Kaplan et~al.(1999{\natexlab{a}})Kaplan, Savage, Springer, and
  Wise}]{Kaplan:1998xi}
Kaplan, D.~B., Savage, M.~J., Springer, R.~P., and Wise, M.~B.
  (1999{\natexlab{a}}).
\newblock {An Effective field theory calculation of the parity violating
  asymmetry in $\vec n + p \rightarrow d + \gamma$}.
\newblock \emph{Phys. Lett.} B449, 1--5.
\newblock \doi{10.1016/S0370-2693(99)00032-5}
\bibAnnoteFile{Kaplan:1998xi}

\bibitem[{Manohar and Georgi(1984)}]{Manohar:1983md}
Manohar, A. and Georgi, H. (1984).
\newblock {Chiral Quarks and the Nonrelativistic Quark Model}.
\newblock \emph{Nucl. Phys.} B234, 189--212.
\newblock \doi{10.1016/0550-3213(84)90231-1}
\bibAnnoteFile{Manohar:1983md}

\bibitem[{Michel(1964)}]{Michel:1964zz}
Michel, F.~C. (1964).
\newblock {Parity Nonconservation in Nuclei}.
\newblock \emph{Phys. Rev.} 133, B329--B349.
\newblock \doi{10.1103/PhysRev.133.B329}
\bibAnnoteFile{Michel:1964zz}

\bibitem[{Donoghue et~al.(1992)Donoghue, Golowich, and
  Holstein}]{Donoghue:1992dd}
Donoghue, J.~F., Golowich, E., and Holstein, B.~R. (1992).
\newblock {Dynamics of the standard model}.
\newblock \emph{Camb. Monogr. Part. Phys. Nucl. Phys. Cosmol.} 2, 1--540.
\newblock \doi{10.1017/CBO9780511524370}.
\newblock [Camb. Monogr. Part. Phys. Nucl. Phys. Cosmol.35(2014)]
\bibAnnoteFile{Donoghue:1992dd}

\bibitem[{McKellar(1967)}]{McKellar:1967mxj}
McKellar, B. H.~J. (1967).
\newblock {The one pion exchange contribution to the weak parity violating
  nucleon-nucleon potential}.
\newblock \emph{Phys. Lett.} 26B, 107--108.
\newblock \doi{10.1016/0370-2693(67)90561-8}
\bibAnnoteFile{McKellar:1967mxj}

\bibitem[{Fischbach(1968)}]{Fischbach:1968zz}
Fischbach, E. (1968).
\newblock {Application of Current Algebra and Partially Conserved Axial-Vector
  Current to the Weak BBpi Vertex}.
\newblock \emph{Phys. Rev.} 170, 1398--1400.
\newblock \doi{10.1103/PhysRev.170.1398}
\bibAnnoteFile{Fischbach:1968zz}

\bibitem[{Tadic(1968)}]{Tadic:1969xx}
Tadic, D. (1968).
\newblock {Weak parity-nonconserving potentials}.
\newblock \emph{Phys. Rev.} 174, 1694--1703.
\newblock \doi{10.1103/PhysRev.174.1694}
\bibAnnoteFile{Tadic:1969xx}

\bibitem[{Kummer and Schweda(1968)}]{Kummer:1968ra}
Kummer, W. and Schweda, M. (1968).
\newblock {An SU(6) estimate of the effective nonleptonic parity-violating
  coupling in strangeness-conserving weak processes}.
\newblock \emph{Acta Phys. Austriaca} 28, 303--308
\bibAnnoteFile{Kummer:1968ra}

\bibitem[{Mckellar and Pick(1972)}]{Mckellar:1973rr}
Mckellar, B. H.~J. and Pick, P. (1972).
\newblock {Pion-pole dominance of the divergence of the weak
  parity-nonconserving n n rho amplitude}.
\newblock \emph{Phys. Rev.} D6, 2184--2188.
\newblock \doi{10.1103/PhysRevD.6.2184}
\bibAnnoteFile{Mckellar:1973rr}

\bibitem[{Dubovik and Zenkin(1986)}]{Dubovik:1986pj}
Dubovik, V.~M. and Zenkin, S.~V. (1986).
\newblock {Formation of Parity Nonconserving Nuclear Forces in the Standard
  Model SU(2)(l) X U(1) X SU(3)(c)}.
\newblock \emph{Annals Phys.} 172, 100--135.
\newblock \doi{10.1016/0003-4916(86)90021-7}
\bibAnnoteFile{Dubovik:1986pj}

\bibitem[{Feldman et~al.(1991)Feldman, Crawford, Dubach, and
  Holstein}]{Feldman:1991tj}
Feldman, G.~B., Crawford, G.~A., Dubach, J., and Holstein, B.~R. (1991).
\newblock {Delta contributions to the parity violating nuclear interaction}.
\newblock \emph{Phys. Rev.} C43, 863--874.
\newblock \doi{10.1103/PhysRevC.43.863}
\bibAnnoteFile{Feldman:1991tj}

\bibitem[{Mei{\ss}ner and Weigel(1999)}]{Meissner:1998pu}
Mei{\ss}ner, U.~G. and Weigel, H. (1999).
\newblock {The Parity violating pion nucleon coupling constant from a realistic
  three flavor Skyrme model}.
\newblock \emph{Phys. Lett.} B447, 1--7.
\newblock \doi{10.1016/S0370-2693(98)01569-X}
\bibAnnoteFile{Meissner:1998pu}

\bibitem[{Haeberli and Holstein(1995)}]{Haeberli:1995uz}
Haeberli, W. and Holstein, B.~R. (1995).
\newblock {Parity violation and the nucleon-nucleon system,
  arXiv:nucl-th/9510062} \doi{10.1142/9789812831446_0002}
\bibAnnoteFile{Haeberli:1995uz}

\bibitem[{Wasem(2012)}]{Wasem:2011zz}
Wasem, J. (2012).
\newblock {Lattice QCD Calculation of Nuclear Parity Violation}.
\newblock \emph{Phys. Rev.} C85, 022501.
\newblock \doi{10.1103/PhysRevC.85.022501}
\bibAnnoteFile{Wasem:2011zz}

\bibitem[{Cheng(1991)}]{Cheng:1990pi}
Cheng, H.-Y. (1991).
\newblock {Reanalysis of strong CP violating effects in chiral perturbation
  theory}.
\newblock \emph{Phys. Rev.} D44, 166--174.
\newblock \doi{10.1103/PhysRevD.44.166}
\bibAnnoteFile{Cheng:1990pi}

\bibitem[{Pich and de~Rafael(1991)}]{Pich:1991fq}
Pich, A. and de~Rafael, E. (1991).
\newblock {Strong CP violation in an effective chiral Lagrangian approach}.
\newblock \emph{Nucl. Phys.} B367, 313--333.
\newblock \doi{10.1016/0550-3213(91)90019-T}
\bibAnnoteFile{Pich:1991fq}

\bibitem[{Cho(1993)}]{Cho:1992rv}
Cho, P.~L. (1993).
\newblock {Chiral estimates of strong CP violation revisited}.
\newblock \emph{Phys. Rev.} D48, 3304--3309.
\newblock \doi{10.1103/PhysRevD.48.3304}
\bibAnnoteFile{Cho:1992rv}

\bibitem[{Borasoy(2000)}]{Borasoy:2000pq}
Borasoy, B. (2000).
\newblock {The Electric dipole moment of the neutron in chiral perturbation
  theory}.
\newblock \emph{Phys. Rev.} D61, 114017.
\newblock \doi{10.1103/PhysRevD.61.114017}
\bibAnnoteFile{Borasoy:2000pq}

\bibitem[{Ottnad et~al.(2010)Ottnad, Kubis, Mei{\ss}ner, and
  Guo}]{Ottnad:2009jw}
Ottnad, K., Kubis, B., Mei{\ss}ner, U.-G., and Guo, F.~K. (2010).
\newblock {New insights into the neutron electric dipole moment}.
\newblock \emph{Phys. Lett.} B687, 42--47.
\newblock \doi{10.1016/j.physletb.2010.03.005}
\bibAnnoteFile{Ottnad:2009jw}

\bibitem[{Liu et~al.(2012)Liu, de~Vries, Mereghetti, Timmermans, and van
  Kolck}]{Liu:2012tra}
Liu, C.~P., de~Vries, J., Mereghetti, E., Timmermans, R. G.~E., and van Kolck,
  U. (2012).
\newblock {Deuteron Magnetic Quadrupole Moment From Chiral Effective Field
  Theory}.
\newblock \emph{Phys. Lett.} B713, 447--452.
\newblock \doi{10.1016/j.physletb.2012.06.024}
\bibAnnoteFile{Liu:2012tra}

\bibitem[{Peccei and Quinn(1977)}]{Peccei:1977hh}
Peccei, R.~D. and Quinn, H.~R. (1977).
\newblock {CP Conservation in the Presence of Instantons}.
\newblock \emph{Phys. Rev. Lett.} 38, 1440--1443.
\newblock \doi{10.1103/PhysRevLett.38.1440}.
\newblock [,328(1977)]
\bibAnnoteFile{Peccei:1977hh}

\bibitem[{Mereghetti and van Kolck(2015)}]{Mereghetti:2015rra}
Mereghetti, E. and van Kolck, U. (2015).
\newblock {Effective Field Theory and Time-Reversal Violation in Light Nuclei}.
\newblock \emph{Ann. Rev. Nucl. Part. Sci.} 65, 215--243.
\newblock \doi{10.1146/annurev-nucl-102014-022344}
\bibAnnoteFile{Mereghetti:2015rra}

\bibitem[{Cirigliano et~al.(2017)Cirigliano, Dekens, de~Vries, and
  Mereghetti}]{Cirigliano:2016yhc}
Cirigliano, V., Dekens, W., de~Vries, J., and Mereghetti, E. (2017).
\newblock {An $\epsilon'$ improvement from right-handed currents}.
\newblock \emph{Phys. Lett.} B767, 1--9.
\newblock \doi{10.1016/j.physletb.2017.01.037}
\bibAnnoteFile{Cirigliano:2016yhc}

\bibitem[{Liu and Timmermans(2004)}]{Liu:2004tq}
Liu, C.~P. and Timmermans, R. G.~E. (2004).
\newblock {P- and T-odd two-nucleon interaction and the deuteron electric
  dipole moment}.
\newblock \emph{Phys. Rev.} C70, 055501.
\newblock \doi{10.1103/PhysRevC.70.055501}
\bibAnnoteFile{Liu:2004tq}

\bibitem[{de~Vries et~al.(2011{\natexlab{b}})de~Vries, Higa, Liu, Mereghetti,
  Stetcu, Timmermans et~al.}]{deVries:2011an}
de~Vries, J., Higa, R., Liu, C.~P., Mereghetti, E., Stetcu, I., Timmermans, R.
  G.~E., et~al. (2011{\natexlab{b}}).
\newblock {Electric Dipole Moments of Light Nuclei From Chiral Effective Field
  Theory}.
\newblock \emph{Phys. Rev.} C84, 065501.
\newblock \doi{10.1103/PhysRevC.84.065501}
\bibAnnoteFile{deVries:2011an}

\bibitem[{de~Vries et~al.(2015{\natexlab{a}})de~Vries, Mereghetti, and
  Walker-Loud}]{deVries:2015una}
de~Vries, J., Mereghetti, E., and Walker-Loud, A. (2015{\natexlab{a}}).
\newblock {Baryon mass splittings and strong CP violation in SU(3) Chiral
  Perturbation Theory}.
\newblock \emph{Phys. Rev.} C92, 045201.
\newblock \doi{10.1103/PhysRevC.92.045201}
\bibAnnoteFile{deVries:2015una}

\bibitem[{Borsanyi et~al.(2015)}]{Borsanyi:2014jba}
Borsanyi, S. et~al. (2015).
\newblock {Ab initio calculation of the neutron-proton mass difference}.
\newblock \emph{Science} 347, 1452--1455.
\newblock \doi{10.1126/science.1257050}
\bibAnnoteFile{Borsanyi:2014jba}

\bibitem[{Brantley et~al.(2016)Brantley, Joo, Mastropas, Mereghetti,
  Monge-Camacho, Tiburzi et~al.}]{Brantley:2016our}
Brantley, D.~A., Joo, B., Mastropas, E.~V., Mereghetti, E., Monge-Camacho, H.,
  Tiburzi, B.~C., et~al. (2016).
\newblock {Strong isospin violation and chiral logarithms in the baryon
  spectrum}
\bibAnnoteFile{Brantley:2016our}

\bibitem[{de~Vries et~al.(2017)de~Vries, Mereghetti, Seng, and
  Walker-Loud}]{deVries:2016jox}
de~Vries, J., Mereghetti, E., Seng, C.-Y., and Walker-Loud, A. (2017).
\newblock {Lattice QCD spectroscopy for hadronic CP violation}.
\newblock \emph{Phys. Lett.} B766, 254--262.
\newblock \doi{10.1016/j.physletb.2017.01.017}
\bibAnnoteFile{deVries:2016jox}

\bibitem[{Seng and Ramsey-Musolf(2017)}]{Seng:2016pfd}
Seng, C.-Y. and Ramsey-Musolf, M. (2017).
\newblock {Parity-violating and time-reversal-violating pion-nucleon couplings:
  Higher order chiral matching relations}.
\newblock \emph{Phys. Rev.} C96, 065204.
\newblock \doi{10.1103/PhysRevC.96.065204}
\bibAnnoteFile{Seng:2016pfd}

\bibitem[{Pospelov and Ritz(2001)}]{Pospelov:2000bw}
Pospelov, M. and Ritz, A. (2001).
\newblock {Neutron EDM from electric and chromoelectric dipole moments of
  quarks}.
\newblock \emph{Phys. Rev.} D63, 073015.
\newblock \doi{10.1103/PhysRevD.63.073015}
\bibAnnoteFile{Pospelov:2000bw}

\bibitem[{Samart et~al.(2016)Samart, Schat, Schindler, and
  Phillips}]{Samart:2016ufg}
Samart, D., Schat, C., Schindler, M.~R., and Phillips, D.~R. (2016).
\newblock {Time-reversal-invariance-violating nucleon-nucleon potential in the
  $1/N_c$ expansion}.
\newblock \emph{Phys. Rev.} C94, 024001.
\newblock \doi{10.1103/PhysRevC.94.024001}
\bibAnnoteFile{Samart:2016ufg}

\bibitem[{Guo et~al.(2015)Guo, Horsley, Mei{\ss}ner, Nakamura, Perlt, Rakow
  et~al.}]{Guo:2015tla}
Guo, F.~K., Horsley, R., Mei{\ss}ner, U.-G., Nakamura, Y., Perlt, H., Rakow, P.
  E.~L., et~al. (2015).
\newblock {The electric dipole moment of the neutron from 2+1 flavor lattice
  QCD}.
\newblock \emph{Phys. Rev. Lett.} 115, 062001.
\newblock \doi{10.1103/PhysRevLett.115.062001}
\bibAnnoteFile{Guo:2015tla}

\bibitem[{Abramczyk et~al.(2017)Abramczyk, Aoki, Blum, Izubuchi, Ohki, and
  Syritsyn}]{Abramczyk:2017oxr}
Abramczyk, M., Aoki, S., Blum, T., Izubuchi, T., Ohki, H., and Syritsyn, S.
  (2017).
\newblock {Lattice calculation of electric dipole moments and form factors of
  the nucleon}.
\newblock \emph{Phys. Rev.} D96, 014501.
\newblock \doi{10.1103/PhysRevD.96.014501}
\bibAnnoteFile{Abramczyk:2017oxr}

\bibitem[{Gupta et~al.(2018)Gupta, Jang, Yoon, Lin, Cirigliano, and
  Bhattacharya}]{Gupta:2018qil}
Gupta, R., Jang, Y.-C., Yoon, B., Lin, H.-W., Cirigliano, V., and Bhattacharya,
  T. (2018).
\newblock {Isovector Charges of the Nucleon from 2+1+1-flavor Lattice QCD}.
\newblock \emph{Phys. Rev.} D98, 034503.
\newblock \doi{10.1103/PhysRevD.98.034503}
\bibAnnoteFile{Gupta:2018qil}

\bibitem[{Aoki et~al.(2019)}]{Aoki:2019cca}
Aoki, S. et~al. (2019).
\newblock {FLAG Review 2019, arXiv:1902.08191}
\bibAnnoteFile{Aoki:2019cca}

\bibitem[{Kim et~al.(2019)Kim, Dragos, Shindler, Luu, and
  de~Vries}]{Kim:2018rce}
Kim, J., Dragos, J., Shindler, A., Luu, T., and de~Vries, J. (2019).
\newblock {Towards a determination of the nucleon EDM from the quark chromo-EDM
  operator with the gradient flow}.
\newblock \emph{PoS} LATTICE2018, 260.
\newblock \doi{10.22323/1.334.0260}
\bibAnnoteFile{Kim:2018rce}

\bibitem[{Rizik et~al.(2018)Rizik, Monahan, and Shindler}]{Rizik:2018lrz}
Rizik, M., Monahan, C., and Shindler, A. (2018).
\newblock {Renormalization of CP-Violating Pure Gauge Operators in Perturbative
  QCD Using the Gradient Flow}.
\newblock \emph{PoS} LATTICE2018, 215.
\newblock \doi{10.22323/1.334.0215}
\bibAnnoteFile{Rizik:2018lrz}

\bibitem[{Mereghetti et~al.(2011)Mereghetti, de~Vries, Hockings, Maekawa, and
  van Kolck}]{Mereghetti:2010kp}
Mereghetti, E., de~Vries, J., Hockings, W.~H., Maekawa, C.~M., and van Kolck,
  U. (2011).
\newblock {The Electric Dipole Form Factor of the Nucleon in Chiral
  Perturbation Theory to Sub-leading Order}.
\newblock \emph{Phys. Lett.} B696, 97--102.
\newblock \doi{10.1016/j.physletb.2010.12.018}
\bibAnnoteFile{Mereghetti:2010kp}

\bibitem[{Haisch and Hala(2019)}]{Haisch:2019bml}
Haisch, U. and Hala, A. (2019).
\newblock {Sum rules for CP-violating operators of Weinberg type,
  arXiv:1909.08955}
\bibAnnoteFile{Haisch:2019bml}

\bibitem[{Epelbaum et~al.(2003)Epelbaum, Mei{\ss}ner, and
  Gl{\"o}ckle}]{Epelbaum:2002gb}
Epelbaum, E., Mei{\ss}ner, U.-G., and Gl{\"o}ckle, W. (2003).
\newblock {Nuclear forces in the chiral limit}.
\newblock \emph{Nucl. Phys.} A714, 535--574.
\newblock \doi{10.1016/S0375-9474(02)01393-3}
\bibAnnoteFile{Epelbaum:2002gb}

\bibitem[{Krebs et~al.(2017)Krebs, Epelbaum, and Mei{\ss}ner}]{Krebs:2016rqz}
Krebs, H., Epelbaum, E., and Mei{\ss}ner, U.~G. (2017).
\newblock {Nuclear axial current operators to fourth order in chiral effective
  field theory}.
\newblock \emph{Annals Phys.} 378, 317--395.
\newblock \doi{10.1016/j.aop.2017.01.021}
\bibAnnoteFile{Krebs:2016rqz}

\bibitem[{Pastore et~al.(2008)Pastore, Schiavilla, and Goity}]{Pastore:2008ui}
Pastore, S., Schiavilla, R., and Goity, J.~L. (2008).
\newblock {Electromagnetic two-body currents of one- and two-pion range}.
\newblock \emph{Phys. Rev.} C78, 064002.
\newblock \doi{10.1103/PhysRevC.78.064002}
\bibAnnoteFile{Pastore:2008ui}

\bibitem[{Pastore et~al.(2011)Pastore, Girlanda, Schiavilla, and
  Viviani}]{Pastore:2011ip}
Pastore, S., Girlanda, L., Schiavilla, R., and Viviani, M. (2011).
\newblock {The two-nucleon electromagnetic charge operator in chiral effective
  field theory ($\chi$EFT) up to one loop}.
\newblock \emph{Phys. Rev.} C84, 024001.
\newblock \doi{10.1103/PhysRevC.84.024001}
\bibAnnoteFile{Pastore:2011ip}

\bibitem[{Krebs et~al.(2020)Krebs, Epelbaum, and Mei{\ss}ner}]{Krebs:2020rms}
Krebs, H., Epelbaum, E., and Mei{\ss}ner, U.-G. (2020).
\newblock {Box diagram contribution to the axial two-nucleon current,
  arXiv:2001.03904}
\bibAnnoteFile{Krebs:2020rms}

\bibitem[{Fukuda et~al.(1954)Fukuda, Sawada, and Taketani}]{FST}
Fukuda, N., Sawada, K., and Taketani, M. (1954).
\newblock {On the construction of potential in field theory}.
\newblock \emph{Prog. Theor. Phys.} 12, 156.
\newblock \doi{10.1143/PTP.12.156}
\bibAnnoteFile{FST}

\bibitem[{Okubo(1954)}]{Okubo:1954zz}
Okubo, S. (1954).
\newblock {Diagonalization of Hamiltonian and Tamm-Dancoff Equation}.
\newblock \emph{Prog. Theor. Phys.} 12, 603.
\newblock \doi{10.1143/PTP.12.603}
\bibAnnoteFile{Okubo:1954zz}

\bibitem[{Epelbaum et~al.(1998)Epelbaum, Gl{\"o}ckle, and
  Mei{\ss}ner}]{Epelbaum:1998ka}
Epelbaum, E., Gl{\"o}ckle, W., and Mei{\ss}ner, U.-G. (1998).
\newblock {Nuclear forces from chiral Lagrangians using the method of unitary
  transformation. 1. Formalism}.
\newblock \emph{Nucl. Phys.} A637, 107--134.
\newblock \doi{10.1016/S0375-9474(98)00220-6}
\bibAnnoteFile{Epelbaum:1998ka}

\bibitem[{Epelbaum(2007)}]{Epelbaum:2007us}
Epelbaum, E. (2007).
\newblock {Four-nucleon force using the method of unitary transformation}.
\newblock \emph{Eur. Phys. J.} A34, 197--214.
\newblock \doi{10.1140/epja/i2007-10496-0}
\bibAnnoteFile{Epelbaum:2007us}

\bibitem[{Epelbaum(2006)}]{Epelbaum:2006eu}
Epelbaum, E. (2006).
\newblock {Four-nucleon force in chiral effective field theory}.
\newblock \emph{Phys. Lett.} B639, 456--461.
\newblock \doi{10.1016/j.physletb.2006.06.046}
\bibAnnoteFile{Epelbaum:2006eu}

\bibitem[{Bernard et~al.(2008)Bernard, Epelbaum, Krebs, and
  Mei{\ss}ner}]{Bernard:2007sp}
Bernard, V., Epelbaum, E., Krebs, H., and Mei{\ss}ner, U.-G. (2008).
\newblock {Subleading contributions to the chiral three-nucleon force. I.
  Long-range terms}.
\newblock \emph{Phys. Rev.} C77, 064004.
\newblock \doi{10.1103/PhysRevC.77.064004}
\bibAnnoteFile{Bernard:2007sp}

\bibitem[{Bernard et~al.(2011)Bernard, Epelbaum, Krebs, and
  Mei{\ss}ner}]{Bernard:2011zr}
Bernard, V., Epelbaum, E., Krebs, H., and Mei{\ss}ner, U.-G. (2011).
\newblock {Subleading contributions to the chiral three-nucleon force II:
  Short-range terms and relativistic corrections}.
\newblock \emph{Phys. Rev.} C84, 054001.
\newblock \doi{10.1103/PhysRevC.84.054001}
\bibAnnoteFile{Bernard:2011zr}

\bibitem[{Krebs et~al.(2012)Krebs, Gasparyan, and Epelbaum}]{Krebs:2012yv}
Krebs, H., Gasparyan, A., and Epelbaum, E. (2012).
\newblock {Chiral three-nucleon force at N$^4$LO I: Longest-range
  contributions}.
\newblock \emph{Phys. Rev.} C85, 054006.
\newblock \doi{10.1103/PhysRevC.85.054006}
\bibAnnoteFile{Krebs:2012yv}

\bibitem[{Krebs et~al.(2013)Krebs, Gasparyan, and Epelbaum}]{Krebs:2013kha}
Krebs, H., Gasparyan, A., and Epelbaum, E. (2013).
\newblock {Chiral three-nucleon force at N$^4$LO II: Intermediate-range
  contributions}.
\newblock \emph{Phys. Rev.} C87, 054007.
\newblock \doi{10.1103/PhysRevC.87.054007}
\bibAnnoteFile{Krebs:2013kha}

\bibitem[{de~Vries et~al.(2015{\natexlab{b}})de~Vries, Li, Mei{\ss}ner, Nogga,
  Epelbaum, and Kaiser}]{deVries:2015pza}
de~Vries, J., Li, N., Mei{\ss}ner, U.-G., Nogga, A., Epelbaum, E., and Kaiser,
  N. (2015{\natexlab{b}}).
\newblock {Parity violation in neutron capture on the proton: Determining the
  weak pion–nucleon coupling}.
\newblock \emph{Phys. Lett.} B747, 299--304.
\newblock \doi{10.1016/j.physletb.2015.05.074}
\bibAnnoteFile{deVries:2015pza}

\bibitem[{Gnech(2016)}]{Gnech:Thesis:2016}
Gnech, A. (2016).
\newblock \emph{Parity and Time Reversal Violation in Two Nucleons Systems}.
\newblock Master's thesis, University of Pisa, Pisa, Italy
\bibAnnoteFile{Gnech:Thesis:2016}

\bibitem[{Kaiser(2007)}]{Kaiser:2007zzb}
Kaiser, N. (2007).
\newblock {Parity-violating two-pion exchange nucleon-nucleon interaction}.
\newblock \emph{Phys. Rev.} C76, 047001.
\newblock \doi{10.1103/PhysRevC.76.047001}
\bibAnnoteFile{Kaiser:2007zzb}

\bibitem[{Epelbaum et~al.(2004)Epelbaum, Gl{\"o}ckle, and
  Mei{\ss}ner}]{Epelbaum:2003gr}
Epelbaum, E., Gl{\"o}ckle, W., and Mei{\ss}ner, U.-G. (2004).
\newblock {Improving the convergence of the chiral expansion for nuclear
  forces. 1. Peripheral phases}.
\newblock \emph{Eur. Phys. J.} A19, 125--137.
\newblock \doi{10.1140/epja/i2003-10096-0}
\bibAnnoteFile{Epelbaum:2003gr}

\bibitem[{Friar(1996)}]{Friar:1996tj}
Friar, J.~L. (1996).
\newblock {Dimensional regularization and nuclear potentials}.
\newblock \emph{Mod. Phys. Lett.} A11, 3043--3048.
\newblock \doi{10.1142/S0217732396003027}
\bibAnnoteFile{Friar:1996tj}

\bibitem[{Kaiser et~al.(1997)Kaiser, Brockmann, and Weise}]{Kaiser:1997mw}
Kaiser, N., Brockmann, R., and Weise, W. (1997).
\newblock {Peripheral nucleon-nucleon phase shifts and chiral symmetry}.
\newblock \emph{Nucl. Phys.} A625, 758--788.
\newblock \doi{10.1016/S0375-9474(97)00586-1}
\bibAnnoteFile{Kaiser:1997mw}

\bibitem[{Epelbaum et~al.(2015{\natexlab{b}})Epelbaum, Krebs, and
  Mei{\ss}ner}]{Epelbaum:2014efa}
Epelbaum, E., Krebs, H., and Mei{\ss}ner, U.-G. (2015{\natexlab{b}}).
\newblock {Improved chiral nucleon-nucleon potential up to
  next-to-next-to-next-to-leading order}.
\newblock \emph{Eur. Phys. J.} A51, 53.
\newblock \doi{10.1140/epja/i2015-15053-8}
\bibAnnoteFile{Epelbaum:2014efa}

\bibitem[{Krebs(2019)}]{Krebs:2019uvm}
Krebs, H. (2019).
\newblock {Electroweak Current Operators in Chiral Effective Field Theory,
  arXiv:1908.01538}.
\newblock In \emph{{9th International Workshop on Chiral Dynamics (CD18)
  Durham, NC, USA, September 17-21, 2018}}
\bibAnnoteFile{Krebs:2019uvm}

\bibitem[{Epelbaum(2019)}]{Epelbaum:2019jbv}
Epelbaum, E. (2019).
\newblock {Towards high-precision nuclear forces from chiral effective field
  theory, arXiv:1908.09349}.
\newblock In \emph{{6th International Conference Nuclear Theory in the
  Supercomputing Era (NTSE-2018) Daejeon, Korea, October 29-November 2, 2018}}
\bibAnnoteFile{Epelbaum:2019jbv}

\bibitem[{Epelbaum et~al.(2019)Epelbaum, Krebs, and Reinert}]{Epelbaum:2019kcf}
Epelbaum, E., Krebs, H., and Reinert, P. (2019).
\newblock {High-precision nuclear forces from chiral EFT: State-of-the-art,
  challenges and outlook, arXiv:1911.11875}
\bibAnnoteFile{Epelbaum:2019kcf}

\bibitem[{Maekawa et~al.(2000)Maekawa, Veiga, and van Kolck}]{Maekawa:2000bd}
Maekawa, C.~M., Veiga, J.~S., and van Kolck, U. (2000).
\newblock {The Nucleon anapole form-factor in chiral perturbation theory to
  subleading order}.
\newblock \emph{Phys. Lett.} B488, 167--174.
\newblock \doi{10.1016/S0370-2693(00)00851-0}
\bibAnnoteFile{Maekawa:2000bd}

\bibitem[{Epelbaum et~al.(2000)Epelbaum, Gl{\"o}ckle, and
  Mei{\ss}ner}]{Epelbaum:1999dj}
Epelbaum, E., Gl{\"o}ckle, W., and Mei{\ss}ner, U.-G. (2000).
\newblock {Nuclear forces from chiral Lagrangians using the method of unitary
  transformation. 2. The two nucleon system}.
\newblock \emph{Nucl. Phys.} A671, 295--331.
\newblock \doi{10.1016/S0375-9474(99)00821-0}
\bibAnnoteFile{Epelbaum:1999dj}

\bibitem[{Girlanda et~al.(2010)Girlanda, Pastore, Schiavilla, and
  Viviani}]{Girlanda:2010ya}
Girlanda, L., Pastore, S., Schiavilla, R., and Viviani, M. (2010).
\newblock {Relativity constraints on the two-nucleon contact interaction}.
\newblock \emph{Phys. Rev.} C81, 034005.
\newblock \doi{10.1103/PhysRevC.81.034005}
\bibAnnoteFile{Girlanda:2010ya}

\bibitem[{Kaplan et~al.(1999{\natexlab{b}})Kaplan, Savage, and
  Wise}]{Kaplan:1998sz}
Kaplan, D.~B., Savage, M.~J., and Wise, M.~B. (1999{\natexlab{b}}).
\newblock {A Perturbative calculation of the electromagnetic form-factors of
  the deuteron}.
\newblock \emph{Phys. Rev.} C59, 617--629.
\newblock \doi{10.1103/PhysRevC.59.617}
\bibAnnoteFile{Kaplan:1998sz}

\bibitem[{Savage and Springer(1998)}]{Savage:1998rx}
Savage, M.~J. and Springer, R.~P. (1998).
\newblock {Parity violation in effective field theory and the deuteron anapole
  moment}.
\newblock \emph{Nucl. Phys.} A644, 235--244.
\newblock \doi{10.1016/S0375-9474(98)80013-4, 10.1016/S0375-9474(99)00341-3}.
\newblock [Erratum: Nucl. Phys.A657,457(1999)]
\bibAnnoteFile{Savage:1998rx}

\bibitem[{Phillips et~al.(2009)Phillips, Schindler, and
  Springer}]{Phillips:2008hn}
Phillips, D.~R., Schindler, M.~R., and Springer, R.~P. (2009).
\newblock {An Effective-field-theory analysis of low-energy parity-violation in
  nucleon-nucleon scattering}.
\newblock \emph{Nucl. Phys.} A822, 1--19.
\newblock \doi{10.1016/j.nuclphysa.2009.02.011}
\bibAnnoteFile{Phillips:2008hn}

\bibitem[{Vanasse and David(2019)}]{Vanasse:2019fzl}
Vanasse, J. and David, A. (2019).
\newblock {Time-Reversal-Invariance Violation in the $N\!d$ System and
  Large-$N_C$, arXiv:1910.03133}
\bibAnnoteFile{Vanasse:2019fzl}

\bibitem[{'t~Hooft(1974)}]{tHooft:1973alw}
't~Hooft, G. (1974).
\newblock {A Planar Diagram Theory for Strong Interactions}.
\newblock \emph{Nucl. Phys.} B72, 461.
\newblock \doi{10.1016/0550-3213(74)90154-0}.
\newblock [,337(1973)]
\bibAnnoteFile{tHooft:1973alw}

\bibitem[{Witten(1979)}]{Witten:1979kh}
Witten, E. (1979).
\newblock {Baryons in the 1/n Expansion}.
\newblock \emph{Nucl. Phys.} B160, 57--115.
\newblock \doi{10.1016/0550-3213(79)90232-3}
\bibAnnoteFile{Witten:1979kh}

\bibitem[{Dashen et~al.(1995)Dashen, Jenkins, and Manohar}]{Dashen:1994qi}
Dashen, R.~F., Jenkins, E.~E., and Manohar, A.~V. (1995).
\newblock {Spin flavor structure of large N(c) baryons}.
\newblock \emph{Phys. Rev.} D51, 3697--3727.
\newblock \doi{10.1103/PhysRevD.51.3697}
\bibAnnoteFile{Dashen:1994qi}

\bibitem[{Kaplan and Savage(1996)}]{Kaplan:1995yg}
Kaplan, D.~B. and Savage, M.~J. (1996).
\newblock {The Spin flavor dependence of nuclear forces from large n QCD}.
\newblock \emph{Phys. Lett.} B365, 244--251.
\newblock \doi{10.1016/0370-2693(95)01277-X}
\bibAnnoteFile{Kaplan:1995yg}

\bibitem[{Kaplan and Manohar(1997)}]{Kaplan:1996rk}
Kaplan, D.~B. and Manohar, A.~V. (1997).
\newblock {The Nucleon-nucleon potential in the 1/N(c) expansion}.
\newblock \emph{Phys. Rev.} C56, 76--83.
\newblock \doi{10.1103/PhysRevC.56.76}
\bibAnnoteFile{Kaplan:1996rk}

\bibitem[{Nagels et~al.(1975)Nagels, Rijken, and de~Swart}]{Nagels:1975fb}
Nagels, M.~M., Rijken, T.~A., and de~Swart, J.~J. (1975).
\newblock {Baryon Baryon Scattering in an OBEP Approach. 1. Nucleon-Nucleon
  Scattering}.
\newblock \emph{Phys. Rev.} D12, 744.
\newblock \doi{10.1103/PhysRevD.12.744}
\bibAnnoteFile{Nagels:1975fb}

\bibitem[{Nagels et~al.(1977)Nagels, Rijken, and de~Swart}]{Nagels:1976xq}
Nagels, M.~M., Rijken, T.~A., and de~Swart, J.~J. (1977).
\newblock {Baryon Baryon Scattering in a One Boson Exchange Potential Approach.
  2. Hyperon-Nucleon Scattering}.
\newblock \emph{Phys. Rev.} D15, 2547.
\newblock \doi{10.1103/PhysRevD.15.2547}
\bibAnnoteFile{Nagels:1976xq}

\bibitem[{Sakurai(1960)}]{Sakurai:1960ju}
Sakurai, J.~J. (1960).
\newblock {Theory of strong interactions}.
\newblock \emph{Annals Phys.} 11, 1--48.
\newblock \doi{10.1016/0003-4916(60)90126-3}
\bibAnnoteFile{Sakurai:1960ju}

\bibitem[{Barton(1961)}]{Barton:1961eg}
Barton, G. (1961).
\newblock {Notes on the static parity nonconserving internucleon potential}.
\newblock \emph{Nuovo Cim.} 19, 512--527.
\newblock \doi{10.1007/BF02733247}
\bibAnnoteFile{Barton:1961eg}

\bibitem[{Holstein(1981)}]{Holstein:1981cg}
Holstein, B.~R. (1981).
\newblock {Nuclear parity violation parameter $h_\rho^{(1)\prime}$}.
\newblock \emph{Phys. Rev.} D23, 1618--1623.
\newblock \doi{10.1103/PhysRevD.23.1618}
\bibAnnoteFile{Holstein:1981cg}

\bibitem[{Herczeg(1987)}]{Herczeg:1987gp}
Herczeg, P. (1987).
\newblock {T violating effects in neutron physics and cp violation in gauge
  models}.
\newblock In \emph{{Workshop on Time Reversal Invariance in Neutron Physics
  Chapel Hill, North Carolina, April 17-19, 1987}}
\bibAnnoteFile{Herczeg:1987gp}

\bibitem[{Schiavilla et~al.(2008)Schiavilla, Viviani, Girlanda, Kievsky, and
  Marcucci}]{Schiavilla:2008ic}
Schiavilla, R., Viviani, M., Girlanda, L., Kievsky, A., and Marcucci, L.~E.
  (2008).
\newblock {Neutron spin rotation in n-polarized - d scattering}.
\newblock \emph{Phys. Rev.} C78, 014002.
\newblock \doi{10.1103/PhysRevC.78.014002, 10.1103/PhysRevC.83.029902}.
\newblock [Erratum: Phys. Rev.C83,029902(2011)]
\bibAnnoteFile{Schiavilla:2008ic}

\bibitem[{Haxton and Henley(1983)}]{Haxton:1983dq}
Haxton, W.~C. and Henley, E.~M. (1983).
\newblock {ENHANCED T VIOLATING NUCLEAR MOMENTS}.
\newblock \emph{Phys. Rev. Lett.} 51, 1937.
\newblock \doi{10.1103/PhysRevLett.51.1937}
\bibAnnoteFile{Haxton:1983dq}

\bibitem[{Gudkov et~al.(1993)Gudkov, He, and McKellar}]{Gudkov:1992yc}
Gudkov, V.~P., He, X.-G., and McKellar, B. H.~J. (1993).
\newblock {On the CP odd nucleon potential}.
\newblock \emph{Phys. Rev.} C47, 2365--2368.
\newblock \doi{10.1103/PhysRevC.47.2365}
\bibAnnoteFile{Gudkov:1992yc}

\bibitem[{Towner and Hayes(1994)}]{Towner:1994qe}
Towner, I.~S. and Hayes, A.~C. (1994).
\newblock {P, T violating nuclear matrix elements in the one meson exchange
  approximation}.
\newblock \emph{Phys. Rev.} C49, 2391--2397.
\newblock \doi{10.1103/PhysRevC.49.2391}
\bibAnnoteFile{Towner:1994qe}

\bibitem[{Song et~al.(2013)Song, Lazauskas, and Gudkov}]{Song:2012yh}
Song, Y.-H., Lazauskas, R., and Gudkov, V. (2013).
\newblock {Nuclear electric dipole moment of three-body systems}.
\newblock \emph{Phys. Rev.} C87, 015501.
\newblock \doi{10.1103/PhysRevC.87.015501}
\bibAnnoteFile{Song:2012yh}

\bibitem[{Yamanaka(2017)}]{Yamanaka:2016umw}
Yamanaka, N. (2017).
\newblock {Review of the electric dipole moment of light nuclei}.
\newblock \emph{Int. J. Mod. Phys.} E26, 1730002.
\newblock \doi{10.1142/S0218301317300028}
\bibAnnoteFile{Yamanaka:2016umw}

\bibitem[{Blyth et~al.(2018)}]{Blyth:2018aon}
Blyth, D. et~al. (2018).
\newblock {First Observation of $P$-odd $\gamma$ Asymmetry in Polarized Neutron
  Capture on Hydrogen}.
\newblock \emph{Phys. Rev. Lett.} 121, 242002.
\newblock \doi{10.1103/PhysRevLett.121.242002}
\bibAnnoteFile{Blyth:2018aon}

\bibitem[{Schiavilla et~al.(2004)Schiavilla, Carlson, and
  Paris}]{Schiavilla:2004wn}
Schiavilla, R., Carlson, J., and Paris, M.~W. (2004).
\newblock {Parity violating interaction effects in the np system}.
\newblock \emph{Phys. Rev.} C70, 044007.
\newblock \doi{10.1103/PhysRevC.70.044007}
\bibAnnoteFile{Schiavilla:2004wn}

\bibitem[{Hyun et~al.(2001)Hyun, Park, and Min}]{Hyun:2001yg}
Hyun, C.~H., Park, T.-S., and Min, D.-P. (2001).
\newblock {Asymmetry in polarized n + p ---> d + gamma}.
\newblock \emph{Phys. Lett.} B516, 321--326.
\newblock \doi{10.1016/S0370-2693(01)00917-0}
\bibAnnoteFile{Hyun:2001yg}

\bibitem[{Schiavilla et~al.(2003)Schiavilla, Carlson, and
  Paris}]{Schiavilla:2002uc}
Schiavilla, R., Carlson, J., and Paris, M.~W. (2003).
\newblock {Parity violating interactions and currents in the deuteron}.
\newblock \emph{Phys. Rev.} C67, 032501.
\newblock \doi{10.1103/PhysRevC.67.032501}
\bibAnnoteFile{Schiavilla:2002uc}

\bibitem[{Desplanques(1975)}]{Desplanques:1975vle}
Desplanques, B. (1975).
\newblock {Study of parity-violating effects in the neutron capture $n + p
  \rightarrow d + \gamma$}.
\newblock \emph{Nucl. Phys.} A242, 423--428.
\newblock \doi{10.1016/0375-9474(75)90105-0}
\bibAnnoteFile{Desplanques:1975vle}

\bibitem[{McKellar(1975)}]{McKellar:1975iy}
McKellar, B. H.~J. (1975).
\newblock {Parity Nonconservation in the Low-Energy Nucleon-Nucleon System:
  Evidence for an Isotensor Weak Interaction?}
\newblock \emph{Nucl. Phys.} A254, 349--352.
\newblock \doi{10.1016/0375-9474(75)90221-3}
\bibAnnoteFile{McKellar:1975iy}

\bibitem[{Desplanques(1980)}]{Desplanques:1980pa}
Desplanques, B. (1980).
\newblock {Parity nonconserving nuclear forces}.
\newblock \emph{Nucl. Phys.} A335, 147--167.
\newblock \doi{10.1016/0375-9474(80)90174-8}
\bibAnnoteFile{Desplanques:1980pa}

\bibitem[{Eversheim et~al.(1991)}]{Eversheim:1991tg}
Eversheim, P.~D. et~al. (1991).
\newblock {Parity violation in proton proton scattering at 13.6-MeV}.
\newblock \emph{Phys. Lett.} B256, 11--14.
\newblock \doi{10.1016/0370-2693(91)90209-9}
\bibAnnoteFile{Eversheim:1991tg}

\bibitem[{Kistryn et~al.(1987)}]{Kistryn:1987tq}
Kistryn, S. et~al. (1987).
\newblock {Precision Measurement of Parity Violation in Proton Proton
  Scattering at 45-{MeV}}.
\newblock \emph{Phys. Rev. Lett.} 58, 1616.
\newblock \doi{10.1103/PhysRevLett.58.1616}
\bibAnnoteFile{Kistryn:1987tq}

\bibitem[{Berdoz et~al.(2003)}]{Berdoz:2002sn}
Berdoz, A.~R. et~al. (2003).
\newblock {Parity violation in proton proton scattering at 221-MeV}.
\newblock \emph{Phys. Rev.} C68, 034004.
\newblock \doi{10.1103/PhysRevC.68.034004}
\bibAnnoteFile{Berdoz:2002sn}

\bibitem[{Hoferichter et~al.(2015)Hoferichter, Ruiz~de Elvira, Kubis, and
  Mei{\ss}ner}]{Hoferichter:2015tha}
Hoferichter, M., Ruiz~de Elvira, J., Kubis, B., and Mei{\ss}ner, U.-G. (2015).
\newblock {Matching pion-nucleon Roy-Steiner equations to chiral perturbation
  theory}.
\newblock \emph{Phys. Rev. Lett.} 115, 192301.
\newblock \doi{10.1103/PhysRevLett.115.192301}
\bibAnnoteFile{Hoferichter:2015tha}

\bibitem[{Carlson et~al.(2002)Carlson, Schiavilla, Brown, and
  Gibson}]{Carlson:2001ma}
Carlson, J., Schiavilla, R., Brown, V.~R., and Gibson, B.~F. (2002).
\newblock {Parity violating interaction effects 1: The Longitudinal asymmetry
  in pp elastic scattering}.
\newblock \emph{Phys. Rev.} C65, 035502.
\newblock \doi{10.1103/PhysRevC.65.035502}
\bibAnnoteFile{Carlson:2001ma}

\bibitem[{Desplanques(2001)}]{Desplanques:2000ej}
Desplanques, B. (2001).
\newblock {About the parity nonconserving asymmetry in $n + p \rightarrow d +
  \gamma$}.
\newblock \emph{Phys. Lett.} B512, 305--313.
\newblock \doi{10.1016/S0370-2693(01)00713-4}
\bibAnnoteFile{Desplanques:2000ej}

\bibitem[{Song et~al.(2011)Song, Lazauskas, and Gudkov}]{Song:2010sz}
Song, Y.-H., Lazauskas, R., and Gudkov, V. (2011).
\newblock {Parity violation in low energy neutron deuteron scattering}.
\newblock \emph{Phys. Rev.} C83, 015501.
\newblock \doi{10.1103/PhysRevC.83.015501}
\bibAnnoteFile{Song:2010sz}

\bibitem[{Walecka(1995)}]{Walecka:1995mi}
Walecka, J.~D. (1995).
\newblock \emph{{Theoretical nuclear and subnuclear physics}}, vol.~16
\bibAnnoteFile{Walecka:1995mi}

\bibitem[{Yamanaka and Hiyama(2015)}]{Yamanaka:2015qfa}
Yamanaka, N. and Hiyama, E. (2015).
\newblock {Enhancement of the CP-odd effect in the nuclear electric dipole
  moment of $^6$Li}.
\newblock \emph{Phys. Rev.} C91, 054005.
\newblock \doi{10.1103/PhysRevC.91.054005}
\bibAnnoteFile{Yamanaka:2015qfa}

\bibitem[{Lebedev et~al.(2004)Lebedev, Olive, Pospelov, and
  Ritz}]{Lebedev:2004va}
Lebedev, O., Olive, K.~A., Pospelov, M., and Ritz, A. (2004).
\newblock {Probing CP violation with the deuteron electric dipole moment}.
\newblock \emph{Phys. Rev.} D70, 016003.
\newblock \doi{10.1103/PhysRevD.70.016003}
\bibAnnoteFile{Lebedev:2004va}

\bibitem[{Hoferichter et~al.(2016)Hoferichter, Ruiz~de Elvira, Kubis, and
  Mei{\ss}ner}]{Hoferichter:2015hva}
Hoferichter, M., Ruiz~de Elvira, J., Kubis, B., and Mei{\ss}ner, U.-G. (2016).
\newblock {Roy–Steiner-equation analysis of pion–nucleon scattering}.
\newblock \emph{Phys. Rept.} 625, 1--88.
\newblock \doi{10.1016/j.physrep.2016.02.002}
\bibAnnoteFile{Hoferichter:2015hva}

\bibitem[{Siemens et~al.(2017)Siemens, Ruiz~de Elvira, Epelbaum, Hoferichter,
  Krebs, Kubis et~al.}]{Siemens:2016jwj}
Siemens, D., Ruiz~de Elvira, J., Epelbaum, E., Hoferichter, M., Krebs, H.,
  Kubis, B., et~al. (2017).
\newblock {Reconciling threshold and subthreshold expansions for pion–nucleon
  scattering}.
\newblock \emph{Phys. Lett.} B770, 27--34.
\newblock \doi{10.1016/j.physletb.2017.04.039}
\bibAnnoteFile{Siemens:2016jwj}

\bibitem[{Guo and Seng(2019)}]{Guo:2018aiq}
Guo, F.-K. and Seng, C.-Y. (2019).
\newblock {Effective Field Theory in The Study of Long Range Nuclear Parity
  Violation on Lattice}.
\newblock \emph{Eur. Phys. J.} C79, 22.
\newblock \doi{10.1140/epjc/s10052-018-6529-y}
\bibAnnoteFile{Guo:2018aiq}

\bibitem[{Bass et~al.(2009)}]{Bass:2009fs}
Bass, C.~D. et~al. (2009).
\newblock {A Liquid helium target system for a measurement of parity violation
  in neutron spin rotation}.
\newblock \emph{Nucl. Instrum. Meth.} A612, 69--82.
\newblock \doi{10.1016/j.nima.2009.10.055}
\bibAnnoteFile{Bass:2009fs}

\bibitem[{Lang et~al.(1985)Lang, Maier, Muller, Nessi-Tedaldi, Roser, Simonius
  et~al.}]{Lang:1985jv}
Lang, J., Maier, T., Muller, R., Nessi-Tedaldi, F., Roser, T., Simonius, M.,
  et~al. (1985).
\newblock {Parity nonconservation in elastic p alpha scattering and the
  determination of the weak meson - nucleon coupling constants}.
\newblock \emph{Phys. Rev. Lett.} 54, 170--173.
\newblock \doi{10.1103/PhysRevLett.54.170}
\bibAnnoteFile{Lang:1985jv}

\bibitem[{Guidoboni et~al.(2016)}]{Guidoboni:2016bdn}
Guidoboni, G. et~al. (2016).
\newblock {How to Reach a Thousand-Second in-Plane Polarization Lifetime with
  0.97-GeV/c Deuterons in a Storage Ring}.
\newblock \emph{Phys. Rev. Lett.} 117, 054801.
\newblock \doi{10.1103/PhysRevLett.117.054801}
\bibAnnoteFile{Guidoboni:2016bdn}

\bibitem[{Yamanaka et~al.(2019)Yamanaka, Yamada, and Funaki}]{Yamanaka:2019vec}
Yamanaka, N., Yamada, T., and Funaki, Y. (2019).
\newblock {Nuclear electric dipole moment in the cluster model with a triton:
  $^7$Li and $^{11}$B}.
\newblock \emph{Phys. Rev.} C100, 055501.
\newblock \doi{10.1103/PhysRevC.100.055501}
\bibAnnoteFile{Yamanaka:2019vec}

\bibitem[{Kabir(1982)}]{Kabir:1981tp}
Kabir, P.~K. (1982).
\newblock {Test of $T$ Invariance in Neutron Optics}.
\newblock \emph{Phys. Rev.} D25, 2013.
\newblock \doi{10.1103/PhysRevD.25.2013}
\bibAnnoteFile{Kabir:1981tp}

\bibitem[{Stodolsky(1982)}]{Stodolsky:1981vn}
Stodolsky, L. (1982).
\newblock {Parity Violation in Threshold Neutron Scattering}.
\newblock \emph{Nucl. Phys.} B197, 213--227.
\newblock \doi{10.1016/0550-3213(82)90287-5}
\bibAnnoteFile{Stodolsky:1981vn}

\bibitem[{Bunakov and Gudkov(1983)}]{Bunakov:1982is}
Bunakov, V.~E. and Gudkov, V.~P. (1983).
\newblock {Parity Violation and Related Effects in Neutron Induced Reactions}.
\newblock \emph{Nucl. Phys.} A401, 93--116.
\newblock \doi{10.1016/0375-9474(83)90338-X}
\bibAnnoteFile{Bunakov:1982is}

\bibitem[{Gudkov(1992)}]{Gudkov:1991qg}
Gudkov, V.~P. (1992).
\newblock {On CP violation in nuclear reactions}.
\newblock \emph{Phys. Rept.} 212, 77--105.
\newblock \doi{10.1016/0370-1573(92)90121-F}
\bibAnnoteFile{Gudkov:1991qg}

\bibitem[{Bowman and Gudkov(2014)}]{Bowman:2014fca}
Bowman, J.~D. and Gudkov, V. (2014).
\newblock {Search for time reversal invariance violation in neutron
  transmission}.
\newblock \emph{Phys. Rev.} C90, 065503.
\newblock \doi{10.1103/PhysRevC.90.065503}
\bibAnnoteFile{Bowman:2014fca}

\bibitem[{Shimizu et~al.(2017)}]{Shimizu:2017zan}
Shimizu, H. et~al. (2017).
\newblock {Discrete Symmetry Tests In Neutron-induced Compound States}.
\newblock \emph{PoS} INPC2016, 187.
\newblock \doi{10.22323/1.281.0187}
\bibAnnoteFile{Shimizu:2017zan}

\bibitem[{Gudkov and Shimizu(2018)}]{Gudkov:2017vqn}
Gudkov, V. and Shimizu, H.~M. (2018).
\newblock {Nuclear spin dependence of time reversal invariance violating
  effects in neutron scattering}.
\newblock \emph{Phys. Rev.} C97, 065502.
\newblock \doi{10.1103/PhysRevC.97.065502}
\bibAnnoteFile{Gudkov:2017vqn}

\bibitem[{Uzikov and Haidenbauer(2016)}]{Uzikov:2016lsc}
Uzikov, {\relax Yu}.~N. and Haidenbauer, J. (2016).
\newblock {Polarized proton-deuteron scattering as a test of time-reversal
  invariance}.
\newblock \emph{Phys. Rev.} C94, 035501.
\newblock \doi{10.1103/PhysRevC.94.035501}
\bibAnnoteFile{Uzikov:2016lsc}

\bibitem[{Murata et~al.(2017)}]{Murata:2017ifq}
Murata, J. et~al. (2017).
\newblock {The MTV Experiment: from T-violation to Lorentz-violation}.
\newblock \emph{PoS} INPC2016, 185.
\newblock \doi{10.22323/1.281.0185}
\bibAnnoteFile{Murata:2017ifq}

\end{thebibliography}

\bibliographystyle{frontiersinSCNS_ENG_HUMS_sort_by_appearance}

\end{document}